\documentclass[seceq,preprint]{ptptex}
\usepackage{wrapft}
\usepackage{graphicx}
\usepackage{mathrsfs}

\newcommand{\be}{\begin{equation}}
\newcommand{\ee}{\end{equation}}
\newcommand{\bea}{\begin{eqnarray}}
\newcommand{\eea}{\end{eqnarray}}
\newcommand{\bean}{\begin{eqnarray*}}
\newcommand{\eean}{\end{eqnarray*}}
\newcommand{\ba}{\begin{array}}
\newcommand{\ea}{\end{array}}

\newcommand{\slashm}[1]{\ooalign{\hfil/\hfil\crcr$#1$}}
\newcommand{\non}{\nonumber}
\newcommand{\ra}{\rightarrow}

\newcommand{\aqt}{{\mathscr{A}_{TT}(Q_T)}}
\newcommand{\aqtn}{\mathscr{A}_{TT}}
\newcommand{\aqtnn}{\mathscr{A}_{TT}^{\rm NLL+LO}(Q_T)} 
\newcommand{\mR}{{\cal R}}
\newcommand{\mD}{\mathscr{D}}
\newcommand{\mW}{\mathscr{W}}
\newcommand{\mS}{\mathcal{S}}
\newcommand{\mU}{\mathcal{U}}
\newcommand{\mI}{\mathcal{I}}
\newcommand{\mN}{\mathscr{N}}
\newcommand{\lsim}{\raisebox{-0.07cm   }
{$\, \stackrel{<}{{\scriptstyle\sim}}\, $}}
\newcommand{\grtsim}{\raisebox{-0.07cm   }
{$\, \stackrel{>}{{\scriptstyle\sim}}\, $}}
\newcommand{\bc}{\begin{center}}
\newcommand{\ec}{\end{center}}
\pubinfo{Vol.118, No.4, October 2007 }
\notypesetlogo  

\title{
Transversely Polarized Drell-Yan Process\\ 
and Soft Gluon Resummation in QCD
}

\author{
Hiroyuki \textsc{Kawamura},$^{1}$
Jiro \textsc{Kodaira}$^{2,}$\footnote{Deceased.}
and
Kazuhiro \textsc{Tanaka}$^{3}$
}

\inst{
$^1$Radiation Laboratory, RIKEN, Wako 351-0198, Japan\\
$^2$Theory Division, KEK, Tsukuba 305-0801, Japan\\
$^3$Department of Physics, Juntendo University, Inba, Chiba 270-1695, Japan
}

\recdate{
August 20, 2007}

\abst{
We calculate the transverse-momentum $Q_T$ spectrum of the dilepton in 
the transversely polarized Drell-Yan process
on the basis of the factorization theorem in QCD. 
We take into account universal logarithmically enhanced corrections
in edge region of phase space by resumming multiple soft-gluon 
emissions to all orders in the small $Q_T$ region.
}

\begin{document}
\maketitle
\section{Introduction}

The nucleon structure appearing 
in high energy processes has been studied for a long time 
on the basis of quantum chromodynamics (QCD).  
The basic theoretical framework has been developed as the 
\lq\lq factorization theorem\rq\rq\ in QCD~\cite{CSS:89}.
As a result of factorization, the physical quantities (cross sections)
are obtained as a convolution of the short- and long-distance parts.
The former contains all the dependence on the hard scale, and the latter 
is controlled by the nonperturbative dynamics of QCD.
We can apply perturbation theory to the short-distance part,
thanks to the asymptotically free nature of QCD.
The long-distance parts, however, can be determined only by experiments
or some nonperturbative method, such as lattice QCD.
The importance and advantage of perturbative QCD based on the
factorization theorem reside in the fact that this theorem allows us to
define the long-distance parts 
as process-independent universal objects, which are 
represented unambiguously as nucleon matrix elements
of the operators of quarks and/or gluons and
are often much simpler than the original quantities. 

For spin-independent processes, the many experiments performed 
to this time have helped 
us to determine the long-distance, nonperturbative parts as
``parton distributions'' inside the nucleon~\cite{Alt94},
and we have obtained a consistent understanding of 
the perturbative and nonperturbative dynamics, 
except in \lq\lq edge regions\rq\rq of phase space~\cite{CSS:89}. 
For spin-dependent processes, however, our understanding 
is still poor, and many questions remain unanswered\cite{Bass,WV07}. 
Therefore, to understand spin-dependent processes and 
the spin structure of nucleons through them is an 
important problem.
Furthermore, spin-dependent quantities are, in general,
believed to be quite sensitive to the structure of interactions among particles. 
These are the reasons for the great amount of activity recently
in high-energy spin physics.

It is expected that a number of ongoing polarization
experiments, such as RHIC-Spin~\cite{PHENIX,STAR}, using
polarized proton-proton collisions~\cite{RHIC}, 
HERMES~\cite{HERMES} and COMPASS~\cite{COMPASS}, 
using lepton scattering off polarized protons, etc.,
will provide important experimental data to reveal spin-dependent phenomena
associated with the structure of nucleons. 
We also expect that data from future polarization
experiments using proton-proton collisions 
at J-PARC~\cite{Dutta}, proton-antiproton collisions at GSI,~\cite{PAX} etc
will provide useful information.
Therefore, it is important and interesting to investigate various processes 
to be studied in those experiments.

Among the many spin-dependent processes, the polarized Drell-Yan (DY) process
plays a unique role. Indeed,  
transversity~\cite{RS,JJ:92,BDR:02} is one of the characteristic
observables which can be measured in DY using a transversely polarized beam. 
The transversity $\delta q(x)$ is a twist-2 parton distribution
associated with the probability distribution of 
transversely polarized quarks inside transversely polarized nucleons, 
i.e., the partonic structure of nucleons which is complementary to that
represented by the other twist-2 distributions, such as the familiar 
density and helicity distributions $q(x)$ and $\Delta q(x)$.
However, $\delta q(x)$ is not yet well understood,
because $\delta q(x)$ cannot be measured in inclusive deep inelastic scattering (DIS) 
due to its ``chiral-odd'' nature~\cite{RS,JJ:92,BDR:02}. 
In this paper, we focus on the transversely polarized Drell-Yan (tDY) process,
$p^{\uparrow} p^{\uparrow}\rightarrow l^+l^-X$,
producing the dilepton $l^+l^-$ with an invariant mass $Q \gg
\Lambda_{\rm QCD}$ in the final state,
as a process that is likely to allow investigation of 
the transversity $\delta q(x)$. 
Our aim is to calculate the QCD corrections to the tDY cross section 
on the basis of the QCD factorization framework, establishing control over 
the large higher-order corrections near edge region of phase space.

Simpleminded 
calculations of QCD corrections to the DY cross section 
suffer from ultraviolet (UV) and infrared (IR) divergences due to 
the loops associated with (massless) quarks and gluons. 
The UV divergence can be regularized and renormalized straightforwardly 
using a standard procedure, and therefore it does not pose any problem. 
By contrast, the IR divergence is more intricate and 
must be treated using 
the factorization theorem~\cite{CSS:89}: 
Introducing an appropriate IR regulator, one has to confirm that the IR divergences 
from the different diagrams cancel  
(``Kinoshita-Lee-Nauenberg cancellation''~\cite{KLN}) or 
are completely factorized into parton distribution functions. 
Such a calculation has been performed for ${\cal O}(\alpha_s)$ 
corrections to tDY using various IR regularization schemes.
For example, the one-loop calculation of the tDY cross section 
was done by Vogelsang and Weber~\cite{VW} using the massive gluon scheme.
The same calculation was also done using the dimensional reduction scheme~\cite{CKM}. 
The relations 
among the results obtained with different schemes are discussed in Ref.~\citen{K1}.
The result in the dimensional regularization scheme was obtained~\cite{V} 
by using the scheme transformation relation (see also Ref.~\citen{MSSV}).
We also note that all these works~\cite{VW,CKM,K1,V,MSSV} 
investigating tDY treat
the case in which the transverse momentum $Q_T$ of the final
dilepton $l^+l^-$ with respect to the beam axis is unobserved 
(integrated).\footnote{
Ref.~\citen{VW} treats also the case with $Q_T$ observed, restricting $Q_T$ to 
large values, which makes the IR regulator irrelevant for calculation 
of the ${\cal O}(\alpha_s)$ contributions to the cross section.
}

When a calculation of such QCD corrections is performed, one 
encounters an interesting technical problem in the case of transverse polarization. 
As is well known~\cite{RS}, the cross section depends on the azimuthal 
angle, $\phi$, of the observed particle, and its dependence in tDY 
is of the form $\cos (2 \phi)$.  
Therefore, we must keep the azimuthal angle dependence of the cross
section in the case of transverse polarization,
which makes it difficult to perform the phase space integrals in
the higher-order calculations.
This difficulty becomes much more severe if dimensional regularization
is employed. One cannot use the techniques developed for the unpolarized and
longitudinally polarized DY.
A related problem in the dimensional regularization scheme 
has been discussed by Kamal~\cite{K2}, 
but an explicit result was not given.
Recently, Mukherjee et al.~\cite{MSV} proposed a new technique 
which allows one to overcome a similar problem in their calculation 
of prompt photon production.

In this paper, we employ the dimensional regularization scheme
and discuss our approach to directly integrate out the phase space in 
$D$ dimensions. 
The final result for the tDY cross section up to ${\cal O}(\alpha_s)$ 
accuracy in the $\overline{\rm MS}$ scheme has been reported in 
a previous paper~\cite{KKST06}, and here we present the details of our calculation.
We believe that our explicit calculation in the dimensional regularization 
is interesting and useful. 
Once the above-mentioned complication associated with the transverse
polarization is worked out, the dimensional regularization provides us with
a manifestly gauge-invariant and the most transparent framework to calculate
the QCD corrections for both the $Q_T$-unobserved and $Q_T$-observed cases. 
For the latter case, in particular, we are able to derive the tDY cross section up 
to ${\cal O}(\alpha_s)$ without any restriction on $Q_T$, explicitly isolating the terms 
that diverge
as $\ln (Q^2/Q_T^2)/Q_T^2$, $1 /Q_T^2$, and $\delta (Q_T^2)$, as $Q_T \rightarrow 0$.
The resulting ``$Q_T$-differential'' cross section
gives the leading order (LO) QCD prediction in the large $Q_T$ region, 
in which $Q_T \sim Q$, where the lepton-pair production via the DY 
mechanism has to be accompanied by the radiation of at least one recoiling ``hard'' 
gluon, and for this reason the fixed-order truncation of perturbation theory 
is effective.
On the other hand, the unlimited growth of the singular terms of the form 
$\ln (Q^2/Q_T^2)/Q_T^2$
in the cross section at small $Q_T$ is associated with the recoil 
from ``soft'' gluon radiation, and this implies that we cannot truncate 
perturbation theory 
and have to calculate the higher-order QCD corrections beyond the one-loop 
level in this edge region of phase space.

The higher-order corrections that control the small $Q_T$ behavior of the cross section
can be taken into account through the soft gluon resummation 
(``transverse-momentum ($Q_T$) resummation''). 
As the next step, we derive the soft gluon resummation for our
$Q_T$-differential tDY cross section,
so that we can extend our LO QCD prediction at large $Q_T$ to the entire range of $Q_T$.
Below, we summarize the development of the soft gluon resummation.

A small value of $Q_T$ ($\ll Q$)
implies that there exists a new scale in the problem,
and as a result, in the perturbative calculation, there appear terms containing
large logarithms of $Q^2/Q_T^2$:
The coefficient of $\alpha_s^n (Q^2)$ includes a factor of $1/Q_T^2$ multiplying 
a series of logarithms of the form $\ln^m (Q^2/Q_T^2)$, with $m=0,1, \ldots, 2n-1$.
The pattern of these logarithmic terms is characteristic of a theory 
with massless vector bosons, such as QCD and QED,
and is produced by recoil from the radiation of gluons and photons.
The first work dealing with these enhanced ``recoil logarithms'' in QCD 
was carried out by Dokshitzer, Dyakonov and Troyan~\cite{DDT}.
Their result corresponds to the leading logarithmic (LL) resummation in momentum space, 
i.e., the resummation of the terms $\alpha_s^n \ln^{2n-1} (Q^2/Q_T^2)$
to all orders in $\alpha_s$. The level of this approximation is the same
as that in Sudakov's QED analysis~\cite{Sudakov}, and the result was derived by imposing 
the ``strong ordering'' of the gluon's transverse momenta in the relevant 
Feynman diagrams; the strong ordering excessively constrains the phase
space of the emitted soft gluons, and thus results in 
the transverse-momentum conservation being ignored.
To take into account the transverse-momentum conservation, 
Parisi and Petronzio~\cite{PP} developed a formulation in the space of 
the impact parameter, $b$, which is the Fourier conjugate of the $Q_T$ space.
The relation between the $Q_T$-space and the $b$-space approaches
is analyzed in Ref.~\citen{ESRW}.
That work clarifies the impact of the transverse-momentum conservation 
on the subleading logarithmic terms.
The general form of the soft gluon resummation to all orders of logarithms 
was proved and formulated in the $b$ space approach by Collins and Soper~\cite{CS}.
The universal two-loop anomalous dimension in the resummed cross sections,
which is necessary for the next-to-leading logarithmic (NLL)
analysis of {\it all} relevant processes, was first calculated in Ref.~\citen{KT}. 
This result was based on a plausible assumption, and
Davies et al.~\cite{DWS} confirmed the result using an explicit
calculation of the unpolarized DY process up to order $\alpha_s^2$.
They also carried out a phenomenological analysis, but only for small $Q_T$.

Advanced $b$-space formulations of the soft gluon resummation,
which are suitable for phenomenological analyses of DY in all $Q_T$ regions, 
were developed in the following works. Altarelli et al.~\cite{AEGM} 
proposed a recipe to include the NLL resummation effects into 
the unpolarized DY cross section, but their approach 
is somewhat naive; an application of this approach to the longitudinally
polarized DY process is considered in Ref.~\citen{W}. 
Presently, the formulation which is valid to all orders of 
logarithms developed by Collins, Soper and Sterman (CSS)~\cite{CSS} 
is regarded as standard. Recently, de Florian and Grazzini~\cite{dG} 
derived a universal expression for the CSS resummation formula up to 
next-to-next-to-leading logarithmic (NNLL) accuracy, 
which is applicable to DY production, electroweak boson production, 
Higgs boson production, etc. 
Further developments of formulations and their applications to phenomenology 
are underway
(see Refs.~\citen{BCFG,KS,ResBos} and references therein).\footnote{
Besides the $Q_T$ resummation discussed in this paper,
another kind of soft gluon resummation, the so-called ``threshold resummation,''
has been developed 
for the purpose of resumming the large higher-order corrections 
near the threshold of partonic scattering~\cite{THRESH,CMNT,THRESH2,SSVY:05}.
Also, the joint resummation formalism has been devised to perform
the $Q_T$ and threshold resummations simultaneously~\cite{Li,KLSV,KSVKSV}.}
The extension of the formulations to DY with transversely polarized beams
has been carried out by the present authors, and the main results are reported in 
Refs.~\citen{KKST06} and \citen{KKT07}.\footnote{See Ref.~\citen{Boer} for a treatment 
within the LL-level resummation. The extension to polarized semi-inclusive DIS
has recently been discussed~\cite{KNV:06}.
}

The general formulation of soft gluon resummation used to obtain our results 
is only briefly described in Refs.~\citen{KKST06} and \citen{KKT07}.
In this paper, we discuss the general form of the $b$-space resummation formula 
in detail from a modern viewpoint, emphasizing its theoretical basis as
well as its physical content. We also demonstrate that the $b$ space can 
be divided into three distinct regions, associated with different distance scales; 
each of these three regions has to be treated differently,
and the results for the three regions must eventually be combined in a consistent manner.
This point is addressed in the original papers on the $b$-space approach~\cite{PP,CSS},
but here we present a more systematic explanation of this point, which is
important to ensure the maximal applicability of the resummation formalisms to a
wide range of processes, including spin-dependent processes at moderate
and high energies.
Then, we apply the resummation formalism to our tDY $Q_T$-differential cross section,
which allows us to include all orders resummation of
the logarithmically-enhanced contributions for small $Q_T$ due to multiple 
emissions of soft gluons in QCD.
We perform the corresponding resummation up to NLL accuracy, 
and the result is combined with the fixed-order LO cross section 
that controls the large $Q_T$ region, yielding 
a tDY cross section with uniform accuracy over the entire range of $Q_T$.
We also explicitly derive the important properties of our cross section, 
which were briefly addressed in previous papers~\cite{KKST06,KKT07}.
For example, our cross section satisfies 
the unitarity constraint exactly; the NLL resummation 
is controlled completely by the ``saddle point'' in the $b$ space 
in the asymptotic regime, $Q \gg \Lambda_{\rm QCD}$, $Q_T \approx 0$.
As an application of our results, we calculate the dilepton $Q_T$ spectrum and 
the cross section asymmetry as functions of $Q_T$. 
These quantities are to be observed in polarized 
$pp$ collisions with large CM energy $\sqrt{S} \gtrsim 200$ GeV at RHIC. 
We also present the results to be observed in polarized 
$pp$ collisions with moderate $\sqrt{S} = 10$ GeV at J-PARC. 
We demonstrate that, for both RHIC and J-PARC energies,
the soft gluon resummation is crucial for making a reliable 
QCD prediction for the small $Q_T$ region, where the bulk of dileptons is produced. 

To our great sorrow, one of the present authors, Jiro Kodaira, 
died on September 16, 2006.
The present work was performed by the three authors jointly, and
many parts of this paper are based on our ``collaboration notes,'' 
originally written by Jiro. To complete this paper, we have had to 
reorganize and expand those notes without the direct assistance of Jiro, but 
we have been guided by his style in the approach to 
the problem, which we learned through our association with him for more than ten years.

The rest of the paper is organized as follows. 
Sections 2-4 are devoted to the calculation of the tDY cross section up to 
${\cal O}(\alpha_s)$ in the dimensional regularization.
Section 2 is mainly introductory, explaining the factorization theorem 
for the tDY cross section, and the relevant partonic mechanism to ${\cal O}(\alpha_s)$. 
The total and $Q_T$-differential cross sections are derived
in \S\S3 and 4, respectively, by performing 
a collinear factorization in the $\overline{\rm MS}$ scheme. 
Sections 5-9 contain discussion of the soft gluon resummation.
In \S5 we introduce the general formalism of soft gluon resummation,
and in \S6 we elaborate on its $b$-space structure.
We perform the soft gluon resummation for tDY up to NLL accuracy in \S7.
In \S8, we also derive an asymptotic formula for our NLL resummed
cross section in the $Q_T \approx 0$ region.
Section 9 contains numerical results for tDY processes to be observed  
at RHIC and J-PARC.
We demonstrate the roles of QCD soft gluon effects in the cross sections 
and asymmetries as functions of $Q_T$, and discuss a new approach 
to extracting the transversity from experimental data.
The final section, \S10, is reserved for conclusions. 
This paper contains three appendices.
In Appendix~A, we collect the operator definitions and basic properties
of transversity, in Appendix~B we collect the formulae for the tDY cross sections 
integrated over the rapidity of the dilepton, and Appendix~C supplements 
the discussion given in \S7.

\section{
Drell-Yan mechanism to order $\alpha_s$}

\subsection{Factorization formula for the transversely polarized Drell-Yan process}

The process we consider is the tDY process,
\begin{equation}
h_1 (P_1\,,\,S_1) + h_2 (P_2\,,\,S_2) \to
     l (k_1) + \bar{l} (k_2) + X , 
\label{DYprocess}
\end{equation}
where $h_1$ and $h_2$ denote spin-1/2 hadrons with momenta $P_1$ and $P_2$ 
and transverse spins $S_1$ and $S_2$, and $Q=k_1+k_2$ is the 4-momentum of the DY pair.  
The tDY (\ref{DYprocess}) may be induced by partonic subprocesses, 
such as quark-antiquark annihilation,
\begin{equation}
  q (p_1 , s_1) + \bar{q} (p_2 , s_2) 
              \to l (k_1) + \bar{l} (k_2) + X ,
\label{partonic}
\end{equation}
where $q$ and $\bar{q}$ have momenta $p_{1}$ and $p_2$ and transverse spins 
$s_1$ and $s_2$. This can be formulated precisely 
using QCD factorization~\cite{CSS:89}, which 
is valid when $Q^2$ is large, i.e., $Q^2 \gg \Lambda_{\rm QCD}^2$, 
with $\tau \equiv Q^2 / S~(S\equiv(P_1 +P_2 )^2)$ fixed.
In this case, (\ref{partonic}) and 
$\bar{q} (p_1 , s_1) + q (p_2 , s_2) \to l (k_1) + \bar{l} (k_2) + X$
provide the dominant mechanism for tDY in the region with $p_i = x_i P_i$ ($i=1,2$),
where $x_1$ and $x_2$ denote the longitudinal-momentum fractions 
of partons inside the parent hadrons $h_1$ and $h_2$, respectively,
while the contributions from the other regions or other partonic processes
give subleading corrections.
The values that $x_1$ or $x_2$ actually takes are determined by the
parton distributions, due to the long-distance, nonperturbative 
dynamics inside each hadron $h_{1,2}$, whose typical distance scale 
$\sim 1/\Lambda_{\rm QCD}$ is well-separated from the short-distance 
scale of order $1/Q$, within which partonic annihilation processes like 
(\ref{partonic}) occur.
Therefore, the short- and long-distance contributions can be treated separately,
and the spin-dependent cross section, $\Delta_T d \sigma \equiv 
(d \sigma (S_1 , S_2) - d \sigma (S_1 , - S_2))/2$, for tDY (\ref{DYprocess})
is given as a convolution of the short- and long-distance contributions, 
over each of the partonic momentum fractions $x_1$ and $x_2$ for the two
hadrons $h_{1,2}$~\cite{RS,JJ:92},
\begin{equation}
  \Delta_T d \sigma =
\int d x_1 d x_2\,  
    \delta H (x_1 ,\,x_2;  \mu_F^2)\, 
      \left[   \Delta_T d \hat{\sigma} ( p_1, p_2 ; Q^2; \mu_F^2 ) \right]_{p_i=x_i P_i},
  \label{convo}
\end{equation}
up to the higher-twist corrections suppressed by the powers of $1/Q$.
Here,  $\mu_F$ is the factorization scale, 
$\Delta_T d \hat{\sigma}(p_1, p_2 ; Q^2; \mu_F^2 )$ 
is the transversely polarized partonic cross section for (\ref{partonic})
representing the short-distance contribution, and 
\begin{equation}
\delta H(x_1,x_2;\mu_F^2)=\sum_{q} e_q^2 
\left[
\delta q_{h_1}(x_1 ,\mu_F^2)\delta\bar{q}_{h_2}(x_2,\mu_F^2)
+\delta \bar{q}_{h_1}(x_1 ,\mu_F^2)\delta q_{h_2}(x_2, \mu_F^2)
\right]
\label{tPDF}
\end{equation} 
is the long-distance contribution as a product of the transversity 
distributions~\cite{RS,JJ:92}
$\delta q_{h_1} (x_1 , \mu_F^2)$ and $\delta q_{h_2} (x_2 , \mu_F^2)$
for the two colliding hadrons, $h_1$ and $h_2$,
summed over the massless quark flavors $q$ with their charge squared $e_q^2$.

Intuitively, the leading-twist factorization formula (\ref{convo}) 
can be interpreted using the language of the parton model, 
similarly to the factorization formula for the twist-2 structure functions in DIS.
In particular, the transversity distribution $\delta q_h (x , \mu_F^2)$
is one of the three independent twist-2 quark distribution functions 
for a spin-1/2 hadron $h$, which are associated with the probability 
distribution of finding a quark with 
a longitudinal momentum fraction $x$ inside the parent hadron $h$.
In fact, $\delta q_h (x , \mu_F^2)$ is defined by~\cite{RS,JJ:92,BDR:02}
\[  \delta q_h (x , \mu_F^2) \equiv q_h^{\uparrow} (x , \mu_F^2)
      - q_h^{\downarrow} (x , \mu_F^2)\ , \]
where $q_h^{\uparrow}$ ($q_h^{\downarrow}$) denotes the probability distribution of
transversely polarized quarks with 
spin parallel (antiparallel) to the spin of their transversely-polarized 
parent hadron~\cite{BDR:02}.
The spin-dependent partonic cross section participating in (\ref{convo}) is defined as
\bea
\Delta_T d \hat{\sigma} (p_1, p_2 ; Q^2; \mu_F^2)
     &\equiv& \frac{1}{2} \left[ 
d \hat{\sigma} (p_1, p_2 ; s_1\,,\,s_2 ; Q^2; \mu_F^2 )\right.
\nonumber\\
            &  &- \left. 
d \hat{\sigma} (p_1, p_2 ; s_1\,,\, - s_2 ; Q^2; \mu_F^2 ) \right] ,
\label{pcs}
\eea
in terms of the partonic hard cross section
$d \hat{\sigma} (p_1, p_2 ; s_1\,,\,s_2 ; Q^2; \mu_F^2 )$ which describes 
the annihilation process (\ref{partonic}) of the quark and antiquark 
with the transverse spins $s_1$ and $s_2$, respectively. 
Note that at the leading twist level, the gluon distributions do not contribute
to (\ref{convo}) for the transversely polarized process, due to its
chiral-odd nature~\cite{RS,JJ:92,BDR:02}.
Therefore, only the $q\bar{q}$ annihilation subprocesses give
the relevant twist-2 mechanism for the spin-dependent cross section in tDY.

The factorization formula (\ref{convo}) provides not only 
a realization of the parton model based on QCD but also 
a systematic framework for calculating the QCD corrections beyond the parton model. 
As mentioned above, the meaning of the factorization formula (\ref{convo}) 
is the separation of the short- and long-distance contributions
contained in the process (\ref{DYprocess}) at the factorization scale $\mu_F$.
In formal field theoretical language, (\ref{DYprocess}) can be represented by 
Feynman diagrams, assuming that the hadrons $h_{1,2}$ can be described 
in terms of virtual partonic states, as if they were a beam of quarks and gluons.
Then, in a corresponding generic diagram, 
the long-distance contributions may give rise to infrared (IR) divergence,
but, considering the set of diagrams contributing at the same order of accuracy, 
those IR-divergent contributions can be completely factorized into
universal hadron matrix elements [transversity distributions, see (\ref{pdf})],
or cancel out, and thus the partonic cross section (\ref{pcs})
possesses purely short-distance contributions containing all the dependence on 
the relevant hard scale $Q$~\cite{CSS:89}.
Therefore, QCD perturbation theory can provide a prediction for (\ref{pcs}) 
in a systematic way, while the transversity distributions as the
universal hadron matrix elements 
can only be determined by taking into account nonperturbative effects.

A traditional approach for the separation of short- and long-distance contributions
is the operator product expansion (OPE), and in fact, the OPE has been 
successfully applied to, e.g., deep inelastic lepton-hadron scattering.
But, in the case of DY processes, as well as various hard processes
in which two or more hadrons participate,
the conventional OPE in terms of local operators is useless for this purpose,
and the factorization theorems associated with formulae like (\ref{convo}) provide 
a generalization of the OPE suitable for those processes~\cite{CSS:89}.
The partonic cross section (\ref{pcs}) in the factorization formula (\ref{convo})
plays the role of the Wilson coefficient functions in the conventional OPE,
and (\ref{pcs}) can be determined by a ``matching'' procedure similar to that used 
to obtain the Wilson coefficient functions in the OPE:
To obtain the prediction for (\ref{pcs}), we apply the factorization
formula (\ref{convo}) to the parton-level process (\ref{partonic}), 
employing an appropriate IR regularization scheme.
In this case, (\ref{tPDF}) can be calculated explicitly as 
the ``quark distribution inside a quark,''
$\delta q_{h_1 =q}(x_1 , \mu_F^2) =\delta(x_1 -1)+ \cdots$ and 
$\delta \bar{q}_{h_2 =\bar{q}}(x_2 , \mu_F^2) =\delta(x_2 -1)+ \cdots$, 
where the ellipses represent the terms of ${\cal O}(\alpha_s)$ or higher, 
depending on the specific IR regulator [see (\ref{pdf_1-loop}) in Appendix~A]. 
The IR-regulator-dependent contributions on the RHS of (\ref{convo}) 
actually coincide with the IR-regulator-dependent contributions  
appearing on the LHS for the parton-level cross section,
and thus 
a comparison of both sides of (\ref{convo}), order-by-order in $\alpha_s$,
yields $\Delta_T d \hat{\sigma}(p_1, p_2 ; Q^2; \mu_F^2)$ of (\ref{pcs}) 
as an ``IR safe'' power series in $\alpha_s$.
Through this matching procedure, we confirm that the partonic cross section,  
$\Delta_T d \hat{\sigma}(p_1, p_2 ; Q^2; \mu_F^2)$, indeed represents 
a purely short-distance component in the cross section (\ref{convo}). 
In our calculation presented below, we employ the dimensional regularization scheme to 
regularize IR and UV divergences.

\subsection{Momentum and spin for parton-level processes in $D$ dimensions}

\begin{figure}
\begin{center}
\begin{tabular}{cc}
\includegraphics[height=2.4cm,clip]{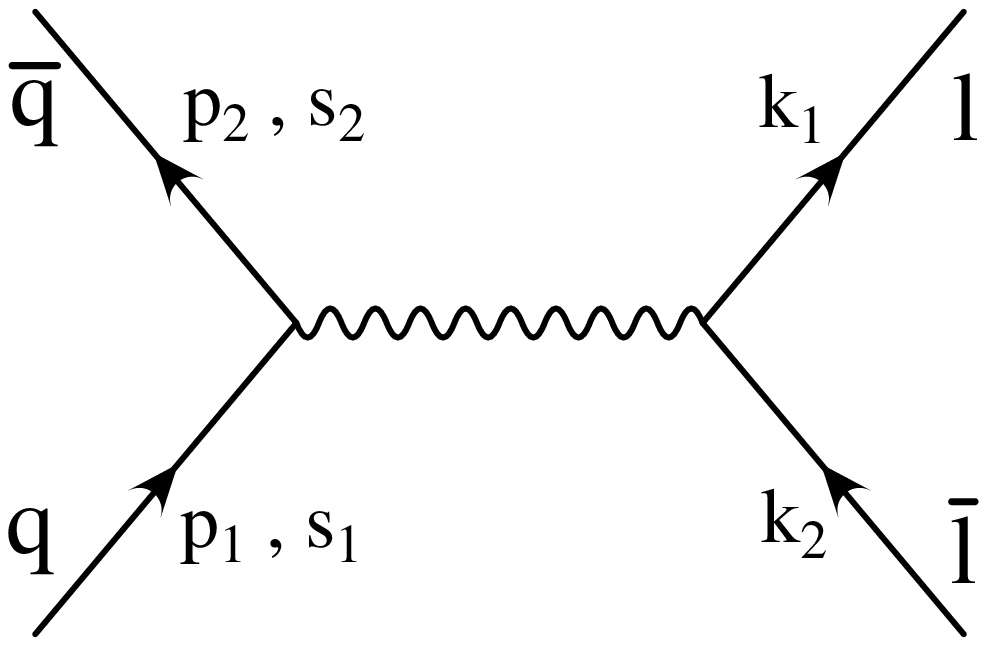} \quad & \quad 
   \includegraphics[height=2.4cm,clip]{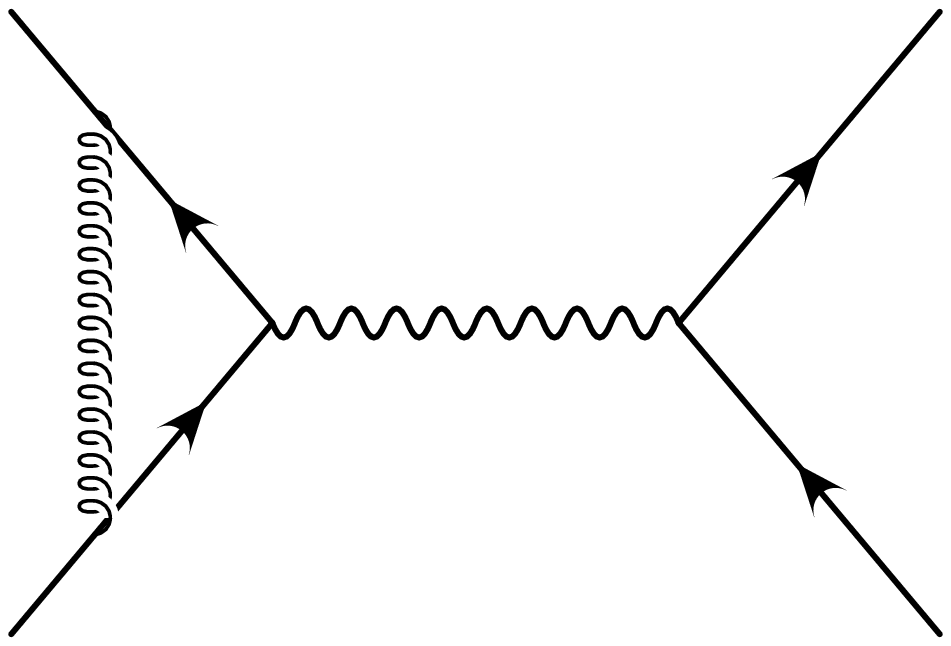}\\
 & \\
\includegraphics[height=3.2cm,clip]{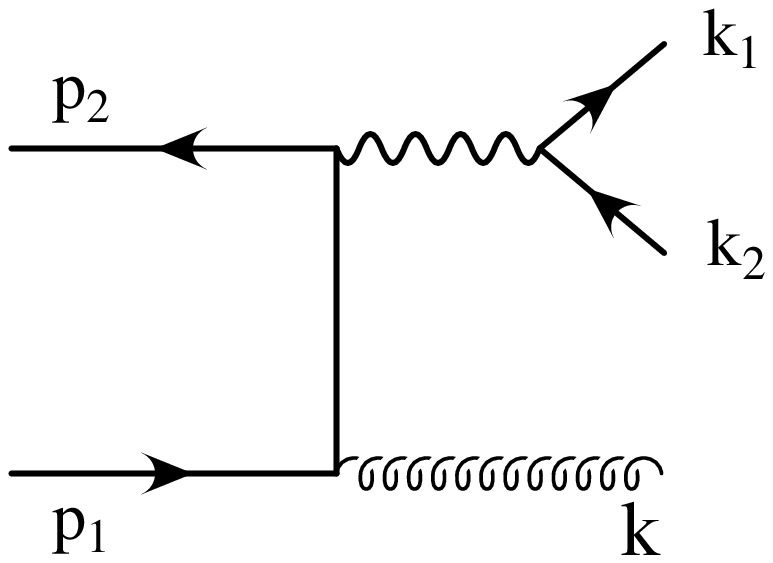} \quad & \quad
\includegraphics[height=2.5cm,clip]{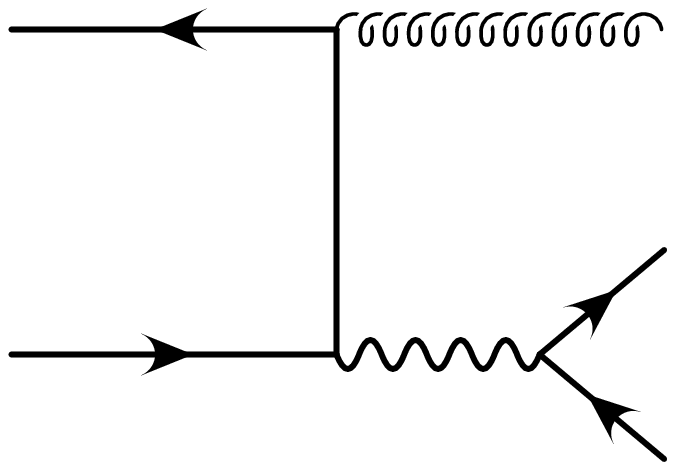} 
\end{tabular}
\caption{Parton-level processes contributing to the spin-dependent cross section in tDY
to $\mathcal{O} (\alpha_s)$.}
\label{fig1}
\end{center}
\end{figure}

We now perform the above-mentioned matching at the one-loop level and derive
$\Delta_T d \hat{\sigma}(p_1, p_2 ; Q^2; \mu_F^2)$ in (\ref{pcs}) to 
${\cal O}(\alpha_s)$.
For this purpose, we calculate the LHS of (\ref{convo}) for the parton-level 
process (\ref{partonic}), which we write as $\Delta_T d \sigma^{q\bar{q}}$.
At the one-loop level, the diagrams to be calculated are given in Fig.~\ref{fig1}. 
We calculate all these contributions in $D = 4 - 2 \epsilon$ dimensions 
to regulate the IR and UV divergences.
For these parton-level processes, we define the invariants as 
\begin{equation}
   s = (p_1 + p_2 )^2 \ , \  Q^2 = (k_1 + k_2)^2 \equiv z\, s \ , \  
   t = (p_1 - k_1)^2\ , \ u = (p_2 - k_1)^2 ,
\label{mandel}
\end{equation}
in terms of the momenta assigned in (\ref{partonic}).
In the case of one-gluon emission corresponding to the lower diagrams 
in Fig.~\ref{fig1}, i.e., (\ref{partonic}) with $X \equiv g (k)$,
we also define
\begin{equation}
 \hat{t} = (p_1 - k)^2\ , \quad \hat{u} = (p_2 - k)^2 . 
\label{eq:tuhat}
\end{equation}
The momentum of a massless particle, $p=\{p_1 , p_2 , k_1 , k_2 , k\}$, in
$D$ dimensions is generically expressed as
\begin{equation}
   p^{\mu}=(p^0\, , \, \boldsymbol{p})= (|\boldsymbol{p}| \, , \, \boldsymbol{p}), 
\;\;\;\;\;\;\;\;\;\;\;\;
\boldsymbol{p} \equiv |\boldsymbol{p}| ( 
\boldsymbol{n}_p^{(D-2)} \sin \theta_p \, , \, \cos \theta_p ) ,
\label{momentump}
\end{equation}
where $\boldsymbol{n}_p^{(D-2)}$ denotes a $(D-2)$-dimensional unit vector 
which is parametrized by $D-3$ angular coordinates, 
$\phi_p^{(D-3)} , \phi_p^{(D-4)}, \ldots$, as
\bea
\boldsymbol{n}_p^{(D-2)}& =&
                   ( 
\cos \phi_p^{(D-3)}  ,\ \boldsymbol{n}_p^{(D-3)} \sin \phi_p^{(D-3)}  ) 
\nonumber \\
&=& (
 \cos \phi _p^{(D-3)}  ,\ \sin \phi_p^{(D-3)}  \cos  \phi_p^{(D-4)},\ 
\boldsymbol{n}_p^{(D-4)} \sin \phi_p^{(D-3)}  \sin  \phi_p^{(D-4)}) 
\nonumber\\
&=& \ldots ,
\label{anglep}
\eea
such that in 4 dimensions, 
defining $\phi_p^{(1)} \equiv \phi_p$, we have
\be
\boldsymbol{n}_p^{(2)}  = ( \cos \phi_p ,\  \sin \phi_p).
\label{anglep4}
\ee
%

When we calculate the LHS of (\ref{convo}) for the parton-level process 
(\ref{partonic}), we always choose a frame in which the initial quark
and antiquark in (\ref{partonic}) possess momenta 
$\boldsymbol{p}_1$ and $\boldsymbol{p}_2$ that are collinear along 
the $z$-axis, with $\theta_{p_1}=0$ and $\theta_{p_2}=\pi$, and
possess transverse spins, satisfying $p_1 \cdot s_1 =0$, $p_2 \cdot s_2 =0$, 
and $s_{1,2}^2 = -1$,  that are expressed by
\begin{equation}
s_{1,2}^{\mu} = (0,\  \boldsymbol{s}_{1,2}^{(D-2)},\ 0) ,
\label{spinvec}
\end{equation}
with the $(D-2)$-dimensional unit vectors $\boldsymbol{s}_{1,2}^{(D-2)}$,
which reduce in 4 dimensions to
\be
\boldsymbol{s}_{1}^{(2)} =( \cos \phi_1 ,\  \sin \phi_1), \;\;\;\;\;\;\;\;\;\;\;\;
\boldsymbol{s}_{2}^{(2)} =( \cos \phi_2 ,\  \sin \phi_2).
\label{spinvec4}
\ee

\subsection{Parton-level amplitude squared}

The contribution of each diagram in Fig.~\ref{fig1} to the spin-dependent cross section 
$\Delta_T d \sigma^{q\bar{q}}$, the LHS of (\ref{convo}), is generically given by 
\be
\frac{1}{2 N_c^2 s}\, \Delta_T |M|^2
            d \Phi \ ,\label{pxsection}
\ee
where $d \Phi$ denotes the differential element of the $n$-body phase space
for the corresponding $2\rightarrow n$ process ($n=2,3$), 
$1/N_c$ is the color-averaging factor for the incoming quark (antiquark),
and the spin-dependent amplitude squared, $\Delta_T |M|^2$,
is defined in terms of the relevant $2\rightarrow n$ Feynman amplitude
$M(s_1 , s_2)$, associated with the transverse spins $s_1$ and $s_2$ 
for the colliding quark and antiquark in (\ref{partonic}), as
\be
\Delta_T |M|^2 \equiv \frac{1}{2} 
       \left[ |M(s_1 , s_2) |^2 - |M(s_1 , - s_2) |^2 \right] , 
\label{eq:sqa}
\ee
where the sum over the spins of the final-state leptons is implicit.
We calculate (\ref{pxsection}) and (\ref{eq:sqa}) in the dimensional regularization. 
We rewrite the QCD coupling constant as $g \rightarrow g \mu^{\epsilon}$
with the mass scale $\mu$. This enables us to keep the redefined coupling constant
dimensionless in $D=4-2 \epsilon$ dimensions.\footnote{In principle, one 
can rewrite the QED coupling constant similarly. This results in the multiplication of
the Born cross section and the higher-order cross section by a common factor 
of $\mu^{4\epsilon}$, and it allows us to keep the dimensions of the cross section 
unchanged. But this is irrelevant for the present purpose of matching to obtain 
the partonic cross section (\ref{pcs}).}

In the spin-dependent amplitude squared of (\ref{eq:sqa}), we encounter 
traces of the gamma matrices involving $\gamma_5$.
We employ the naive anticommuting-$\gamma_5$ scheme, 
$\{ \gamma_5, \gamma^{\mu} \}=0$ ($\mu=0, 1, \ldots D-1$),
which is the usual prescription in the transverse-spin channel:
For the transverse-spin case, it appears that these traces involve only 
even numbers of $\gamma_5$.
Therefore the matrices $\gamma_5$ eventually disappear due to $\gamma_5^2 = 1$, 
and we do not anticipate any inconsistencies related to $\gamma_5$.
This suggests that the naive anticommuting-$\gamma_5$ prescription will work
(see Ref.~\citen{V}).

The upper two diagrams in Fig.~\ref{fig1} represent the $2 \rightarrow 2$ processes
up to ${\cal O}(\alpha_s)$.
The corresponding tree-plus-virtual 
contributions to (\ref{eq:sqa}) can be calculated straightforwardly, and they read 
\bea
  \Delta_T |M|^2_{\rm T + V}
    &=& 4\,N_c \, \left( \frac{4 \pi \alpha  e_q }{s} \right)^2
         \left[ 2 \,s \, (k_1 \cdot s_1) (k_1 \cdot s_2)
           + t \, u \, (s_1 \cdot s_2) 
      + \epsilon \frac{s^2}{2} \, (s_1 \cdot s_2) \right] \nonumber\\
   &\times & \left[ 1 + \frac{\alpha_s}{\pi} C_F
            \left( \frac{4 \pi \mu^2}{s} \right)^{\epsilon}
           \frac{1}{\Gamma (1 - \epsilon)} 
         \left( - \frac{1}{\epsilon^2} - \frac{3}{2 \epsilon}
           - 4 + \frac{\pi^2}{2} + {\cal O}(\epsilon) \right) \right] , 
\non\\
\label{tandv}
\eea
where $C_F = (N_c^2 -1 )/(2N_c )$ [also see (\ref{mandel})].
Equation (\ref{tandv}) includes the contribution due to the wave-function 
renormalization factors for the incoming quark and antiquark legs, which cancels 
the quark-photon vertex renormalization constant, 
as guaranteed by the Ward identity.
Therefore, (\ref{tandv}) is UV finite, and it is also gauge invariant
in a gauge-invariant regularization scheme.
Note that the ratio of the ${\cal O}(\alpha_s )$ term to the ${\cal O}(1)$ 
term in (\ref{tandv}) is the same as the corresponding ratio 
in the unpolarized case~\cite{AEM}.

The lower two diagrams in Fig.~\ref{fig1} represent 
the $2 \rightarrow 3$ processes up to ${\cal O}(\alpha_s)$.
The relevant polarization sum for the final-state gluon, 
${\cal P}^{\alpha \beta}(k) \equiv \sum_\lambda 
e_{(\lambda)}^{\alpha}(k)e_{(\lambda)}^{\ast \beta}(k)$,
where $e_{(\lambda)}^{\alpha}(k)$ is the polarization vector for the
real gluon with momentum $k$ and the physical polarization $\lambda$,
can be replaced as ${\cal P}^{\alpha \beta}(k) \rightarrow -g^{\alpha\beta}$, 
using the Ward identity, in the present case of {\it one} real external
massless gluon~\cite{F:89}.
We write the corresponding result for the one-gluon-radiation contributions as
\be
   \Delta_T |M|_{\rm R}^2 \equiv N_c C_F \,
         \left( \frac{4 \pi \alpha e_q g \mu^{\epsilon}}{Q^2} \right)^2 
    \Delta_T \mR ,
\label{eq:realg}
\ee
where $\Delta_T \mR$ is formally given by a second-order polynomial 
in $\epsilon$, after working out the traces of the Dirac matrices.
Actually, the terms proportional to $\epsilon^2$ in $\Delta_T \mR$
do not contribute to the cross section in the limit of 
$\epsilon \to 0$, and therefore we drop them.
Similarly, we drop some of the terms proportional to $\epsilon$,
which clearly do not contribute to the cross section, and 
we thereby obtain
\bea
   \lefteqn{ \Delta_T \mR 
=
     \frac{16 \epsilon}{\hat{t}^2}
       \left[ s \, (z \, t + u + Q^2) (k_1 \cdot s_2)( k \cdot s_1)
        - t \, u \, (k \cdot s_1)( k \cdot s_2) \right]} \nonumber\\
    && + \frac{8 \epsilon}{\hat{t}} \ 
         \left[ 2 \, (s - Q^2) \, (k_1 \cdot s_1)( k_1 \cdot s_2) 
          + u \, (t + u + Q^2)\ (s_1 \cdot s_2) \right]\nonumber\\
    && + \frac{16 \epsilon}{\hat{u}^2}
       \left[ s \, (z \, u + t + Q^2) (k_1 \cdot s_1)( k \cdot s_2)
        - t \, u \, (k \cdot s_1)( k \cdot s_2) \right]\nonumber\\
    && + \frac{8 \epsilon}{\hat{u}} \ 
         \left[ 2 \, (s - Q^2) \, (k_1 \cdot s_1)( k_1 \cdot s_2) 
          + t \, (u + t + Q^2) \ (s_1 \cdot s_2) \right] \nonumber\\
    && + \frac{16 Q^2}{\hat{t} \hat{u}}
        \left[ 2 \, s \, (k_1 \cdot s_1)( k_1 \cdot s_2)
       + t \, u \, (s_1 \cdot s_2)
                 + \epsilon \frac{Q^2 s}{2} (s_1 \cdot s_2) \right]\nonumber\\
    && + \frac{8 Q^2}{\hat{t} \hat{u}}
            \left[ 2 \, s \, \{ (k_1 \cdot s_2)( k \cdot s_1) 
      +  (k_1 \cdot s_1)( k \cdot s_2) + (k \cdot s_1)(k \cdot s_2  )\}
                               \right.\nonumber\\
    && \quad\qquad\qquad 
          + \left. \{Q^2 \, (s + t + u) - t \, \hat{t} - u \, \hat{u}
          + \hat{t} \, \hat{u} \} \,(s_1 \cdot s_2) \right]\nonumber\\
    && + \frac{8 \epsilon}{\hat{t} \hat{u}}
          s\, (t + u + Q^2)\, (s + t + u)\ (s_1 \cdot s_2)\nonumber\\
    && + \frac{4 \epsilon}{\hat{t}}
          \left[ t \, (s + t + u) + (s + t)\,( t + u + Q^2)\right]\ 
                  (s_1 \cdot s_2)\nonumber\\
    && + \frac{4 \epsilon}{\hat{u}}
          \left[ u \, (s + t + u) + (s + u)\,( u + t + Q^2)\right]\ 
                  (s_1 \cdot s_2) , \label{gemission}
\eea
using the variables in (\ref{mandel}) and (\ref{eq:tuhat}).
Note that the terms in (\ref{gemission}) may receive additional 
factors of $1/\epsilon$ as IR-divergent poles through the integration
over the phase space in (\ref{pxsection}).

\section{NLO matching for the $Q_T$-integrated (total) cross section
}

First, we consider the case in which the transverse momentum, $Q_T$, of the 
final dilepton is unobserved (integrated) in the parton-level process (\ref{partonic}).
We calculate the corresponding cross section (\ref{convo}) using  
(\ref{pxsection})-(\ref{gemission}) and determine the $Q_T$-integrated 
partonic cross section, 
$\Delta_T d \hat{\sigma}(p_1, p_2 ; Q^2; \mu_F^2)$, to ${\cal O}(\alpha_s)$.

\subsection{Tree-plus-virtual contributions}

The phase space for an outgoing massless particle with momentum $p$
is given by [see (\ref{momentump}) and (\ref{anglep})]
\[ \frac{d^{D-1} p}{(2 \pi)^{D-1} 2 p^0}
    = \frac{1}{2 (2 \pi)^{D-1}}\, |\boldsymbol{p}|^{D-3} \, 
d |\boldsymbol{p}| \, d \Omega_p^{(D-1)} \ ,\]
where the differential element for the $(D-2)$-dimensional angular
integration is expressed as
\be
  d \Omega_p^{(D-1)} = ( 1 - \cos^2 \theta_p )^{- \epsilon}
        \, d \cos \theta_p \ d \Omega_p^{(D-2)} \ .
\label{eq:angmeas}
\ee
The two-particle phase space for the $2\rightarrow 2$ processes of Fig.~\ref{fig1}
is given by
\be
  d \Phi_2 =
\frac{d^{D-1} k_1}{(2 \pi)^{D-1} 2 k_1^0}
               \frac{d^{D-1} k_2}{(2 \pi)^{D-1} 2 k_2^0}
               \ (2 \pi)^D\ \delta^{(D)}\, (p_1 + p_2 - k_1 - k_2) , 
\label{eq:phase2}
\ee
and we obtain for the tree-plus-virtual contributions to (\ref{pxsection}) 
the expression
\[ \frac{\Delta_T \, d \sigma_{\rm T + V}^{q\bar{q}}}{d Q^2 d
                \Omega_{k_1}^{(D-1)}}
       = \frac{1}{64 \pi^2 N_c^2 s^2} 
     \left( \frac{\sqrt{s}}{4 \pi} \right)^{- 2 \, \epsilon}
         \, \Delta_T |M|^2_{\rm T + V} \, \delta \, (1 - z) , \]
using the invariants of (\ref{mandel}).
Inserting the expression for $\Delta_T |M|^2_{\rm T + V}$, (\ref{tandv}), 
and using the explicit forms (\ref{momentump})-(\ref{spinvec}) 
for the lepton's momentum $k_1$ and the quark's transverse spin $s_{1,2}$,
we obtain
\bea
   \frac{\Delta_T \, d \sigma_{\rm T + V}^{q\bar{q}}}{d Q^2 d
                \Omega_{k_1}^{(D-1)}}
   &=& \frac{1}{16 \pi^2 N_c} 
       \left( \frac{4 \pi \alpha e_q}{Q^2} \right)^2
   \left( \frac{Q^2}{16 \pi^2} \right)^{- \, \epsilon}
          \, \delta \, (1 - z) \nonumber\\
    & \times & \left[ \, \frac{1}{4} \,  \alpha_{\Phi} \sin^2 \theta_{k_1} 
           + \frac{\epsilon}{2} \, (s_1 \cdot s_2) \right] \non\\
    & \times& \left[ \, 1 + \frac{\alpha_s}{\pi} C_F
            \left( \frac{4 \pi \mu^2}{Q^2} \right)^{\epsilon}
           \frac{1}{\Gamma (1 - \epsilon)} 
         \left( - \frac{1}{\epsilon^2} - \frac{3}{2 \epsilon}
           - 4 + \frac{\pi^2}{2} + {\cal O}(\epsilon ) \right) \right] , 
\non\\
\label{eq:tplusv}
\eea 
where
\be
 \alpha_{\Phi} =  2\, (\boldsymbol{n}_{k_1}^{(D-2)} \cdot \boldsymbol{s}_1^{(D-2)})\,
     (\boldsymbol{n}_{k_1}^{(D-2)} \cdot \boldsymbol{s}_2^{(D-2)}) + s_1 \cdot s_2 ,
\label{eq:alphaphi}
\ee
with
\be
    s_1 \cdot s_2 = - \, \boldsymbol{s}_1^{(D-2)} \cdot \boldsymbol{s}_2^{(D-2)} . 
\label{eq:s1s2}
\ee
From (\ref{eq:angmeas}),
the corresponding \lq\lq azimuthal\rq\rq\ angular distribution of the lepton
is obtained by integrating (\ref{eq:tplusv}) over the
\lq\lq scattering\rq\rq\ angle $\theta_{k_1}$, with the measure
$( 1 - \cos^2 \theta_{k_1})^{- \epsilon} \, d \cos \theta_{k_1}$, as 
\bea
   \frac{\Delta_T \, d \sigma_{\rm T + V}^{q\bar{q}}}{d Q^2 d
                \Omega_{k_1}^{(D-2)}}
   &=& \frac{1}{16 \pi^2 N_c} 
       \left( \frac{4 \pi \alpha e_q}{Q^2} \right)^2
   \left( \frac{Q^2}{16 \pi^2} \right)^{- \, \epsilon}
          \, \delta \, (1 - z) \nonumber\\
    &\times& \left[ \, 2^{1 - 2 \epsilon}\, B(2-\epsilon\,,\,2-\epsilon) \ 
       \alpha_{\Phi}
           + \epsilon \, 2^{ - \, 2 \epsilon}\ B(1-\epsilon\,,\,1-\epsilon)
          \  (s_1 \cdot s_2) 
                 \right] \non\\
    & \times& \left[ \, 1 + \frac{\alpha_s}{\pi} C_F
            \left( \frac{4 \pi \mu^2}{Q^2} \right)^{\epsilon}
           \frac{1}{\Gamma (1 - \epsilon)} 
         \left( - \frac{1}{\epsilon^2} - \frac{3}{2 \epsilon}
           - 4 + \frac{\pi^2}{2}  + {\cal O}(\epsilon ) \right) \right]
   , \nonumber\\
\label{tvx}
\eea 
where $B(x,y) = \Gamma(x)\Gamma(y)/\Gamma(x+y)$.

\subsection{Real gluon emission}

For the $2 \rightarrow 3$ processes depicted in Fig.~\ref{fig1}, the real-gluon emission 
contributions, the corresponding amplitude squared seems rather 
complicated [see (\ref{gemission})].
Fortunately, miraculous cancellations occur among the first four lines
and among the last three lines of (\ref{gemission})~\cite{K2}:
Because the corresponding terms are proportional to $\epsilon$, these terms survive 
only in the configuration obtained in the \ $\hat{t}, \hat{u} \to 0$ limit, 
in which the divergences associated with the collinear 
($k^\mu \propto p_1^\mu$, $k^\mu \propto p_2^\mu$)
or soft ($k^\mu \approx 0$) gluon are produced [see (\ref{eq:tuhat})].
An explicit calculation shows that the collinear configuration is indeed 
relevant in the present case,
and in this configuration an additional $1/\epsilon$ factor
multiplying these terms appears.
However, for such collinear configurations, it turns out that
the first (third) line cancels the second (fourth) line in (\ref{gemission}), 
and also, the last three lines cancel.
Therefore, (\ref{gemission}) can be reduced to the very simple form 
\be
\Delta_T \mR = \Delta_T \mR_S + \Delta_T \mR_{NS} ,
\label{eq:simpleR}
\ee
with
\bea
 \Delta_T \mR_S &=&
    \frac{16 Q^2}{\hat{t} \hat{u}}
        \left[ 2 \, s \, (k_1 \cdot s_1)( k_1 \cdot s_2)
       + t \, u \, (s_1 \cdot s_2)
                 + \epsilon \frac{Q^2 s}{2} (s_1 \cdot s_2) \right] ,
      \nonumber\\
 \Delta_T \mR_{NS}     &=&  \, \frac{8 Q^2}{\hat{t} \hat{u}}
            \biggl[ 2 \, s \, \{ (k_1 \cdot s_2)( k \cdot s_1) 
      +  (k_1 \cdot s_1)( k \cdot s_2) + (k \cdot s_1)(k \cdot s_2  )\}
                               \nonumber\\
    && + \{Q^2 \, (s + t + u) - t \, \hat{t} - u \, \hat{u}
          + \hat{t} \, \hat{u} \} \,(s_1 \cdot s_2) \biggr] .
\label{reducedm}
\eea
We have divided the RHS of (\ref{eq:simpleR}) into two terms;
as will be discussed below, only $\Delta_T \mR_S$, given in (\ref{reducedm}), 
receives the $1/\epsilon$ singularity through the integration over the phase space 
in (\ref{pxsection}), while $\Delta_T \mR_{NS}$ yields the result finite  
in the limit $\epsilon \to 0$.\footnote{Note that we have $k\cdot
s_{1,2}\ra 0$ and $Q^2(s+t+u)-t\hat{t}-u\hat{u}+\hat{t}\hat{u}\ra 0$ 
in the collinear limit, that is,
$k^\mu \parallel p_{1}^\mu$ or $k^\mu \parallel p_{2}^\mu$.}

The three-particle phase space for the $2\rightarrow 3$ processes
depicted in Fig.~\ref{fig1} is given by
\be
  d \Phi_3 = \frac{d^{D-1} k_1}{(2 \pi)^{D-1} 2 k_1^0}
               \frac{d^{D-1} k_2}{(2 \pi)^{D-1} 2 k_2^0}
               \frac{d^{D-1} k}{(2 \pi)^{D-1} 2 k^0}
           \ (2 \pi)^D\ \delta^{(D)}\, (p_1 + p_2 - k - k_1 - k_2) .
\label{phase3old}
\ee
To use this in (\ref{pxsection}), it is convenient to employ the $q \bar{q}$ 
CM frame and first determine 
the phase space factor corresponding to the differential elements, 
$d Q^2\ d \Omega_k^{(D-1)}\ d \Omega_{k_1}^{(D-1)}$, integrating over the other
degrees of freedom [see (\ref{mandel})]. 
This results in the replacement
\be
  d \Phi_3 \rightarrow 
\frac{1}{8 \sqrt{s} (2 \pi)^{5 - 4 \epsilon}}
          \frac{Q^{2 (1 - 2 \epsilon)}}{2^{2 - 2 \epsilon}}
          \frac{|\boldsymbol{k}|^{1 - 2 \epsilon}}{\left(Q^0 + 
\boldsymbol{k}\cdot \boldsymbol{k}_1/|\boldsymbol{k}_1 |
\right)^{2 - 2 \epsilon}}\ 
          d Q^2\ d \Omega_k^{(D-1)}\ d \Omega_{k_1}^{(D-1)}\ ,
\label{eq:phase3}
\ee
where
\[ Q^0 = \frac{\sqrt{s}}{2}\, (1 + z)  , \quad
                 |\boldsymbol{k}| = \frac{\sqrt{s}}{2}\, (1 - z) \ ,\]
and the magnitude of the lepton momentum, 
$k_1^{\mu} \equiv (|\boldsymbol{k}_1| \, , \, \boldsymbol{k}_1)$, is fixed 
by the relevant kinematical constraints as
\be
 |\boldsymbol{k}_1| =
\frac{Q^2}{2}\, \frac{1}{Q^0 + 
\boldsymbol{k}\cdot \boldsymbol{k}_1/|\boldsymbol{k}_1 |}
\equiv
      \frac{Q^2}{\sqrt{s}}\ \frac{1}{\mD_I} ,
\label{eq:DI}
\ee
with [see (\ref{momentump})]
\be
 \mD_I = 1 + z + (1 - z)
     \left( \cos \theta_k\, \cos \theta_{k_1} 
        + \sin \theta_k\, \sin \theta_{k_1}\ 
         \boldsymbol{n}_k^{(D-2)} \cdot \boldsymbol{n}_{k_1}^{(D-2)} \right) ,
\label{eq:DI2}
\ee
which is positive definite. The invariants in (\ref{reducedm}) are now given by 
[see (\ref{mandel}), (\ref{eq:tuhat})]
\bea
  \hat{t} &=& - \, \frac{s}{2}\, (1 - z)\, (1 - \cos \theta_k)  ,
     \qquad  \hat{u} = - \, \frac{s}{2}\, (1 - z)\, (1 + \cos \theta_k) ,\nonumber \\
   t &=& - \, \sqrt{s} \, |\boldsymbol{k}_1| \, (1 - \cos \theta_{k_1})  ,\ \
     \qquad u = - \, \sqrt{s} \, |\boldsymbol{k}_1| \, (1 + \cos \theta_{k_1}) . 
\label{eq:mandel2}
\eea
Substituting the above results, the cross section (\ref{pxsection})
for one-gluon emission in the $q \bar{q}$ CM frame reads
\be
  \frac{\Delta_T\, d \sigma_{\rm R}^{q\bar{q}}}{d Q^2 d \Omega_{k_1}^{(D-1)}}
     = G (\epsilon )\,
      \int \Delta_T \mR\
      \frac{(1 - z)^{1 - 2 \epsilon}}
            {\mD_I{}^{2 - 2\epsilon}}\ d \Omega_k^{(D-1)} \ ,
\label{eq:oneglue}
\ee
where we have 
\[  G(\epsilon) = \frac{Q^2}{64 \pi^2 N_c s^2}\ 
        \left( \frac{4 \pi \alpha e_q}{Q^2} \right)^2\ 
        \left( \frac{Q^2}{16 \pi^2} \right)^{- \epsilon}\ 
        \frac{\alpha_s}{\pi}\ C_F \ 
        \left( \frac{4 \pi \mu^2}{Q^2}\right)^{\epsilon}\ 
        \frac{\pi^{\epsilon}}{4 \pi} . \]

With (\ref{eq:DI}), (\ref{eq:mandel2}), (\ref{eq:alphaphi}), and (\ref{eq:s1s2}),
the singular part of $\Delta_T \mR$, i.e., 
$\Delta_T \mR_S$ given in (\ref{reducedm}), reads
\bean
\Delta_T \mR_S
     &=& \frac{16\, Q^2}{\hat{t}\hat{u}} 
    \left[ \frac{Q^4}{\mD_I{}^2} \alpha_{\Phi} \sin^2 \theta_{k_1}
        + \epsilon\, \frac{Q^2 s}{2}\, s_1 \cdot s_2 \right]\\
   &=&  \frac{64\, Q^2}{s^2}\, \frac{1}{(1 - z)^2\, \sin^2 \theta_k}
    \left[ \frac{Q^4}{\mD_I{}^2}  \alpha_{\Phi} \sin^2 \theta_{k_1}
        + \epsilon\, \frac{Q^2 s}{2}\, s_1 \cdot s_2 \right] .
\eean
Using (\ref{eq:angmeas}) for the angular integration of the gluon in (\ref{eq:oneglue}),
and changing the integration variable $\cos \theta_k$ to 
$z_1 \equiv (1 + \cos \theta_k)\,/\,2$,
we obtain the contribution to (\ref{eq:oneglue}) from the singular part
of one-gluon emission as
\bea
  \frac{\Delta_T\, d \sigma_{{\rm R}(S)}^{q\bar{q}}}{d Q^2 d \Omega_{k_1}^{(D-1)}}
      &=& G (\epsilon ) \frac{32\, Q^2}{s^2}
         \frac{2^{- 2 \epsilon}}{(1 - z)^{1 + 2 \epsilon}}
          \int_0^1 d z_1  \int d \Omega_k^{(D-2)}
     \left[ \frac{z_1^{- \epsilon}}{(1 - z_1)^{1 + \epsilon}}
       + \frac{(1 - z_1)^{- \epsilon}}{z_1^{1 + \epsilon}} \right]
                 \nonumber\\
     && \qquad\qquad\qquad \times\ 
       \left[ \frac{Q^4}{\mD_I{}^{4 - 2 \epsilon}}\alpha_{\Phi}  \sin^2 \theta_{k_1}
                + \epsilon\, 
     \frac{Q^2 s}{2\, \mD_I{}^{2 - 2 \epsilon}} \, s_1 \cdot s_2 \right]
       \label{totals}\ .
\eea        
Because $\mD_I$ depends on the angles of both the gluon and lepton as 
expressed in (\ref{eq:DI2}), it is difficult to work out the integration
of (\ref{totals}) in $D=4-2\epsilon$ dimensions.
However, fortunately we can carry out the integration by noting the well-known formula
\be
    \frac{1}{(1 - z)^{1 + \epsilon}} =
           - \, \frac{1}{\epsilon} \, \delta (1 - z) +
            \frac{1}{(1 - z)_+} - \epsilon\, 
      \left( \frac{\ln (1 - z)}{1 - z} \right)_+ + {\mathcal O}
    (\epsilon^2) ,
\label{eq:wellknown}
\ee
and the similar formula obtained through the replacement $z \rightarrow 1-z$ 
in (\ref{eq:wellknown}); these formulae allow us to make manifest 
the singularities appearing in the limit $\epsilon \rightarrow 0$.
Here the ``+'' distributions are defined, as usual, as
\bea
\int_{0}^{1}dz \frac{f(z)}{(1-z )_+} &\equiv& \int_{0}^{1}dz \frac{f(z)-f(1)}{1-z },
\nonumber\\ 
\int_{0}^{1}dz f(z)  \left( \frac{\ln (1 - z)}{1 - z} \right)_+ 
&\equiv& \int_{0}^{1}dz \left[f(z)-f(1)  \right] \frac{\ln (1 - z)}{1-z},
\label{eq:plusdist}
\eea
for a smooth test function $f(z)$. 
Note that the $1/\epsilon$ poles proportional to $\delta (1-z_1 )$ and 
$\delta (z_1 )$, when employing the formula (\ref{eq:wellknown}), 
are associated with the radiation of the collinear gluon from an initial particle as 
$z_1 = (1 + \cos \theta_k)\,/\,2 = 1$ or $0$,
while the $1/\epsilon$ pole proportional to $\delta (1-z)$ 
is associated with the radiation of the soft gluon at the threshold for the reaction, 
$z=Q^2 /s = 1$.
After inserting (\ref{eq:wellknown}) into (\ref{totals}), 
we need to carry out the integration in arbitrary $D=4-2\epsilon$ dimensions 
only for the terms proportional to $1/\epsilon^2$ 
and $1/\epsilon$, associated with the following limiting cases for $\mD_I$:
\bea
   \mD_I (z \rightarrow 1) &=& 2 \ ,\label{eq:DIeps}\\
   \mD_I (z_1 \rightarrow  1) &=& 1 + z + (1 - z)\, \cos \theta_{k_1} \ , \\
   \mD_I (z_1 \rightarrow 0) &=& 1 + z - (1 - z)\, \cos \theta_{k_1}  \ .
\eea 
Then, the corresponding angular integrals in (\ref{totals}) can be performed exactly.
The integration of the remaining terms in (\ref{totals}), 
which are finite as $\epsilon \rightarrow 0$,
can be performed in $D=4$ dimensions, using [see (\ref{anglep4})]
\be
\mD_I ( D\rightarrow 4) = 1 + z + (1 - z)\, 
     \left( \cos \theta_k\, \cos \theta_{k_1} 
        + \sin \theta_k\, \sin \theta_{k_1} \cos\, (\phi_k -
             \phi_{k_1} ) \right) \ .
\label{eq:DI4}
\ee
The corresponding angular integrals 
can be done straightforwardly, by using the formulae in Appendix~A of Ref.~\citen{VW}.
Therefore, the phase space integrals in (\ref{totals}) are tractable 
exactly up to corrections that vanish as $\epsilon \rightarrow 0$.

We write the contribution to (\ref{eq:oneglue}) from the nonsingular part of 
$\Delta_T \mR$, i.e., $\Delta_T \mR_{NS}$ of (\ref{reducedm}), as 
$\Delta_T\, d \sigma_{{\rm R}(NS)}^{q\bar{q}}/d Q^2 d \Omega_{k_1}^{(D-1)}$.
Similarly to the finite terms in (\ref{totals}) discussed just above, 
the relevant angular integrals 
in $\Delta_T\, d \sigma_{{\rm R}(NS)}^{q\bar{q}}/d Q^2 d \Omega_{k_1}^{(D-1)}$ 
can be calculated in $D=4$ dimensions, where $\Delta_T \mR_{NS}$ reduces to 
\bea
&&  \Delta_T \mR_{NS} 
\non\\&& \hspace{1.cm}    
=\frac{32\, Q^2}{(1-z)^2\sin^2{\theta_k}} 
    \biggl[ \frac{(1-z)^2}{2}\sin^2{\theta_k}
        \cos(\phi_k - \phi_1)\, \cos (\phi_k - \phi_2) \nonumber \\
&&\hspace{1.cm}
  + z (1-z) \, \frac{\sin{\theta_k}\sin{\theta_{k_1}}}{\mD_I} 
        \left\{\cos(\phi_k-\phi_1)\cos(\phi_{k_1}-\phi_2)
          + ( 1 \leftrightarrow 2) \right\} \non\\
&&\hspace{1.cm}
  - \left\{z+\frac{(1-z)^2 }{4}\sin^2{\theta_k}-z(1+z)\frac{1}{\mD_I} 
           -z(1-z)\frac{\cos{\theta_k}\cos{\theta_{k_1}}}{\mD_I}\right\}
   \cos(\phi_1-\phi_2) \, \biggr]
\non\\
 \label{eq:RNS}
\eea
with $\phi_{1,2}$ parameterizing the spin vectors $s_{1,2}$ in 4
dimensions as in (\ref{spinvec4}).
Using (\ref{eq:RNS}), (\ref{eq:DI4}) and the formulae in Appendix~A of Ref.~\citen{VW},
the calculation is straightforward, and gives 
$\Delta_T\, d \sigma_{{\rm R}(NS)}^{q\bar{q}}/d Q^2 d \Omega_{k_1}^{(D-1)}$, 
up to corrections that vanish as $\epsilon \rightarrow 0$.

Combining the above results, we obtain  
$\Delta_T\, d \sigma_{\rm R}^{q\bar{q}}/d Q^2 d \Omega_{k_1}^{(D-1)}=\\
\Delta_T\, d \sigma_{{\rm R}(S)}^{q\bar{q}}/d Q^2 d \Omega_{k_1}^{(D-1)}+
\Delta_T\, d \sigma_{{\rm R}(NS)}^{q\bar{q}}/d Q^2 d \Omega_{k_1}^{(D-1)}$.
Then, using (\ref{eq:angmeas}) with $p \rightarrow k_1$ and 
integrating over $\theta_{k_1}$,
we finally derive the \lq\lq azimuthal\rq\rq\ angular distribution of 
one of the leptons for one-gluon radiation:
\bea
    \frac{\Delta_T\, d \sigma_{\rm R}^{q\bar{q}}}{d Q^2 d \Omega_{k_1}^{(D-2)}}
      &=& \frac{16\, \pi^{1 - \epsilon}}{\Gamma (1 - \epsilon)}
          \, G (\epsilon)\, Q^2 
      \, \left[\, \frac{1}{\epsilon^2} \, \delta (1 - z)
              - \frac{1}{\epsilon}\, \frac{2}{(1 - z)_+}
           \right] \nonumber\\
      && \times \left[ \, 2^{1 - 2 \epsilon} \,
               B (2-\epsilon\,,\,2 - \epsilon) \ \alpha_{\Phi}
       + \epsilon\, 2^{- 2 \epsilon} B (1 - \epsilon\,,\,1 - \epsilon)
                     (s_1 \cdot s_2) \right] \nonumber\\
     &+&  16\, \pi \, G (0)\, Q^2\ \frac{2}{3} \, 
                \cos (2 \phi_{k_1} - \phi_1 - \phi_2) \non\\
     &&  \times \left[ \, - \frac{\pi^2}{12} \ \delta (1 - z)
           + 2 \left( \frac{\ln (1 - z)}{1 - z} \right)_+
              + \frac{1 - z}{z} - \frac{3}{2} \ \frac{\ln^2 z}{1 - z}
               - \frac{\ln z}{1 - z} \right] \ .\nonumber\\
          \label{rx}
\eea
Apparently, this final form is valid in any frame in which 
the momenta $\boldsymbol{p}_1$ and $\boldsymbol{p}_2$ of 
the initial quark and antiquark in (\ref{partonic}) are collinear along the $z$-axis.

\subsection{Partonic cross section to ${\cal O}(\alpha_s)$ 
in the $\overline{\rm MS}$ factorization scheme}

With the tree-plus-virtual contribution (\ref{tvx}) added to (\ref{rx}),
the double poles in $\epsilon$ cancel out, while the single-pole terms remain. 
The result gives the azimuthal angular distribution to ${\cal O}(\alpha_s )$ as
\be
\frac{\Delta_T\, d \sigma^{q\bar{q}}}{d Q^2 d \Omega_{k_1}^{(D-2)}} =
\frac{\Delta_T\, d \sigma_{\rm T+V}^{q\bar{q}}}{d Q^2 d \Omega_{k_1}^{(D-2)}}
+ \frac{\Delta_T\, d \sigma_{\rm R}^{q\bar{q}}}{d Q^2 d \Omega_{k_1}^{(D-2)}}
=\frac{\Delta_T\, d \sigma_{\rm T}^{q\bar{q}}}{d Q^2 d \Omega_{k_1}^{(D-2)}}
+ \frac{\Delta_T\, d \sigma_{\rm V+R}^{q\bar{q}}}{d Q^2 d
\Omega_{k_1}^{(D-2)}} \ ,
\label{eq:qqbartotal}
\ee
for the parton-level process (\ref{partonic}) in the dimensional 
regularization. Here, in the second equality, 
we have rearranged $\Delta_T\, d \sigma^{q\bar{q}}/ d Q^2 d \Omega_{k_1}^{(D-2)}$
into a sum of the ${\cal O}(\alpha_s^0 )$ and ${\cal O}(\alpha_s^1 )$ contributions.
The remaining single-pole terms in 
$\Delta_T\, d \sigma_{\rm V+R}^{q\bar{q}}/ d Q^2 d \Omega_{k_1}^{(D-2)}$
represent the mass singularity associated with the emission of the collinear 
gluon. Indeed, these single-pole terms are completely absorbed into the 
parton distributions as the corresponding mass singularities:
We substitute the result (\ref{eq:qqbartotal}) 
into the LHS of the factorization formula (\ref{convo})
applied to the parton-level process (\ref{partonic}) and (\ref{pdf_1-loop}) in Appendix~A
into $\delta H(x_1, x_2 ; \mu_F^2)$ in the RHS as the transversity distributions 
in the $\overline{\rm MS}$ scheme for
the incoming quark and antiquark [see (\ref{tPDF})]. 
Matching both sides of the resulting formula, we determine the partonic cross section
$\Delta_T\, d \hat{\sigma}/ d Q^2 d \Omega_{k_1}^{(D-2)}$ on the RHS
order-by-order in $\alpha_s$. 
The matching at LO shows that the ${\cal O}(\alpha_s^0)$ term of
$\Delta_T\, d \hat{\sigma}/ d Q^2 d \Omega_{k_1}^{(D-2)}$ reads
\bea
\frac{\Delta_T \, d \hat{\sigma}^{(0)} (p_1, p_2 ; Q^2; \mu_F^2 )}{d Q^2 d
                \Omega_{k_1}^{(D-2)}} 
&&= \frac{1}{e_q^2} \frac{\Delta_T\, d \sigma_{\rm T}^{q\bar{q}}}
{d Q^2 d \Omega_{k_1}^{(D-2)}} 
\nonumber\\
 = \frac{1}{16 \pi^2 N_c} &&
       \left( \frac{4 \pi \alpha}{Q^2} \right)^2
   \left( \frac{Q^2}{16 \pi^2} \right)^{- \, \epsilon}
          \, \delta \, (1 - z ) \nonumber\\
\times 
 \left[ \, 2^{1 - 2 \epsilon}\, B(2-\epsilon\,,\, \right.&&\left. \!
2-\epsilon) \ 
       \alpha_{\Phi} 
          + \epsilon \, 2^{ - \, 2 \epsilon}\ B(1-\epsilon\,,\,1-\epsilon)
          \  (s_1 \cdot s_2) 
                 \right] , \label{ttt}
\eea
exactly for arbitrary $\epsilon$.
Apparently the RHS of (\ref{ttt}) is free from the $1/\epsilon$ singularities 
and does not depend on $\mu_F$.
Similarly, the matching at the NLO level yields the ${\cal O}(\alpha_s^1)$ term of
$\Delta_T\, d \hat{\sigma}/ d Q^2 d \Omega_{k_1}^{(D-2)}$ as
\bea
\lefteqn{\frac{\Delta_T \, d \hat{\sigma}^{(1)} (p_1, p_2 ; Q^2; \mu_F^2 )}{d Q^2 d
                \Omega_{k_1}^{(D-2)}} 
= \frac{1}{e_q^2}\left. 
\frac{\Delta_T\, d \sigma_{\rm V+R}^{q\bar{q}}}
{d Q^2 d \Omega_{k_1}^{(D-2)}}\right|_{\mu \rightarrow \mu_F}
+ \frac{1}{2\hat{\epsilon}}\frac{\alpha_s}{\pi}\int dx \Delta_T P_{qq}(x)}
\nonumber\\
&& \;\;\;\;\;\;\;\;\;\;
\times \left(
\frac{\Delta_T \, d \hat{\sigma}^{(0)} (x p_1, p_2 ; Q^2; \mu_F^2 )}{d Q^2 d
                \Omega_{k_1}^{(D-2)}} 
+ \frac{\Delta_T \, d \hat{\sigma}^{(0)} (p_1, x p_2 ; Q^2; \mu_F^2 )}{d Q^2 d
                \Omega_{k_1}^{(D-2)}} \right) ,
\label{tttt}
\eea
for the $\overline{\rm MS}$ factorization of the mass singularities at the scale $\mu_F$,
where $1/\hat{\epsilon}= 1/\epsilon -\gamma_{E} +\ln(4\pi )$, with $\gamma_E$ 
the Euler constant, and $\Delta_T P_{qq} (x)$ is 
the LO Dokshitzer-Gribov-Lipatov-Altarelli-Parisi (DGLAP) splitting function 
for the transversity distributions, given by (\ref{eq:LOAP}).
The terms proportional to $1/\hat{\epsilon}$ in (\ref{tttt})
are generated by the ${\cal O}(\alpha_s)$ term of the parton
distributions (\ref{pdf}), combined with the ${\cal O}(\alpha_s^0)$ terms 
of the partonic cross section, (\ref{ttt}), and cancel
the $1/\epsilon$ poles in the first term on the RHS of (\ref{tttt}).
As a result, (\ref{tttt}) is also finite as $\epsilon \rightarrow 0$.

Now, the mass singularities for the parton-level process (\ref{partonic}) 
have been completely factorized into the relevant parton distributions, and 
the $\epsilon \rightarrow  0$ limit of the sum of (\ref{ttt}) and (\ref{tttt})
gives the final result for the partonic cross section (\ref{pcs}) to 
${\cal O}(\alpha_s)$ in 4 dimensions,
\bea
  \lefteqn{\frac{\Delta_T d \hat{\sigma} ( 
p_1 , p_2 ; Q^2; \mu_F^2 )
}{d Q^2\, d \phi_{k_1}}
 = \frac{\alpha^2}{3 N_c \, s \, Q^2}\, 
             \cos \, (2 \phi_{k_1} - \phi_1 - \phi_2)}\nonumber\\
 && \;\;\;\;\;\;\;\;\;\;\;\;\;\;\;\;\;\;\;\; 
\times \left( \delta (1 - z) + \frac{\alpha_s}{2 \pi} \left[
2\, \Delta_T P_{qq}(z)\ 
       \ln \frac{Q^2}{\mu_F^2} 
  + C_F \left\{ 
 8 z \left(\frac{\ln (1 - z)}{1 - z}\right)_+ \right. \right.  \right. \non\\
              &  & \;\;\;\;\;\;\;\;\;\;\;\;\;\;\;\;\;\;\;\;
 -  \left. \left. \left. 6\,  \frac{z\,\ln^2z}{1-z} 
              - 4\, \frac{z\, \ln z}{1-z}
                + 4  (1 - z) + \left( \frac{2 \pi^2}{3} - 8 \right)
                 \delta (1 - z) \right\}\right] \right) ,  \label{fans}
\eea
where $\phi_{k_1}$ and $\phi_{1,2}$ are defined as in (\ref{anglep4}) 
and (\ref{spinvec4}), $\mu_F$ is the $\overline{\rm MS}$ factorization scale,
and we have used the fact [see (\ref{eq:alphaphi})]
\[   \alpha_{\Phi} ( D\rightarrow 4 ) = \cos\, (2 \phi_{k_1} - \phi_1 - \phi_2) . \]
The result (\ref{fans}), along with (\ref{eq:LOAP}), coincides with that
obtained in previous works employing the massive gluon scheme~\cite{VW} 
and the dimensional reduction scheme~\cite{CKM}, 
via the scheme transformation relation~\cite{V} (see also Ref.~\citen{MSSV}).
Substituting (\ref{fans}), and also the NLO transversity distributions
for the hadrons $h_{1,2}$ in the $\overline{\rm MS}$ scheme, into the RHS 
of (\ref{convo}), we obtain the NLO QCD prediction 
for the spin-dependent cross section of tDY (\ref{DYprocess})
in hadron-hadron collisions, $\Delta_T\, d \sigma / d Q^2 d \phi_{k_1}$,
as the mass ($Q$) and the azimuthal angular ($\phi_{k_1}$) distribution associated with
the observed lepton pair.

\section{The $Q_T$ differential cross section to ${\cal O}(\alpha_s)$}

In this section, we extend the calculation of the last section to the case
in which the final state in (\ref{DYprocess}) is observed in more detail.
We calculate the spin-dependent cross section for tDY, 
which is differential also in the transverse momentum
$Q_T$ and rapidity $y$ of the produced lepton pair,
\be
   \frac{\Delta_T  d \sigma}{d Q^2 \,d Q_T^2 \, d y \, d \phi_{k_1} } , 
\label{eq:differential}
\ee
taking into account the QCD mechanism up to ${\cal O}(\alpha_s )$ in perturbation theory.
The rapidity of the lepton pair is defined by
\be
   y = \frac{1}{2} \, \ln \frac{Q^0 + Q^3}{Q^0 - Q^3} 
\label{eq:rapidity}
\ee 
in the hadron-hadron CM system for (\ref{DYprocess}), in which we work in this section.
Here, $Q^0$ and $Q^3$ are the components of the dilepton's momentum, 
$Q^\mu =(Q^0, \boldsymbol{Q}_T, Q^3)$, with $Q^3$ denoting the component 
along the direction of the colliding beam, while $Q_T$ is defined 
as $Q_T = |\boldsymbol{Q}_T |$.
Now, in the factorization formula (\ref{convo}) for (\ref{eq:differential}),
the corresponding partonic cross section on the RHS depends also on $Q_T$ and $y$,
as $\Delta_T d \hat{\sigma} ( p_1, p_2 ; Q_T^2, Q^2 , y;  \mu_F^2)
/d Q^2 \,d Q_T^2 \, d y \, d \phi_{k_1}$.
Similarly to the last section, this partonic cross section 
can be obtained by matching the LHS and RHS of the factorization formula (\ref{convo})
in the case of the parton-level process (\ref{partonic}) with $Q_T$ and $y$ observed.
The LHS including the ${\cal O}(\alpha_s)$ corrections is represented by the
diagrams in Fig.~\ref{fig1},
which we calculate using dimensional regularization and obtain
$\Delta_T d \sigma^{q\bar{q}}/ d Q^2 \,d Q_T^2 \, d y \,d \Omega_{k_1}^{(D-2)} 
= \Delta_T d \sigma^{q\bar{q}}_{\rm T+V+R} / d Q^2 \,d Q_T^2 \, d y \,
d \Omega_{k_1}^{(D-2)}$
as the tree-plus-virtual contribution plus the real emission 
[compare with (\ref{eq:qqbartotal})].\footnote{Here and below, 
$\Delta_T d \sigma^{q\bar{q}}_{\rm V} / d Q^2 \,d Q_T^2 \, d y \,d \Omega_{k_1}^{(D-2)}$
denotes the contribution of the virtual-correction diagram in Fig.~\ref{fig1},
combined with the contribution due to the wave-function 
renormalization factors for the incoming quark and antiquark legs 
[see the discussion below (\ref{tandv})].}
As demonstrated in the last section, the corresponding order-by-order matching 
accomplishes the factorization of mass singularities in terms of 
$\delta H (x_1 ,\,x_2)$ of (\ref{tPDF}).
Because $\delta H (x_1 ,\,x_2)$ is universal (process independent), 
the factorization of the mass singularities 
can be expressed by the relation between 
$\Delta_T d \sigma^{q\bar{q}}/ d Q^2 \,d Q_T^2 \, d y \,d \Omega_{k_1}^{(D-2)}$ 
and $\Delta_T d \hat{\sigma}/ d Q^2 \,d Q_T^2 \, d y \,d \Omega_{k_1}^{(D-2)}$,
which is analogous to (\ref{ttt}) and (\ref{tttt}):
\bea
\frac{\Delta_T \, d \hat{\sigma} (p_1, p_2 ;  Q_T^2, Q^2 , y; \mu_F^2 )}
{d Q^2 dQ_T^2 dy d \Omega_{k_1}^{(D-2)}} 
&=& \frac{1}{e_q^2}\left[ 
\left.
\frac{\Delta_T\, d \sigma^{q\bar{q}}_{\rm T+V+R}}
{d Q^2 dQ_T^2 dy d \Omega_{k_1}^{(D-2)}}
\right|_{\mu \rightarrow \mu_F}
\!\!\!\!
+ \frac{\Delta_T\, d \sigma_{\rm CT}^{q\bar{q}}}
{d Q^2 dQ_T^2 dy d \Omega_{k_1}^{(D-2)}} \right]  \nonumber\\
&\equiv&
\frac{\Delta_T \, d \hat{\sigma}_{\rm T+V+R+CT} 
(p_1, p_2 ;  Q_T^2, Q^2 , y; \mu_F^2 )}{d Q^2 dQ_T^2 dy d
                \Omega_{k_1}^{(D-2)}} ,
\label{eq:masssing}
\eea
where the second term on the RHS of the first line plays the role of 
the ``counter-term'' to cancel the mass singularities
in the ``V+R'' contributions contained in the first term, and it reads,
in the $\overline{\rm MS}$ factorization scheme, 
\bea
\frac{\Delta_T\, d \sigma_{\rm CT}^{q\bar{q}}}
{d Q^2 dQ_T^2 dy d \Omega_{k_1}^{(D-2)}} &&=
\frac{1}{2\hat{\epsilon}}\frac{\alpha_s}{\pi}\int dx \Delta_T P_{qq}(x)
\nonumber\\
&&\hspace{-1.5cm} \times \left[ \left. \frac{\Delta_T\, d \sigma_{\rm T}^{q\bar{q}}}
{d Q^2 dQ_T^2 dy d \Omega_{k_1}^{(D-2)}} \right|_{p_1 \rightarrow xp_1}
\!\!\!\!+ \left.  \frac{\Delta_T\, d \sigma_{\rm T}^{q\bar{q}}}
{d Q^2 dQ_T^2 dy d \Omega_{k_1}^{(D-2)}} \right|_{p_2 \rightarrow xp_2}\right] ,
\label{eq:masssing2}
\eea
with the LO splitting function $\Delta_T P_{qq}(x)$ of (\ref{eq:LOAP}). 
From (\ref{pxsection}), (\ref{tandv}), and (\ref{eq:phase2}),
it is straightforward to show 
\be
\frac{\Delta_T\, d \sigma_{\rm T}^{q\bar{q}}}
{d Q^2 dQ_T^2 dy d \Omega_{k_1}^{(D-2)}} =
\delta(Q_T^2) \delta \left(y-\frac{1}{2}\ln \frac{p_1^0+p_1^3}{p_2^0-p_2^3}\right) 
\frac{\Delta_T\, d \sigma_{\rm T}^{q\bar{q}}}
{d Q^2 d \Omega_{k_1}^{(D-2)}} ,
\label{eq:newtree}
\ee
exactly in arbitrary $D=4-2\epsilon$ dimensions, where
$\Delta_T\, d \sigma_{\rm T}^{q\bar{q}}/d Q^2 d \Omega_{k_1}^{(D-2)}$ 
on the RHS is given by (\ref{ttt}).
At the tree level, apparently, it is impossible for the outgoing dilepton to have 
nonzero transverse momentum, and the rapidity is also determined completely 
by the kinematics of the initial state given by (\ref{partonic}).

It is not difficult to identify the $1/\epsilon$ poles in the ``V+R'' 
contributions through a careful treatment of the relevant phase space integration 
in $D$ dimensions, as done in (\ref{eq:wellknown})-(\ref{eq:DI4}) of the last section,
and in so doing to 
verify that those pole contributions are indeed canceled out by the ``CT''
contribution in the partonic cross section (\ref{eq:masssing}).
The result would yield the partonic cross section, which is finite as 
$D \rightarrow 4$ and is expressed in terms of the relevant partonic variables.
However, rather than performing such manipulations directly at the partonic level,
it is actually more convenient to perform the corresponding manipulations 
in the factorization formula (\ref{convo})
for the hadron-hadron collisions, with (\ref{eq:masssing})-(\ref{eq:newtree}) substituted
into the RHS as it is and their partonic variables re-expressed in terms
of the relevant hadronic variables of (\ref{DYprocess}).
Because the finite value of the dilepton's transverse momentum $Q_T$ 
is a consequence of the recoil from the gluon radiation, 
the IR behavior associated with the real gluon emission is controlled by $Q_T$.
Thus, rearranging the convolution integrals in the factorization formula (\ref{convo})
to make explicit its behavior as $Q_T \rightarrow 0$, we confirm the cancellation 
of the $1/\epsilon$ poles and simultaneously get the cross section formula 
for the hadron-hadron collisions, which is organized according to 
the $Q_T$ dependence; such a form of the cross section is particularly suitable for
the calculation of the dilepton $Q_T$ spectrum in hadron-hadron collisions,
and it is also even desirable when we attempt to include the soft gluon resummation 
effects relevant in the small-$Q_T$ region in the next section.

We denote the total CM energy in (\ref{DYprocess}) as
$\sqrt{S}$, and thus we have $P_1 = \frac{1}{2}\sqrt{S}(1, \boldsymbol{0},1)$, 
$P_2 = \frac{1}{2}\sqrt{S}(1, \boldsymbol{0},-1)$, where $\boldsymbol{0}$ denotes 
the  $D-2$ dimensional null vector.
We decompose the cross section (\ref{convo}) into four pieces, 
corresponding to the contributions on the RHS of (\ref{eq:masssing}), as
\be
   \Delta_T d \sigma \equiv \Delta_T d \sigma_{\rm T+V+R+CT} ,
\label{convonew}
\ee
and introduce the following useful hadronic variables according to 
the treatment of the unpolarized DY process in Ref.~\citen{AEGM}:
\bea
\tau =& & \frac{Q^2}{S}\ , \quad x_1^0 = \sqrt{\tau} \, e^y\ ,
     \quad x_2^0 = \sqrt{\tau} \, e^{- y} , \nonumber \\
    x_1^+ =& &
 \frac{Q^0 + Q^3}{\sqrt{S}} 
       = \left( \frac{Q^2 + Q^2_T}{S}\right)^{\frac{1}{2}} e^y  ,\quad
    x_2^+ = \frac{Q^0 - Q^3}{\sqrt{S}} 
       = \left( \frac{Q^2 + Q^2_T}{S}\right)^{\frac{1}{2}} e^{- y} .
\label{eq:hadronvar}
\eea
Here, the definition (\ref{eq:rapidity}) of $y$ is used, 
and $x_{1,2}^0$ are the DY scaling variables, as usual.
First, using (\ref{pxsection}), (\ref{eq:realg}) and (\ref{phase3old}), 
the partonic cross section for the one-gluon radiation of
(\ref{eq:masssing}) can be written for $p_i = x_i P_i $ as 
\bea
\lefteqn{\Delta_T d \hat{\sigma}_{\rm R} 
(x_1 P_1 , x_2 P_2 ) = \frac{1}{e_q^2}\left. \Delta_T d \sigma^{q\bar{q}}_{\rm R}
\right|_{p_i =x_i P_i} }\nonumber\\
     &&\;\;= \frac{1}{2 N_c \, s} C_F 
       \left( \frac{4 \pi \alpha g \mu^{\epsilon}}{Q^2}\right)^2
        \Delta_T \mR\
 d \Phi_3                    \non\\
  & &\;\;=  \frac{G^H (\epsilon)}{x_1 x_2 (Q_T^2)^{\epsilon}}
     \ \frac{\Delta_T \mR
}{16 \, Q^2 \, \mD_T^{2 - 2\epsilon}}
    \ \delta ((x_1 P_1 + x_2 P_2 - Q)^2) d \Omega_Q^{(D-2)} d \Omega_{k_1}^{(D-1)}
       d Q^2 d Q_T^2 d y , \non\\
\label{subx} 
\eea
where, for simplicity, we have denoted the factorization scale as $\mu$
and have suppressed some arguments of the partonic cross section
$\Delta_T d \hat{\sigma}_{\rm R}$. We have also introduced 
the following shorthand notation:
\bea
  G^H (\epsilon) &=& \frac{Q^4}{8 \pi^2 N_c\, S}\ 
        \left( \frac{4 \pi \alpha}{Q^2} \right)^2\ 
        \left( \frac{Q^2}{16 \pi^2} \right)^{- \epsilon}\ 
        \frac{\alpha_s}{\pi}\ C_F \ 
        \left( \frac{4 \pi \mu^2}{Q^2}\right)^{\epsilon}\ 
       \frac{\pi^{\epsilon}}{4 \pi} , \nonumber \\
   \mD_T &=&\frac{Q\cdot k_1}{|\boldsymbol{k}_1 |}= Q^0 - Q^3\ \cos \theta_{k_1} - Q_T\,
    \sin \theta_{k_1}\  \boldsymbol{n}_Q^{(D-2)} \cdot \boldsymbol{n}_{k_1}^{(D-2)} ,
\label{eq:GH}
\eea
with $Q^{0,\,3} = (Q^2 + Q_T^2)^{1/2} (e^y \pm e^{-y})/2$ 
from (\ref{eq:rapidity}). 
The quantity $\mD_T$ is generated in (\ref{subx}) as a phase space 
factor from $d\Phi_3$, while the delta function 
$\delta \left( (x_1 P_1 + x_2 P_2 - Q)^2 \right)$
comes from the on-shell condition for the final-state gluon and 
is given by $\delta(s+\hat{t}+\hat{u}-Q^2)$ in terms of the partonic variables 
in ({\ref{mandel}) and (\ref{eq:tuhat}).
Using the hadronic variables (\ref{eq:hadronvar}), this delta function becomes
\be
  \delta \left( (x_1 P_1 + x_2 P_2 - Q)^2 \right) = \frac{1}{S}\,
      \delta (x_1 x_2 - x_1 x_2^+ - x_2 x_1^+ + \tau) . \label{delta}
\ee
It is important to notice that, in general, a convolution integral like 
(\ref{convo}) with the delta function (\ref{delta}) can be split into
the two symmetric integrals~\cite{AEGM}~,
\footnote{When $\sqrt{\tau_+}e^y\geq 1$ ($\sqrt{\tau_+}e^{-y}\geq1$), 
the first (second) term in the RHS of (\ref{split}) vanishes, 
because of the support property of the parton distributions contained in $f(x_1, x_2)$
[see (\ref{convo})].}
\bea
\lefteqn{\int d x_1 d x_2 \, f (x_1\,,\,x_2)\,   
       \delta (x_1 x_2 - x_1 x_2^+ - x_2 x_1^+ + \tau) }\nonumber\\
  &  &\;\;\;\;\;\;= \int^1_{\sqrt{\tau_+} \, e^y} d x_1\,  
       \frac{f (x_1\,,\,x_2^*)}{x_1 - x_1^+} +
     \int^1_{\sqrt{\tau_+} \, e^{- y}} d x_2\,  
       \frac{f (x_1^*\,,\,x_2)}{x_2 - x_2^+} , \label{split}
\eea
where\footnote{Note that we have $x_{1\,,\,2}^+ = \sqrt{\tau_+} e^{\pm y}=
x_{1\,,\,2}^* = x_{1\,,\,2}^0$, when $Q_T^2 = 0$.}
\be
x_1^* = \frac{x_2 x_1^+ - x_1^0 x_2^0}{x_2 - x_2^+}\ , \quad
    x_2^* = \frac{x_1 x_2^+ - x_1^0 x_2^0}{x_1 - x_1^+}\ , \quad
    \sqrt{\tau_+} = \sqrt{\frac{Q_T^2}{S}} +
                    \sqrt{\tau + \frac{Q_T^2}{S}} .
\label{eq:hadronvar2}
\ee
In $\Delta_T \mR$ of (\ref{subx}), 
we first consider the singular part of (\ref{reducedm}), which now reads
\be
 \frac{\Delta_T \mR_S}{16 \, Q^2}
   = \frac{Q^2}{Q_T^2}
     \left[ \frac{Q^2}{4 \mD_T^2}  \alpha_{\Phi} \sin^2 \theta_{k_1}
          + \epsilon \frac{1}{2}\ (s_1 \cdot s_2) \right] , 
\label{eq:singularp}
\ee
and this yields the following contribution to (\ref{subx}):
\bea
 \frac{\Delta_T d \hat{\sigma}_{{\rm R}(S)} (x_1 P_1\,,\,x_2 P_2)}{d Q^2 d Q_T^2 d y} 
   &=&  G^H (\epsilon)  \, 
       \frac{1}{(Q_T^2)^{1 +\epsilon}} \, \frac{\tau}{x_1 x_2 } 
M_T (\epsilon, Q_T^2 ) \non \\
  &&\hspace{-1cm} \times          
    \delta (x_1 x_2 - x_1 x_2^+ - x_2 x_1^+ + \tau) 
            \ d \Omega_Q^{(D-2)} d \Omega_{k_1}^{(D-1)}  ,    
\label{subx1} 
\eea
where
\be
 M_T (\epsilon, Q_T^2 ) \equiv  \frac{1}{\mD_T^{4 - 2 \epsilon}}\, \frac{Q^2}{4}\,
     \alpha_{\Phi} \sin^2 \theta_{k_1} + \frac{1}{\mD_T^{2 - 2\epsilon}}
      \epsilon \frac{1}{2} (s_1 \cdot s_2)  \ .
\label{eq:MT2}
\ee
Inserting (\ref{subx1}) into (\ref{convo}) and using (\ref{split}), 
we obtain the contribution to $\Delta_T d \sigma_{\rm R}$ in
(\ref{convonew}), associated with the singular part (\ref{eq:singularp}),
\bea
  \frac{\Delta_T d \sigma_{{\rm R}(S)}}{d Q^2 d Q_T^2 d y} &=&
       G^H (\epsilon)\frac{1}{(Q_T^2)^{1 +\epsilon}} \left[ \int^1_{\sqrt{\tau_+} \, e^y}
      \ \frac{d x_1}{x_1 - x_1^+} \, \delta H (x_1\,,\,x_2^*)\, 
       \frac{\tau}{x_1 x_2^*}\right.
\non\\
     &&+
\left. \int^1_{\sqrt{\tau_+} \, e^{- y}}
      \frac{d x_2}{x_2 - x_2^+} \, \delta H (x_1^*\,,\,x_2)\, 
\frac{\tau}{x_1^* x_2}\right]
      M_T(\epsilon, Q_T^2 )  \ d \Omega_Q^{(D-2)} d \Omega_{k_1}^{(D-1)} ,
\non\\
\eea
where we have suppressed the $\mu^2$ dependence of $\delta H (x_1\,,\,x_2)$.
We follow the procedure used for the unpolarized DY process in Ref.~\citen{AEGM}, 
in order to isolate the singular terms in the limit $Q_T^2 \to 0$ of the above equation,
corresponding to the radiation of the collinear gluon:
\bea
&&
\frac{\Delta_T d \sigma_{{\rm R}(S)}}{d Q^2 d Q_T^2 d y} \non\\
&&\hspace{1cm}
=   G^H (\epsilon)\frac{1}{(Q_T^2)^{1 +\epsilon}}  \left[ \int^1_{\sqrt{\tau_+} \, e^y}
      \ \frac{d x_1}{x_1 - x_1^+} 
       \left\{ \delta H (x_1\,,\,x_2^*)
       \, \frac{\tau}{x_1 x_2^*} - \delta H (x_1^0 \,,\,x_2^0) \right\}\right.
\nonumber\\&&\hspace{1cm}
     + 
\int^1_{\sqrt{\tau_+} \, e^{- y}}
      \frac{d x_2}{x_2 - x_2^+} 
\left\{ \delta H (x_1^*\,,\,x_2)
       \, \frac{\tau}{x_1^* x_2} - \delta H (x_1^0 \,,\,x_2^0) \right\}
\nonumber\\&&\hspace{1cm}
     +  \left. 
\delta H (x_1^0 \,,\,x_2^0)\
      \ln \frac{(1 - x_1^+) (1 - x_2^+) S}{Q_T^2} \right] 
   M_T(\epsilon, Q_T^2 ) d \Omega_Q^{(D-2)} d \Omega_{k_1}^{(D-1)} .
\label{tricx}
\eea
We can further isolate the poles in $Q_T^2$ by using the identities
\bea
    \frac{Q^{2 \epsilon}}{(Q_T^2)^{1 + \epsilon}}
      &=& \frac{1}{(Q_T^2)_+}  - \frac{1}{\epsilon} \ \delta (Q_T^2)
               + \mathcal{O} (\epsilon) , \nonumber \\
    \frac{Q^{2 \epsilon}}{(Q_T^2)^{1 + \epsilon}}\ 
         \ln \frac{Q^2}{Q_T^2}
      &=& \left( \frac{\ln ( Q^2 / Q_T^2 )}{Q_T^2}\right)_+ +
                   \frac{1}{\epsilon^2} \  \delta (Q_T^2)
               + \mathcal{O} (\epsilon) , \label{eq:qtid}
\eea
where the ``+'' distributions that regulate the singularity at $Q_T^2 =0$ 
are defined such that~\cite{CSS} 
\bea
\int_0^{p_T^2} d Q_T^2 
     \left( \frac{\ln^n (Q^2 / Q_T^2)}{Q_T^2} \right)_+
     = -\frac{1}{n + 1}  \ln^{n + 1} ( Q^2 /p_T^2).
\label{eq:Q_T-plus}
\eea
Then, analogously to the situation discussed with regard to 
(\ref{eq:DIeps}) -- (\ref{eq:DI4}) for the case of the $Q_T$-integrated 
cross section, we find that only the two limiting cases for $M_T(\epsilon, Q_T^2 )$, 
\[ 
M_T (\epsilon = 0\,,\, Q_T^2 \neq 0) \quad\qquad {\rm and} \quad\qquad 
M_T (\epsilon \neq 0\,,\, Q_T^2 = 0) , \]
are required in order to evaluate the ($D-3$)-dimensional angular 
integration $d \Omega_Q^{(D-2)}$ in (\ref{tricx}),
up to corrections that vanish as $\epsilon \rightarrow 0$.
In these two cases, the integration over the scattering angle
$\theta_{k_1}$ of the lepton can be also performed easily.
After straightforward calculations, we obtain the
\lq\lq azimuthal\rq\rq\ angular distribution of the lepton
for one-gluon radiation as [see (\ref{anglep4}), (\ref{spinvec4})]
\bea
 \lefteqn{\frac{\Delta_T d \sigma_{{\rm R}(S)}}{d Q^2 d Q_T^2 d y d \Omega_{k_1}^{(D-2)}}
      = \frac{2 \pi}{Q^2} \,G^H (0)\, 
            \cos\, (2 \phi_{k_1} - \phi_1 - \phi_2)\,
       \frac{1}{3} \,\left( 1 + \frac{2\, Q_T^2}{Q^2}  \right)}\nonumber\\
    &\times& \left\{ \delta H (x_1^0\,,\,x_2^0) \left[ \,
       \frac{1}{(Q_T^2)_+}\, \ln \frac{(1 - x_1^+) (1 - x_2^+) S}{Q^2}
       + \left( \frac{\ln (Q^2 / Q_T^2 )}{Q_T^2} \right)_+ \right] \right.
                     \nonumber\\
    && \qquad\qquad + \frac{1}{(Q_T^2)_+}\, \int^1_{\sqrt{\tau_+} \, e^y}\ 
      \,  \frac{d x_1}{x_1 - x_1^+} 
        \left[ \delta H (x_1\,,\,x_2^*)
       \, \frac{\tau}{x_1 x_2^*} - \delta H (x_1^0 \,,\,x_2^0) \right] \nonumber\\
    && \qquad\qquad + \left. \frac{1}{(Q_T^2)_+}\, 
                 \int^1_{\sqrt{\tau_+} \, e^{- y}}
      \frac{d x_2}{x_2 - x_2^+} 
        \left[ \delta H (x_1^*\,,\,x_2)
       \, \frac{\tau}{x_1^* x_2} - \delta H (x_1^0 \,,\,x_2^0) \right] \right\}
                    \non\\
   &+& \frac{2 \pi}{Q^2} \, G^H (\epsilon)\, 
      \frac{\pi^{- \epsilon}\,2^{- 2 \epsilon}}{\Gamma (1 - \epsilon)}
      \left[ \, 2\,
               B (2-\epsilon\,,\,2 - \epsilon) \ \alpha_{\Phi}
       + \epsilon\, B (1 - \epsilon\,,\,1 - \epsilon)
                     (s_1 \cdot s_2) \right]\ \delta (Q_T^2) \nonumber\\
    && \times\ \left\{ \delta H (x_1^0\,,\,x_2^0)
          \left[ \frac{1}{\epsilon^2} - \frac{1}{\epsilon}
            \, \ln \frac{(1 - x_1^0) (1 - x_2^0) S}{Q^2} 
              \right] \right. \nonumber\\
    && \qquad - \frac{1}{\epsilon} \
    \left( \int^1_{x_1^0} \, \frac{d x_1}{x_1 - x_1^0} 
        \left[ \delta H (x_1\,,\,x_2^0)
       \, \frac{x_0}{x_1} - \delta H (x_1^0 \,,\,x_2^0) \right] \right.\nonumber\\
    && \hspace{5cm} + \left.\left. \int^1_{x_2^0}
      \frac{d x_2}{x_2 - x_2^0} 
        \left[ \delta H (x_1^0\,,\,x_2)
       \, \frac{x_2^0}{x_2} - \delta H (x_1^0 \,,\,x_2^0) \right] \right) \right\} .
                \nonumber\\\label{shx}
\eea

It is straightforward to show that the partonic cross section for the
virtual correction,
$\Delta_T \, d \hat{\sigma}_{\rm V} (p_1, p_2 )/ d Q^2 dQ_T^2 dy d\Omega_{k_1}^{(D-2)}$ 
of (\ref{eq:masssing}), is given by (\ref{eq:newtree}) multiplied by 
$(\alpha_s C_F/e_q^2 \pi )$ $( 4 \pi \mu^2/Q^2 )^{\epsilon} 
      [1/\Gamma (1 - \epsilon)]
         [ - 1 /\epsilon^2 - 3/ (2 \epsilon)
           - 4 + \pi^2 /2  + {\cal O}(\epsilon ) ]$, which is the factor 
appearing in the last parentheses of (\ref{tvx}). 
Substituting this into (\ref{convo}), the virtual-correction contribution in 
(\ref{convonew}) reads
\bea
   \lefteqn{\frac{\Delta_T d \sigma_{\rm V}}
         {d Q^2 d Q_T^2 d y d \Omega_{k_1}^{(D-2)}}}\nonumber\\
     &=& \frac{2 \pi}{Q^2} \, G^H (\epsilon)\, 
      \frac{\pi^{- \epsilon}\,2^{- 2 \epsilon}}{\Gamma (1 - \epsilon)}
      \left[ \, 2\,
               B (2-\epsilon\,,\,2 - \epsilon) \ \alpha_{\Phi}
       + \epsilon\, B (1 - \epsilon\,,\,1 - \epsilon)
                     (s_1 \cdot s_2) \right]\ \delta (Q_T^2) \nonumber\\
     && \hspace{4cm} \times \ \delta H (x_1^0\, , \,x_2^0)\
       \left[ \, - \frac{1}{\epsilon^2} - \frac{3}{2 \epsilon}
                   - 4 + \frac{\pi^2}{2} + {\cal O}(\epsilon ) \right] .
\label{eq:virtualnew}
\eea 
Combining this result with (\ref{shx}) and using the identity
\[ \int_{x^0}^1 dz \, \frac{f (z) - f (1)}{1 - z}
    = \int_{x^0}^1 dz \, \frac{f (z)}{(1 - z)_+} - f (1)\, \ln (1 - x^0) , \]
we get 
\bea
  \lefteqn{\frac{\Delta_T d \sigma_{{\rm R}(S)+{\rm V}}}
             {d Q^2 d Q_T^2 d y d \Omega_{k_1}^{D-2}}
      = \frac{2 \pi}{Q^2} \,G^H (0)\, 
       \cos\, (2 \phi_{k_1} - \phi_1 - \phi_2)\,
       \frac{1}{3} \,\left( 1 + \frac{2\, Q_T^2}{Q^2}  \right)}\nonumber\\
    &\times& \left\{ \delta H (x_1^0\,,\,x_2^0) \left[ \,
       \frac{1}{(Q_T^2)_+}\, \ln \frac{(1 - x_1^+) (1 - x_2^+) S}{Q^2}
       + \left( \frac{\ln ( Q^2 / Q_T^2 )}{Q_T^2} \right)_+ \right] \right.
                     \nonumber\\
    && \qquad\qquad + \frac{1}{(Q_T^2)_+}\, \int^1_{\sqrt{\tau_+} \, e^y}\ 
      \,  \frac{d x_1}{x_1 - x_1^+} 
        \left[ \delta H (x_1\,,\,x_2^*)
       \, \frac{\tau}{x_1 x_2^*} - \delta H (x_1^0 \,,\,x_2^0) \right] 
                          \nonumber\\
    && \qquad\qquad + \left. \frac{1}{(Q_T^2)_+}\, 
             \int^1_{\sqrt{\tau_+} \, e^{- y}}
      \frac{d x_2}{x_2 - x_2^+} 
        \left[ \delta H (x_1^*\,,\,x_2)
       \, \frac{\tau}{x_1^* x_2} - \delta H (x_1^0 \,,\,x_2^0) \right] \right\}
                  \non\\
   &+& \frac{2 \pi}{Q^2} \, G^H (\epsilon)\, 
      \frac{\pi^{- \epsilon}\,2^{- 2 \epsilon}}{\Gamma (1 - \epsilon)}
      \left[ \, 2\,
               B (2-\epsilon\,,\,2 - \epsilon) \ \alpha_{\Phi}
       + \epsilon\, B (1 - \epsilon\,,\,1 - \epsilon)
                     (s_1 \cdot s_2) \right]\ \delta (Q_T^2) \nonumber\\
    &\times& \left\{ \left( - 4 + \frac{\pi^2}{2} \right) 
                \delta H (x_1^0\,,\,x_2^0)
           \right.  \nonumber\\
    &-& \frac{1}{2\, C_F \epsilon} \left.
    \left[ \int^1_{x_1^0} \, \frac{d z}{z}\  \Delta_T \, P_{qq} (z)\,  
        \delta H \left( \frac{x_1^0}{z} \,,\,x_2^0 \right)
     + \int^1_{x_2^0} \frac{d z}{z}\   \Delta_T \, P_{qq} (z)\, 
         \delta H \left( x_1^0\,,\,\frac{x_2^0}{z} \right)
               \right] \right\} ,\non\\
     \label{svhx}          
\eea
with $\Delta_T P_{qq}(x)$ of (\ref{eq:LOAP}). 
We now derive the ``CT'' contribution of (\ref{convonew}) in the 
$\overline{\rm MS}$ scheme, 
using (\ref{eq:masssing})-(\ref{eq:newtree}) in (\ref{convo}), and
add the result to (\ref{svhx}).
We observe that the mass singularity poles in $\epsilon$ arising in (\ref{svhx})
are completely canceled by the $1/\epsilon$ poles of the CT contributions,
and taking the $\epsilon \rightarrow 0$ limit, we obtain in 4 dimensions 
\bea
  \lefteqn{\frac{\Delta_T d \sigma_{{\rm R}(S){\rm +V+CT}}}
             {d Q^2 d Q_T^2 d y d \phi_{k_1}}
      = \frac{2 \pi}{Q^2} \,G^H (0)\, 
         \cos\, (2 \phi_{k_1} - \phi_1 - \phi_2)\,
       \frac{1}{3} \,\left( 1 + \frac{2\, Q_T^2}{Q^2}  \right)}\nonumber\\
    &\times& \left\{ \delta H (x_1^0\,,\,x_2^0) \left[ \,
       \frac{1}{(Q_T^2)_+}\, \ln \frac{(1 - x_1^+) (1 - x_2^+) S}{Q^2}
       + \left( \frac{\ln (Q^2 / Q_T^2 )}{Q_T^2} \right)_+ \right] \right.
                     \nonumber\\
    && \qquad\qquad + \frac{1}{(Q_T^2)_+}\, \int^1_{\sqrt{\tau_+} \, e^y}\ 
      \,  \frac{d x_1}{x_1 - x_1^+} 
        \left[ \delta H (x_1\,,\,x_2^*)
       \, \frac{\tau}{x_1 x_2^*} - \delta H (x_1^0 \,,\,x_2^0) \right] 
                    \nonumber\\
    && \qquad\qquad + \left. \frac{1}{(Q_T^2)_+}\, 
             \int^1_{\sqrt{\tau_+} \, e^{- y}}
      \frac{d x_2}{x_2 - x_2^+} 
        \left[ \delta H (x_1^*\,,\,x_2)
       \, \frac{\tau}{x_1^* x_2} - \delta H (x_1^0 \,,\,x_2^0) \right] \right.
   \non\\
    &+& \delta (Q_T^2) \ \left( \left[ - 4 + \frac{\pi^2}{2} \right]
           \delta H (x_1^0\,,\,x_2^0) +  \frac{1}{2C_F}\ln \frac{Q^2}{\mu^2} \right. 
                   \nonumber\\
    &&\times \left. \left.       
            \left[ \int^1_{x_1^0} \, \frac{d z}{z}\, \Delta_T \, P_{qq} (z) 
        \, \delta H \left( \frac{x_1^0}{z} \,,\,x_2^0 \right)
     +  \int^1_{x_2^0} \frac{d z}{z}\,  \Delta_T \, P_{qq} (z)
         \, \delta H \left( x_1^0\,,\,\frac{x_2^0}{z} \right)
               \right] \right) \right\} .
                \nonumber               \\\label{fsvhx}
\eea
To completely isolate the growth of the form  $\sim 1/Q_T^2$ as $Q_T^2$ decreases  
from the third and fourth lines of (\ref{fsvhx}), one more step is required. 
We rewrite the integral in the third line as
\bea
  \lefteqn{\int^1_{\sqrt{\tau_+} \, e^y}\ 
      \,  \frac{d x_1}{x_1 - x_1^+} 
        \left[ \delta H (x_1\,,\,x_2^*)
       \, \frac{\tau}{x_1 x_2^*} - \delta H (x_1^0 \,,\,x_2^0) \right]}\non\\
   &=&  \int^1_{\sqrt{\tau_+} \, e^y}\ 
      \,  \frac{d x_1}{x_1 - x_1^+} 
        \left[ \delta H (x_1\,,\,x_2^*)
       \, \frac{\tau}{x_1 x_2^*} - \delta H (x_1^0 \,,\,x_2^0) \right]
   \non\\
   && \hspace{4cm}
    - \int^1_{x_1^0}\ 
      \,  \frac{d x_1}{x_1 - x_1^0} 
        \left[ \delta H (x_1\,,\,x_2^0)
       \, \frac{x_1^0}{x_1} - \delta H (x_1^0 \,,\,x_2^0) \right] \non\\
   && \hspace{2cm}
    +  \int^1_{x_1^0}\ 
      \,  \frac{d x_1}{x_1 - x_1^0} 
        \left[ \delta H (x_1\,,\,x_2^0)
       \, \frac{x_1^0}{x_1} - \delta H (x_1^0 \,,\,x_2^0) \right] \non\\
   &=& \int d x_1 \, \delta H_1
       + \frac{1}{2C_F}\, \int_{x_1^0}^1 \frac{dz}{z} \Delta_T P_{qq} (z)
       \, \delta H \left( \frac{x_1^0}{z} \,,\, x_2^0 \right)
        - \delta H (x_1^0 \,,\, x_2^0) \left( \frac{3}{4} 
         + \ln \frac{1 - x_1^0}{x_1^0} \right) ,
\non\\\label{modification}
\eea
introducing the following shorthand notation for the integral that
vanishes for $Q_T =0$:
\bea
\int dx_1\delta H_1&\equiv&
\int^1_{\sqrt{\tau_+} \, e^y}\ 
      \,  \frac{d x_1}{x_1 - x_1^+} 
        \left[ \delta H (x_1\,,\,x_2^* )
       \, \frac{\tau}{x_1 x_2^*} - \delta H (x_1^0 \,,\,x_2^0 ) \right]
\non\\
   && \hspace{2cm}
    - \int^1_{x_1^0}\ 
      \,  \frac{d x_1}{x_1 - x_1^0} 
        \left[ \delta H (x_1\,,\,x_2^0 )
       \, \frac{x_1^0}{x_1} - \delta H (x_1^0 \,,\,x_2^0 ) \right]. 
\non\\\label{eq:H1H1}
\eea
We also rewrite the integral in the fourth line in the same way,
using 
\bea
\int dx_2\delta H_2&\equiv&
\int^1_{\sqrt{\tau_+} \, e^{-y}}\ 
      \,  \frac{d x_2}{x_2 - x_2^+} 
        \left[ \delta H (x_1^* \,,\,x_2)
       \, \frac{\tau}{x_1^* x_2} - \delta H (x_1^0 \,,\,x_2^0 ) \right]
\non\\
   && \hspace{2cm}
    - \int^1_{x_2^0}\ 
      \,  \frac{d x_2}{x_2 - x_2^0} 
        \left[ \delta H (x_1^0 \,,\,x_2)
       \, \frac{x_2^0}{x_2} - \delta H (x_1^0 \,,\,x_2^0 ) \right]
\nonumber\\
 &=& \left. \int dx_1\delta H_1 \right|_{y \rightarrow -y} .
\eea
In this way, (\ref{fsvhx}) finally becomes
\bea
  \lefteqn{\frac{\Delta_T d \sigma_{{\rm R}(S){\rm +V+CT}}}
             {d Q^2 d Q_T^2 d y d \phi_{k_1}}
      = \frac{2 \pi}{Q^2} \,G^H (0)\, 
               \cos\, (2 \phi_{k_1} - \phi_1 - \phi_2)\,
       \frac{1}{6} \,\left( 1 + \frac{2\, Q_T^2}{Q^2}  \right)}\nonumber\\
    &\times& \left\{ \frac{2}{Q_T^2}\,
     \left( \int d x_1 \, \delta H_1 + \int d x_2 \, \delta H_2 \right)
              + 2\, \delta H (x_1^0\,,\,x_2^0)\, \frac{1}{Q_T^2}
              \, \ln \frac{(1 - x_1^+)(1 - x_2^+)}{(1 - x_1^0)(1 - x_2^0)} 
              \right. \nonumber\\
    && + \, \delta H (x_1^0\,,\,x_2^0)\,
         \left[\, 2\, \left( \frac{\ln ( Q^2 / Q_T^2 )}{Q_T^2} \right)_+ 
              - \frac{3}{(Q_T^2)_+} +
        \left(\, - 8 + \pi^2 \right) \delta (Q_T^2) \right] \non\\
    && + \left( \frac{1}{(Q_T^2)_+} 
        + \delta (Q_T^2)\, \ln \frac{Q^2}{\mu^2} \right)
               \nonumber\\ 
    && \qquad \times \left.
      \frac{1}{C_F} \left[ \int^1_{x_1^0} \, \frac{d z}{z}\, 
           \Delta_T \, P_{qq} (z) 
        \, \delta H \left( \frac{x_1^0}{z} \,,\,x_2^0 \right)
     +  \int^1_{x_2^0} \frac{d z}{z}\,  \Delta_T \, P_{qq} (z)
         \, \delta H \left( x_1^0\,,\,\frac{x_2^0}{z} \right)
               \right]  \right\} . \nonumber\\\label{finalsx}
\eea

Setting $\epsilon = 0$,
the contribution to (\ref{subx}) from the nonsingular part of $\Delta_T \mR$,
(\ref{reducedm}), is easily calculated as
\bean
    \frac{\Delta_T d \hat{\sigma} (x_1 P_1\,,\,x_2 P_2)_{{\rm R}(NS)}}
          {d Q^2 d Q_T^2 d y d \phi_{k_1}}
   &=&  - \, \frac{2 \pi}{Q^2} \, G^H (0)  \,
       \cos\, (2 \phi_{k_1} - \phi_1 - \phi_2)\,
          \frac{1}{Q_T^2}\, \ln \left( 1 + \frac{Q_T^2}{Q^2} \right)\\
  && \qquad\qquad \times\ \frac{\tau}{x_1 x_2}\          
          \delta (x_1 x_2 - x_1 x_2^+ - x_2 x_1^+ + \tau) .
\eean
Thus, the contribution to the hadronic cross section, 
$\Delta_T d \sigma_{\rm R}$ in (\ref{convonew}), is simply
\bea
    \lefteqn{\frac{\Delta_T d \sigma_{{\rm R}(NS)}}
          {d Q^2 d Q_T^2 d y d \phi_{k_1}}
   =  - \, \frac{2 \pi}{Q^2} \, G^H (0)  \,
       \cos\, (2 \phi_{k_1} - \phi_1 - \phi_2)\,
          \frac{1}{Q_T^2}\, \ln \left( 1 + \frac{Q_T^2}{Q^2} \right)}
  \non\\ 
  &\times& \left\{ 
            \int^1_{\sqrt{\tau_+} \, e^y} \, \frac{d x_1}{x_1 - x_1^+} 
         \, \delta H (x_1\,,\,x_2^*) \, \frac{\tau}{x_1 x_2^*}
     + \int^1_{\sqrt{\tau_+} \, e^{-y}} \, \frac{d x_2}{x_2 - x_2^+} 
         \, \delta H (x_1^*\,,\,x_2) \, \frac{\tau}{x_1^* x_2} \right\} .
                 \nonumber\\\label{finalnsx}
\eea

Adding (\ref{finalsx}), (\ref{finalnsx}), and the tree-level
contribution in 4 dimensions [see (\ref{eq:masssing})-(\ref{convonew})
and compare with (\ref{eq:virtualnew})], 
\[   \frac{\Delta_T d \sigma_{\rm T}}
          {d Q^2 d Q_T^2 d y d \phi_{k_1}}
    = \frac{Q^2}{16 \pi^2 N_c S} \left( \frac{4 \pi \alpha}{Q^2} \right)^2
     \, \frac{1}{3}\,
       \cos\, (2 \phi_{k_1} - \phi_1 - \phi_2)\ 
     \delta H (x_1^0\,,\,x_2^0)\, \delta (Q_T^2) , \]
and using
\[  \frac{2 \pi}{Q^2} \, G^H (0) = 
    \frac{Q^2}{16 \pi^2 N_c S} \left( \frac{4 \pi \alpha}{Q^2} \right)^2
     \, \frac{\alpha_s}{\pi}\, C_F ,
\]
we reach the final expression for the differential cross section 
(\ref{eq:differential}) for the tDY (\ref{DYprocess}) as (\ref{convonew}),
including the $\mathcal{O} (\alpha_s)$ QCD mechanism in the $\overline{\rm MS}$ 
factorization scheme:
\be
   \frac{\Delta_T d \sigma}{d Q^2 d Q_T^2 d y d \phi_{k_1}}
    = \mN\, \cos\, (2 \phi_{k_1} - \phi_1 - \phi_2)\
      \left[ \Delta_T X\, (Q_T^2 \,,\, Q^2 \,,\, y)
             + \Delta_TY\, (Q_T^2 \,,\, Q^2 \,,\, y) \right] . \label{owari} 
\ee
Here, we have 
\[ \mN = \frac{Q^2}{16 \pi^2 N_c S} \left( \frac{4 \pi \alpha}{Q^2} \right)^2
     \, \frac{1}{3} =
     \frac{\alpha^2}{3\, N_c\, S\, Q^2} , \] 
$\Delta_TX$ contains all terms that are singular as $Q_T^2 \rightarrow 0$, 
behaving as $Q_T^{-2} \times (\ln(Q^2 /Q_T^2 )$ or $1)$ or $\delta (Q_T^2)$,
and $\Delta_TY$ is the remaining ``finite'' part (containing less singular terms).
Writing $\Delta_TX$ as
\be
 \Delta_T X\, (Q_T^2 \,,\, Q^2 \,,\, y) 
= \Delta_T X^{(0)} (Q_T^2 \,,\, Q^2 \,,\, y) 
+ \Delta_T X^{(1)}(Q_T^2 \,,\, Q^2 \,,\, y)  , 
\label{eq:X00}
\ee
we have
\bea
&&
\Delta_TX^{(0)}  (Q_T^2 \,,\, Q^2 \,,\, y)  
=\delta H (x_1^0\,,\,x_2^0 ; \mu_F^2 )\ \delta (Q_T^2) , 
\label{eq:X0} \\
&&\Delta_TX^{(1)} (Q_T^2 \,,\, Q^2 \,,\, y)
\nonumber\\&&
 = \frac{\alpha_s (\mu_R^2 )}{2 \pi}
       \Biggl\{ 
    C_F\left[\, 2 \left( \frac{\ln (Q^2 / Q_T^2 )}{Q_T^2} \right)_+ 
              - \frac{3}{(Q_T^2)_+}
    + \left( \pi^2 - 8  \right) \delta (Q_T^2) \right] 
                     \delta H (x_1^0\,,\,x_2^0\,;\, \mu_F^2 ) 
\nonumber\\&&
+\left(  \frac{1}{(Q_T^2)_+} + \delta (Q_T^2) \ln \frac{Q^2}{\mu_F^2} \right)
\non\\&&
\times
       \left[ \int^1_{x_1^0} \frac{d z}{z}\right. \left.
\Delta_T \, P_{qq} (z)\,  
        \delta H 
     \left( \frac{x_1^0}{z}, x_2^0 ;\ \mu_F^2 \right)
        +  \int^1_{x_2^0} \frac{d z}{z}\,  \Delta_T \, P_{qq} (z)\, 
         \delta H \left( x_1^0\,,\,\frac{x_2^0}{z};\ \mu_F^2  \right)
 \right] \Biggr\} ,
\nonumber\\\label{eq:X} 
\eea
and
\bea
&&
   \lefteqn{
\Delta_TY\, (Q_T^2 \,,\, Q^2 \,,\, y) 
= 
         \frac{\alpha_s (\mu_R^2 )}{\pi}\, C_F} \nonumber\\
&&\times 
        \left\{ \, 
         \left[\, \frac{2}{Q^2}\,- \, \frac{3}{Q_T^2}\, 
       \ln \left( 1 + \frac{Q_T^2}{Q^2} \right)  \right] \right.
\nonumber\\
&&\times 
\left[ 
            \int^1_{\sqrt{\tau_+} \, e^y} \, \frac{d x_1}{x_1 - x_1^+}\,  
         \delta H (x_1\,,\,x_2^* ; \mu_F^2 ) \, \frac{\tau}{x_1 x_2^*}
\right.\non\\&&\left.\hspace{6.5cm}
+ \int^1_{\sqrt{\tau_+} \, e^{-y}} \, \frac{d x_2}{x_2 - x_2^+}\,  
         \delta H (x_1^*\,,\,x_2 ; \mu_F^2 ) \, \frac{\tau}{x_1^* x_2} \right]
\nonumber\\
    &+& 
     \left. 
\frac{1}{Q_T^2}\,
      \left[ \int d x_1 \, \delta H_1 
+ \int d x_2 \, \delta H_2 
               +
\delta H (x_1^0\,,\,x_2^0 ; \mu_F^2 )
      \, \ln \frac{(1 - x_1^+)(1 - x_2^+)}{(1 - x_1^0)(1 - x_2^0)} 
           \right]
\right\} ,
\label{eq:Y}
\eea
where we have recovered the notation of (\ref{convo}) for the 
factorization scale with the replacement $\mu^2 \rightarrow \mu_F^2$, 
so that $\mu_F$ in (\ref{eq:X}) and (\ref{eq:Y}) denotes 
the $\overline{\rm MS}$ factorization scale.
We have also made explicit the dependence of the QCD coupling constant
$\alpha_s (\mu_R^2 )$ on the corresponding $\overline{\rm MS}$ renormalization 
scale $\mu_{R}$.
When both of the colliding hadrons in (\ref{DYprocess}) are polarized along 
the $x$-axis, we have $\phi_1 = \phi_2 =0$ [see (\ref{spinvec4})].
Note that this result is invariant under the replacement $y \rightarrow -y$,  
as it should be.

Equation (\ref{owari}) with (\ref{eq:X}) and (\ref{eq:Y}) 
was first obtained in Ref.~\citen{KKST06}, but its derivation was not
described in detail there.
Here we give a complete derivation, 
explaining all the necessary techniques to deal with the complications in
the $D$-dimensional calculation involving transverse degrees of freedom.
We note that, integrating (\ref{owari}) over $Q_T^2$,
we obtain the cross section $\Delta_T d \sigma /d Q^2 d y d \phi_{k_1}$
in the $\overline{\rm MS}$ scheme, and the result coincides with that 
found in Ref.\citen{MSSV}. In Appendix~B, we also report the expression 
for $\Delta_T d \sigma /d Q^2 d Q_T^2 d \phi_{k_1}$,
integrating (\ref{owari}) over $y$.
When we integrate (\ref{owari}) over both $Q_T^2$ and $y$, the result 
coincides with the corresponding total cross section associated with the partonic 
cross section (\ref{fans}), which was calculated in the last section.

For $Q_T^2 > 0$, $\Delta_TX^{(0)}$ in (\ref{eq:X}) vanishes. 
Also, the terms proportional to $\delta(Q_T^2 )$ in (\ref{eq:X}),
including those associated with the + distribution,  
$( [ \ln^n (Q^2 / Q_T^2) ]/Q_T^2 )_+$, do not contribute.
The cross section (\ref{owari}) in this case is of ${\cal O}(\alpha_s)$, 
and it gives the LO QCD prediction for the spin-dependent cross section of 
the tDY in the large $Q_T$ region:
\bea
&&
\frac{\Delta_T d \sigma^{\rm LO}}{d Q^2 d Q_T^2 d y d \phi_{k_1}}
\non \\&&
\hspace{1cm}
= 
\mN \cos (2 \phi_{k_1} - \phi_1 - \phi_2)
 \left[ \left. \Delta_TX^{(1)}(Q_T^2 , Q^2 , y)\right|_{Q_T^2 >0} 
+ \Delta_TY(Q_T^2 , Q^2 , y) \right].
\nonumber \\
\label{cross section2}
\eea
%

\section{Transverse-momentum resummation: general formulation}

The $Q_T$-differential cross section of 
the tDY including one-loop corrections, (\ref{owari}), contains the singular part 
$\Delta_{T}X$ of (\ref{eq:X00})-(\ref{eq:X})
that grows as $\sim \alpha_s\ln(Q^2/Q_T^2)/Q_T^2$ and $\sim \alpha_s/Q_T^2$
as $Q_T^2 \rightarrow 0$. The calculations in the last section indicate
that these large contributions associated with the singularities at $Q_T^2 =0$
are induced as the recoil effects from the emission of a soft and/or collinear gluon.
Actually, corrections of this type appear at each order of 
the perturbative calculation and become very large for $Q_T \ll Q$.
Also, in the higher-loop corrections, the radiation of the soft and/or 
collinear partons produces the $1/\epsilon$ poles due to the IR divergences. 
After the factorization of the collinear singularities is accomplished 
in terms of the parton distribution functions,
the remaining IR divergences are found to cancel when
all the diagrams at the same order in $\alpha_s$ are combined.
The Kinoshita-Lee-Nauenberg (KLN) theorem~\cite{KLN} ensures that the corresponding 
cancellation of the IR divergences occurs between the real-emission and 
virtual-loop contributions,  due to the relevant soft-collinear radiations.
However, when the final dilepton of (\ref{DYprocess}) is kinematically 
constrained to have a small transverse-momentum $Q_T$, 
the real emission of the accompanying radiation is strongly inhibited.
In this case, the KLN cancellation of the corresponding IR divergences 
still occurs,  but it is ``incomplete'',  
because the phase space for the multiple real-emission in the final state 
is strongly restricted for small $Q_T$; 
specifically, all $1/\epsilon$ poles still cancel 
(at the leading power in $1/Q$) 
but the remainder depends in a singular manner on the small parameter 
$Q_T$ that constrains the phase space. 
These remnants of the incomplete cancellation at the boundary of 
phase space actually produce a series of enhanced logarithmic contributions,
\begin{equation}
\alpha_s^n \frac{\ln^m(Q^2/Q_T^2)}{Q_T^2}  \ ,   \;\;\;\;\;\;\;\;\;
(m= 2n-1, 2n-2, \ldots, 1, 0)
\label{eq:remnant}
\end{equation} 
at $n$-th order in $\alpha_s$. 
When only soft gluons are radiated and the sum of their transverse momenta 
balances $Q_T$, the contributions corresponding to $m=2n-1$, 
as well as the contributions with $m=2n-2, 2n-3, \ldots$, arise as the IR-finite piece. 
When collinear gluons are also radiated, contributions with $m=2n-2,2n-3, \ldots$ 
arise.\footnote{The higher-order contributions in (\ref{eq:remnant}) can
be produced by the radiation of semi-hard gluons which have $Q_T$
as net transverse momentum.}
These so-called ``recoil logarithms'' make the fixed-order perturbation
theory invalid for $Q_T \ll Q$, and have to be resummed to all orders in $\alpha_s$ 
in order to make a reliable prediction of the cross section at small $Q_T$. 

The corresponding resummation, the ``$Q_T$-resummation,'' can be treated 
on the basis of the general formulation of Collins, Soper and Sterman (CSS)~\cite{CSS}.
After reviewing the CSS formalism, emphasizing its universal structure, 
we use~\cite{KKST06} it to perform 
all-order resummation of the recoil logarithms for the tDY cross section, 
up to NLL accuracy, which corresponds to completely summing the first
three towers of logarithms, i.e., $\alpha_s^n \ln^m(Q^2/Q_T^2)/Q_T^2$ 
with $m=2n-1$, $2n-2$, and $2n-3$, for all $n$.
We also discuss various kinds of elaborations of our NLL resummation formula
beyond the original CSS form,
on the basis of recent developments in the $Q_T$-resummation 
formalism~\cite{dG,CdG,KLSV,BCFG}.
Combining the resulting NLL-resummed cross section with the leading-order 
(LO) cross section (\ref{cross section2}) in a consistent matching procedure,
we obtain the ``NLL+LO'' cross section of 
the tDY, which has a uniform accuracy over the entire range of $Q_T$.    

\subsection{Collins-Soper-Sterman (CSS) resummation formalism}

First, we explain the $Q_T$-resummation formalism in a general form,
such that it is not restricted to the present case of 
the tDY 
(\ref{DYprocess}). 
For this purpose, we consider the process
\be
h_1 (P_1 ) + h_2 (P_2 ) \to F (Q^2 , Q_T^2 ,y ; \phi) + X  ,
\label{eq:genpro}
\ee
where the collisions of the hadrons $h_1$ and $h_2$ with momenta $P_1$ and $P_2$ produce
a system of non-strongly interacting final-state particles, $F$, 
carrying total momentum $Q$, total transverse momentum $Q_T$ and rapidity $y$. 
The additional variable $\phi$ denotes the possible dependence on the kinematics
of the final-state particles in $F$, such as individual transverse momenta.
For the case of the tDY azimuthal angular distribution discussed in \S4, 
we have $\phi \rightarrow \phi_{k_1}$, the azimuthal angle of the final lepton. 
In general, $h_{1,2}$ may be polarized hadrons, but here their 
spins are suppressed for simplicity. 

The CSS resummation formalism yields the cross section for this process,
which is applicable to the small $Q_T$ region as well as the large $Q_T$ region.
The corresponding CSS formula possesses 
the following general structure, decomposed into two types of terms:
\bea  
 \frac{d \sigma^{\rm (CSS)}}{d Q^2 d Q_T^2 d y d \phi}
       = \frac{d \sigma^{\rm (res.)}}{d Q^2 d Q_T^2 d y d \phi}
      + \frac{d \sigma^{\rm (fin.)}}{d Q^2 d Q_T^2 d y d \phi} \ .
\label{cross section}
\eea
Here, both terms on the RHS are expressed as convolutions of the corresponding 
partonic cross sections and the parton distributions, which are formally 
similar to the RHS of (\ref{convo}), but the partonic cross sections 
participating in (\ref{cross section}) are more sophisticated than 
those calculated in \S4. The first term in (\ref{cross section}), 
$d\sigma^{\rm (res.)} / d Q^2 d Q_T^2 d y d \phi$,  
can be evaluated by resumming the ``singular'' terms, 
like $\Delta_T X$ of (\ref{eq:X00})-(\ref{eq:X}), to all orders in $\alpha_s$. 
In particular, $d\sigma^{\rm (res.)} / d Q^2 d Q_T^2 d y d \phi$
collects the logarithmic contributions (\ref{eq:remnant}), which dominate 
the cross section at small $Q_T$.
The second term, $d\sigma^{\rm (fin.)}/ d Q^2 d Q_T^2 d y d \phi$,  
is the ``finite'' component that is not associated with the 
logarithmically-enhanced contributions (\ref{eq:remnant}), 
and thus it can be computed with fixed-order perturbation theory.
Formally, we can define the corresponding finite component
analogously to $\Delta_T Y$ in (\ref{owari}) and (\ref{eq:Y}); i.e., 
$d\sigma^{\rm (fin.)}/ d Q^2 d Q_T^2 d y d \phi$ is obtained 
as the contributions less singular than
$1/Q_T^2$ or $\delta(Q_T^2)$ as $Q_T^2 \rightarrow 0$ among the terms 
in the cross section $d\sigma / d Q^2 d Q_T^2 d y d \phi$, 
which is calculated up to an appropriate order in $\alpha_s$.
In (\ref{cross section}), the first term dominates the cross section 
$d\sigma^{\rm (CSS)}/ d Q^2 d Q_T^2 d y d \phi$ for $Q_T\ll Q$,  
and thus the second term is negligible in this region. 
But the second term becomes important when $Q_T\sim Q$. 

In the CSS resummation formalism, the resummation of the logarithmic 
contributions (\ref{eq:remnant}) is carried out 
with the ``$b$-space'' resummation approach,
introducing a 2-dimensional impact parameter $\boldsymbol{b}$ 
that is the Fourier conjugate of the transverse momentum $\boldsymbol{Q}_T$.
As noted above, multiple-gluon emission induces (\ref{eq:remnant}) as
its IR-finite piece, and the corresponding contributions can be resummed 
most straightforwardly in the $b$ space, to all orders in the perturbation 
theory~\cite{PP}. After the resummation, the $b$-space cross section is
Fourier-transformed back to the $Q_T$ space.
In this way, the first term in (\ref{cross section}) is obtained as~\cite{CSS}
\be
  \frac{d \sigma^{\rm (res.)}}{d Q^2 d Q_T^2 d y d \phi}
    = \frac{1}{4\pi} \int d^2 b\ e^{i \boldsymbol{b}\cdot\boldsymbol{Q}_T}
    \mW(b; Q, y, \phi)
     = \int_0^{\infty} d b \, \frac{b}{2}\,J_0 (b Q_T) \mW(b; Q, y, \phi),
\label{resum00}
\ee
where $J_0(bQ_T)$ is a Bessel function, and the $b$-space representation 
$\mW(b; Q, y, \phi)$ has the general structure 
[using the notation of Ref.~\citen{dG} and the hadronic variables 
(\ref{eq:hadronvar}) also for the present case (\ref{eq:genpro})]
\bea
\mW(b; Q, y, \phi) &=&\sum_{j,i,k} \int_{0}^{1}dx_1 \int_{0}^{1}dx_2 
\frac{d \sigma^{j \bar{j}}_{\rm (LO)}(Q^2)}{d \phi}\delta(Q^2 - x_1 x_2 S) 
\delta\left(y-\frac{1}{2}\ln\frac{x_1}{x_2}\right) \nonumber\\
&& \times
( C_{j i} \otimes f_{i / h_1})\, 
           \left( x_1\,,\, \frac{b_0^2}{b^2} \right)
          ( C_{\bar{j} k} \otimes f_{k / h_2})\, 
           \left( x_2\,,\, \frac{b_0^2}{b^2} \right)e^{\mS_j (b , Q)}
\nonumber\\
&=&  \frac{1}{S} \sum_{j}
\frac{d \sigma^{j \bar{j}}_{\rm (LO)}(Q^2)}{d \phi}W_{j}(b;\, Q, x_1^0, x_2^0) ,
\label{resum0}
\eea
with
\be
 W_j (b;\, Q, x_1, x_2)    = \sum_{i,k}\,   
     e^{\mS_j (b , Q)}  ( C_{j i} \otimes f_{i / h_1})\, 
           \left( x_1 , \frac{b_0^2}{b^2} \right) 
          ( C_{\bar{j} k} \otimes f_{k / h_2})\, 
           \left( x_2 , \frac{b_0^2}{b^2} \right) .
\label{resum}
\ee
Here, the subscripts $j , i$ and $k$ can be $q , \bar{q}$ or $g$, 
including the flavor degrees of freedom, 
$\sigma^{j\bar{j}}_{\rm (LO)}(Q^2)$ is the lowest-order cross section
(integrated over $Q_T^2$)\footnote{
For the spin-dependent cross section in tDY, $j=q$ is relevant, where 
$\sigma^{q\bar{q}}_{\rm (LO)}(Q^2)$ is the LO cross section for 
the $q\bar{q}$ annihilation process $q + \bar{q} \to l\bar{l} (Q^2)$,
which is equal to $\Delta_T \sigma^{q\bar{q}}_{\rm T}$ calculated 
in \S3 [see (\ref{ttt}), (\ref{fans})].} 
for the parton-level process $j + \bar{j} \to F (Q^2)$,
and $b_0=2e^{-\gamma_E}$, with $\gamma_E$ being the Euler constant, 
has a kinematical origin.
The quantity $C_{ji}$ is the coefficient function, and $f_{i/h}(x,\mu^2)$ 
is the unpolarized or polarized distribution function for a parton $i$ inside 
the hadron $h$.
The symbol $\otimes$ denotes their convolution, defined by
\be
 \left( C_{ji} \otimes f_{i/h}\right)  (x, \mu^2 ) = \int_x^1\,
        \frac{d z}{z}\, C_{ji} (z, \alpha_s(\mu^2) )\, f_{i/h}  (x / z, \mu^2)  .
\label{eq:Mcon}
\ee
The large logarithmic corrections are resummed into the Sudakov form
factor $e^{\mS_j(b,Q)}$ in terms of the exponent
\bea
\mS_j (b , Q) = - \, \int^{Q^2}_{b_0^2 / b^2} \frac{d \kappa^2}{\kappa^2}
          \left\{ A_j (\alpha_s (\kappa^2)) \ln \frac{Q^2}{\kappa^2} 
                        + B_j (\alpha_s (\kappa^2)) \right\} ,
\label{sudakov:1}
\eea
with the perturbatively calculable functions $A_j(\alpha_s)$ and $B_j (\alpha_s)$.
Here, for $Q \gg 1/b$, the leading contribution from the $A_j(\alpha_s)$
term is enhanced in comparison with that from the $B_j(\alpha_s)$ term 
due to an explicit logarithm $\ln (Q^2/\kappa^2)$ in the integrand. 
Other large logarithms are also implicit in the integration over $\kappa^2$ 
for both terms. 
In the form factor $e^{\mS_j(b,Q)}$, the $A_j(\alpha_s)$ term of 
(\ref{sudakov:1}) represents exponentiation of large logarithms due to 
soft radiation, while the $B_j(\alpha_s)$ term represents exponentiation 
of large logarithms due to flavor-conserving collinear radiations.~\cite{CdG}

The quantity
$W_j(b;Q, x_1, x_2)$ in (\ref{resum}) depends on $Q^2$ only through 
the Sudakov exponent (\ref{sudakov:1}). This simple structure is a consequence 
of the fact that $W_j(b;Q, x_1, x_2)$ is obtained as the solution of
the following evolution equation~\cite{CSS}:
\be
\frac{\partial}{\partial \ln{Q^2}}W_j (b;Q,x_1, x_2)
=-\left[\int_{b_0^2/b^2}^{Q^2}\frac{d\kappa^2}{\kappa^2}
A_j(\alpha_s(\kappa^2))+B_j(\alpha_s(Q^2)) \right]
W_j(b;Q, x_1, x_2)  .
\label{eq:W}
\ee
Moreover, the large logarithm of $Q^2 b^2$, arising in the integral
$\int_{b_0^2/b^2}^{Q^2}(d\kappa^2 /\kappa^2) A_j(\alpha_s(\kappa^2))$ of (\ref{eq:W})
when $b \sim 1/Q_T \gg 1/Q$, is controlled by the renormalization group (RG)
equations governed by certain anomalous dimensions. 
For $j=q$, which is relevant to the DY and vector boson production,
the corresponding RG equations, as well as (\ref{eq:W}), have been
derived on the basis of the factorization property of the quark form factor 
in QCD into ``hard,'' ``soft,'' and ``jet'' factors to all orders in $\alpha_s$,
and also, exploiting the renormalization property of those individual factors, 
defined as matrix elements of gauge-invariant operators
(see Refs.~\citen{CSS} and \citen{Collins89}).\footnote{
In this approach, $A_q$ and $B_q$ appearing on the RHS of (\ref{eq:W})
are related to another set of two evolution kernels, denoted as 
``$K$'' and ``$G$'' in Ref.~\citen{CSS}. 
$K$ and $G$, respectively, obey the RG equations governed by the so-called 
cusp anomalous dimension. This relation allows us to confirm that 
$A_q$ and $B_q$ contain no $\ln Q^2 b^2$ contributions and indeed 
take the forms of power series in $\alpha_s$.}

The solution of (\ref{eq:W}) is generally expressed as 
\be
W_j (b;Q,x_1 , x_2)=e^{\mS_j(b,Q)}~W_j (b;b_0/b,x_1 , x_2)  ,
\label{eq:W2}
\ee
using the Sudakov exponent $\mS_j(b,Q)$ of (\ref{sudakov:1}).
With the boundary value $W_j (b; b_0/b,x_1 , x_2)$ specified, 
(\ref{eq:W2}) determines the complete behavior of $W_j (b;Q,x_1 , x_2)$.
For this purpose, we note that when $Q_T \ll Q$, and for the relevant region 
$b\sim 1/Q_T$, large logarithms of $Q^2 b^2 \sim Q^2/Q_T^2$ indeed 
arise in the Sudakov factor of (\ref{eq:W2}) via the integral of (\ref{sudakov:1}).
On the other hand, the boundary value $W_j (b; b_0/b,x_1 , x_2)$ depends 
explicitly on only one distance scale, namely $b$,
and thus, for $b \ll 1/\Lambda_{\rm QCD}$, its $b$ dependence is calculable 
by the customary perturbation theory.
Specifically, 
$W_j (b;b_0/b,x_1 , x_2)$ is determined by the singular component of 
the fixed-order differential cross section, which has a form analogous 
to that of $\Delta_T X$ in (\ref{eq:X00})-(\ref{eq:X}) obtained in \S4.
Setting $Q = b_0 /b$ and the factorization scale as $\mu_F = b_0/b$
in formulae like (\ref{eq:X00})-(\ref{eq:X}), 
the Fourier transformation of the corresponding singular component
from the $Q_T$ space to the $b$ space yields a result with the structure
\be
 W_j (b;\, b_0/b, x_1, x_2)    = \sum_{i,k}\,   
     ( C_{j i} \otimes f_{i / h_1})\, 
           \left( x_1 , \frac{b_0^2}{b^2} \right) 
          ( C_{\bar{j} k} \otimes f_{k / h_2})\, 
           \left( x_2 , \frac{b_0^2}{b^2} \right) ,
\label{eq:W3}
\ee
which is factorized into two parts, corresponding to the incoming hadrons, 
$h_1$ and $h_2$. 
This reflects the collinear nature of radiative corrections 
associated with the singular terms in the relevant region, $Q_T \sim 1/b$; i.e., 
the cross section receives logarithms of $b\Lambda_{\rm QCD}$ only 
from collinear radiation, which can be treated with the DGLAP evolution of the 
parton distributions $f_{i/h} (x, b_0^2/b^2)$ associated with each hadron, 
and $C_{ji}$ of (\ref{eq:W3}) represents the remaining perturbative 
corrections as a power series in $\alpha_s$.
Therefore, combining the evolution equation (\ref{eq:W}) with (\ref{eq:W3}) obtained 
from fixed-order perturbation theory, 
the general formula (\ref{resum}) follows~\cite{CSS}. 
This analysis also demonstrates that (\ref{resum}) is 
accurate in the region $b \ll 1/\Lambda_{\rm QCD}$.

We express the expansions of the functions $A_j(\alpha_s)$ and 
$B_j (\alpha_s)$ in (\ref{sudakov:1}) and the coefficient
function $C_{ji} (z , \alpha_s)$ in (\ref{resum}) in powers of $\alpha_s$ as
\bea
    A_j (\alpha_s) &=& \sum_{n=1}^{\infty} 
          \left( \frac{\alpha_s}{2 \pi}\right)^n A_j^{(n)} , \nonumber\\
    B_j (\alpha_s) &=& \sum_{n=1}^{\infty} 
          \left( \frac{\alpha_s}{2 \pi}\right)^n B_j^{(n)} , \nonumber \\
   C_{ji} (z , \alpha_s) &=& \delta_{ji} \, \delta (1 - z) +
           \sum_{n=1}^{\infty} 
          \left( \frac{\alpha_s}{2 \pi}\right)^n C_{ji}^{(n)} (z) \ .
\label{expand}
\eea
The coefficient $A^{(1)}$ yields the LL resummation,
$\{A^{(2)}\,,\,B^{(1)}\,,\,C^{(1)}\}$ give the NLL terms,
$\{A^{(3)}\,,\,B^{(2)}\,,\,C^{(2)}\}$ give the NNLL contributions, and so on.
In particular, among the coefficients $\{ A^{(1)}, A^{(2)}, B^{(1)}, C^{(1)} \}$ 
that are necessary for the NLL resummation, $\{ A^{(1)}\, , \, A^{(2)}\, , \, B^{(1)}\}$ 
are known to be independent of the final states 
(see Refs.~\citen{dG,CdG} and \citen{BCFG}).
For the DY process, (\ref{eq:genpro}) with $F = l\,\bar{l}$, which we consider 
in this paper, and also for other processes, such as $W$ and $Z$ boson production,
these coefficients with $j = q$ are necessary, and they are given by
\bea
A_q^{(1)} = 2\, C_F , \quad A_q^{(2)} = 2\, C_F K  ,
       \quad B_q^{(1)} = - 3\, C_F  , 
\label{coeff_NLL}
\eea
with~\cite{KT}
\bea 
K = \left( \frac{67}{18} - \frac{\pi^2}{6} \right) C_G 
        - \frac{5}{9} N_f  ,
\label{eq:K}
\eea
where $C_G=N_c$, and $N_f$ is the number of QCD massless flavors.
The result (\ref{coeff_NLL}) can be derived directly by evaluating 
certain loop diagrams which represent the evolution kernel of 
(\ref{eq:W})~\cite{CSS,Collins89}.
Apparently, for $A_q^{(2)}$, this requires a two-loop calculation.
Nevertheless, its value is known to be independent of the process as well as the spin. 
In fact, we can confirm that this is indeed the case by using a relation~\cite{KT} 
between $A_q^{(2)}$ and the usual two-loop DGLAP kernels 
for the parton distribution functions.\footnote{
The coefficients $A_q^{(1)}$ and $A_q^{(2)}$  are given by the one- and 
two-loop terms of the universal cusp anomalous dimension, respectively.
The cusp anomalous dimension plays a role~\cite{Korch} 
in the evolution of the quark distribution functions
for large $x$.}
The parton distributions in (\ref{resum}) obey the DGLAP equation~\cite{BDR:02},
\be
\mu^2 \frac{\partial}{\partial \mu^2}f_{i/h}(x, \mu^2)=\frac{\alpha_s(\mu^2)}{2\pi}
\sum_{j} \int_{x}^1\frac{dz}{z} \mathscr{P}_{ij}
\left(\frac{x}{z}, \alpha_s(\mu^2) \right) f_{j/h}(z, \mu^2) ,
\label{eq:DGLAP}
\ee
with the corresponding DGLAP kernel $\mathscr{P}_{ij}(z, \alpha_s(\mu^2) )$
given by a power series in $\alpha_s$:
\begin{equation}
  \mathscr{P}_{ij} (z,\alpha_s)
      =    
                    \mathscr{P}_{ij}^{(0)}(z)                  
         +  \frac{\alpha_s}{2 \pi} \,
                     \mathscr{P}_{ij}^{(1)}(z) + \cdots .
\label{eq:DGLAPk}
\end{equation}
The behavior of $\mathscr{P}_{ij}(z, \alpha_s(\mu^2) )$ for $z \to 1$ 
is dominated by soft gluon emissions, and it is diagonal in $i,j$.
Indeed, the coefficients $A_q^{(1)}$ and $A_q^{(2)}$ are, respectively, 
related to the dominant large-$z$ behavior $\propto 1/(1-z)_+$ in the one- 
and two-loop terms of (\ref{eq:DGLAPk}) for emission of gluons from a quark:
\be
                    \mathscr{P}_{qq}^{(0)}(z)                  
         +  \frac{\alpha_s}{2 \pi} \,
                     \mathscr{P}_{qq}^{(1)}(z) 
\rightarrow
                     \frac{2 \, C_F}{(1 -  z)_+}
         +  \frac{\alpha_s}{2 \pi} \,
                     \frac{2 \, C_F K}{(1 -  z)_+} .
\label{eq:z1}
\ee
The large $z$ behavior (\ref{eq:z1}) is universal to all DGLAP kernels 
associated with the twist-2 quark distributions, i.e., to the DGLAP
                    kernels 
for density distribution $q(x, \mu^2)$, helicity distribution 
$\Delta q(x, \mu^2)$, and transversity distribution $\delta q(x, \mu^2)$.
We also note that the term $\propto \delta(1-z)$ in the one-loop kernel 
$\mathscr{P}_{qq}^{(0)}(z)$ is also universal, and its coefficient determines 
the value of $B_q^{(1)}$ in (\ref{coeff_NLL})~\cite{dG}.
The similar large-$z$ behavior
\begin{displaymath}
                    \mathscr{P}_{gg}^{(0)}(z)                  
         +  \frac{\alpha_s}{2 \pi} \,
                     \mathscr{P}_{gg}^{(1)}(z) 
\rightarrow
                     \frac{2 \, C_G}{(1 -  z)_+}
         +  \frac{\alpha_s}{2 \pi} \,
                     \frac{2 \, C_G K}{(1 -  z)_+} 
\end{displaymath}
and the universal coefficient of the $\delta(1-z)$ term in the one-loop kernel
$\mathscr{P}_{gg}^{(0)}(z)$ lead to~\cite{CDT}
\begin{displaymath}
A_g^{(1)} = 2\, C_G , \quad A_g^{(2)} = 2\, C_G K  ,  \quad B_g^{(1)} =
 - 4\pi \beta_0 \ ,
\end{displaymath}
with $K$ of (\ref{eq:K}).
Here, $\beta_0$ is the first coefficient of the QCD $\beta$ function, and is given by
\be  
\beta_0 = \frac{11 C_G - 2 N_f}{12 \pi} .
\label{eq:beta0}
\ee
Recently, the three-loop term $\mathscr{P}_{ij}^{(2)}(z)$ in (\ref{eq:DGLAPk}) 
for the unpolarized parton distributions has been calculated, and $A_j^{(3)}$ 
has been extracted from its $z\rightarrow 1$ behavior~\cite{MVV}.
We note that $A^{(n)}_j$ ($n \geq 1$) and, thus, the entire function $A_j(\alpha_s)$ 
are actually process independent.
As for the other coefficients in (\ref{expand}), $B^{(n)}_j$ ($n \geq 2$) and
$C_{ji}^{(n)}$ ($n \geq 1$) depend on the process~\cite{dG}. 
General expressions for the coefficients $B_j^{(2)}$ and $C_{ji}^{(1)}$, 
including the process-dependent pieces, are 
derived in Ref.~\citen{dG} in the $\overline{\rm MS}$ scheme.\footnote{
$A_q^{(n)}$ ($n=1,2,\ldots$) and $B_q^{(1)}$ are independent of the factorization
scheme, but $B_q^{(n)}$ ($n \ge 2$) and $C_{ij}^{(n)}(z)$ ($n \ge 1$) 
depend on the factorization scheme (see e.g. Ref.~\citen{BCFG}).}
It is possible to transform the CSS resummation formula given in 
(\ref{resum00})-(\ref{resum}) into a new form, so that the process 
independence of the building block of the resummation formula
is maximal (see Ref.~\citen{CdG}).

The logarithms of $Q^2 b^2$ contained in the Sudakov factor $e^{\mS(b,Q)}$, 
which are large in the region $b \gg 1/Q$, can be made explicit. 
Here we have suppressed the subscript $j$ of the Sudakov exponent for simplicity. 
These contributions can be organized within
a systematic large logarithmic expansion according to consistent order 
counting, where $\alpha_s \ln(Q^2b^2)$ is formally considered
of order unity~\cite{KLSV,BCFG}.
Substituting the expansions (\ref{expand}) and the explicit form of 
the running coupling constant,
\be
\alpha_s(\kappa^2) = \frac{1}{\beta_0 \ln (\kappa^2/\Lambda_{\rm QCD}^2 )}
-\frac{\beta_1 \ln \ln (\kappa^2/\Lambda_{\rm QCD}^2 )}
{\beta_0^3 \ln^2 (\kappa^2/\Lambda_{\rm QCD}^2 ) }+ \cdots ,
\label{eq:run}
\ee 
with $\beta_0$ of (\ref{eq:beta0}) and
\be  
\beta_1 = \frac{17 C_G^2 - 5 C_G N_f - 3 C_F N_f}
              {24 \pi^2}  ,
\label{eq:beta}
\ee
we perform the $\kappa^2$ integral in (\ref{sudakov:1}) and organize the resulting
exponent $\mS(b,Q)$ in terms of
\bea
\lambda = \beta_0\alpha_s(\mu^2_R)\ln  \frac{Q^2b^2}{b_0^2} 
\equiv \beta_0\alpha_s(\mu^2_R)L  ,
\label{eq:lambda}
\eea
where $L$ plays the role of the large logarithmic expansion parameter in
the $b$ space. Considering as 
$\lambda \sim{\cal O}(1)$, $\mS(b,Q)$ can be systematically expanded as
\bea
\mS(b ,\,Q ) =
       \frac{1}{\alpha_s (\mu_R^2)}\  h^{(0)} (\lambda)
      + h^{(1)} (\lambda) + \sum_{n=2}^\infty 
\left(\frac{\alpha_s (\mu_R^2)}{2\pi}\right)^{n-1}h^{(n)}(\lambda) ,
\label{sudakov:11}
\eea
where $h^{(n)}(\lambda)$ ($n=0,1,\ldots$) are functions of $\lambda$,
which vanish at $\lambda=0$, as $h^{(n)}(0)=0$. 
Then, it is straightforward to find
\bea
   h^{(0)} (\lambda ) &=&
       \frac{A^{(1)}}{2\, \pi \beta_0^2} 
                   \, \left[ \lambda + \ln ( 1 - \lambda) \right] , \label{eq:h0}\\
   h^{(1)} (\lambda ) &=&
       \frac{A^{(1)}\, \beta_1}{2\, \pi \beta_0^3}
        \left[ \frac{1}{2}\, \ln^2 (1 - \lambda ) 
         + \frac{\lambda + \ln (1 - \lambda )}{1 - \lambda} \right]
       + \frac{B^{(1)}}{2\, \pi \beta_0} \, \ln (1 - \lambda ) \non\\
    &&\qquad \qquad - \frac{1}{4\, \pi^2 \beta_0^2}
       \left[ A^{(2)} - 2\, \pi \beta_0 A^{(1)} \ln \frac{Q^2}{\mu_R^2} \right]
      \left[ \frac{\lambda }{1 - \lambda} + \ln (1 - \lambda )\right],
             \nonumber\\ \label{eq:h1} 
\eea
and $h^{(2)}(\lambda)$ has a similar form, which can be found in, e.g., 
Ref.~\citen{BCFG}; $h^{(2)}(\lambda)$ involves 
$A^{(3)}$, $B^{(2)}$ and the third coefficient of the QCD $\beta$ function, $\beta_2$,
in addition to the coefficients appearing in (\ref{eq:h1}).
From (\ref{sudakov:11}), one can see that the Sudakov factor
$e^{\mS(b,Q)}$ is generically the 
exponentiation of the logarithmic 
terms $\alpha_s^nL^m$ with $m\leq n+1$ at each order $n$:
The first term, $h^{(0)}/\alpha_s$, in (\ref{sudakov:11}) collects the LL terms,
$\alpha_s^nL^{n+1}~(n\geq 1)$, and the second term, $h^{(1)}(\lambda)$, 
collects the NLL terms, $\alpha_s^nL^n~(n\geq 1)$.
Also, the third term, $(\alpha_s/2\pi)h^{(2)}$, controls the NNLL contributions, 
$\alpha_s^{n+1} L^n~(n\geq 1)$, and so forth.
Note that the ratio of two successive terms in (\ref{sudakov:11}) 
is formally of ${\cal O}(\alpha_s )$, and (\ref{sudakov:11}) is organized as 
a systematic power series in the small expansion parameter 
$\alpha_s(\mu_R^2)$, similarly to 
the customary perturbative expansions.

\subsection{Matching with the fixed-order $\alpha_s$ calculation}

To bring the CSS resummation formula (\ref{resum00})-(\ref{resum}) into contact with
the fixed-order calculation of the cross section, 
we formally expand $\mW(b; Q,y, \phi)$ in (\ref{resum00})
in terms of $\alpha_s(\mu_R^2)$,
and perform, order-by-order, the Fourier transformation to $Q_T$ space.
Here we consider the expansion of $\mW(b; Q,y, \phi)$ up to ${\cal O}(\alpha_s)$. 
Using (\ref{sudakov:11}), we can immediately expand the Sudakov factor 
$e^{\mS(b,Q)}$ in terms of $\alpha_s(\mu_R^2)$ as
\be
e^{\mS(b,Q)}=1- \frac{\alpha_s(\mu_R^2)}{2\pi}\left(
\frac{1}{2}A^{(1)}L^2 + B^{(1)}L \right) +{\cal O}(\alpha_s^2 ),
\label{eq:Sude}
\ee
with $L=\ln (Q^2b^2/b_0^2)$.
Also, the $b$ dependences of the coefficient functions $C_{ji}$
and the parton distributions $f_{i/h}$ in (\ref{resum0}) can be expanded 
perturbatively in terms of $\alpha_s(\mu_R^2)$; 
specifically, (\ref{expand}) and (\ref{eq:DGLAP}) yield
\bea
C_{ji}\left(z, \alpha_s(b_0^2/b^2)\right)&=& \delta_{ji}\delta(1-z)
+\frac{\alpha_s(\mu_R^2)}{2\pi}C_{ji}^{(1)}(z)+{\cal O}(\alpha_s^2 ),
\nonumber \\
f_{i/h}(x, b_0^2/b^2) &=& f_{i/h}(x, \mu_F^2)\non\\
&  -& \frac{\alpha_s(\mu_R^2)}{2\pi}
\ln\frac{\mu_F^2 b^2}{b_0^2} 
\sum_{j} \int_{x}^1\frac{dz}{z} \mathscr{P}^{(0)}_{ij}\left(\frac{x}{z}\right)
f_{j/h}(z, \mu_F^2) +{\cal O}(\alpha_s^2 ) \ .
\nonumber\\
\label{eq:expandcf}
\eea
Substituting these expansions into (\ref{resum00}), 
we encounter an integration of the type
\bea
\mI_n \equiv \int_0^{\infty} db\, \frac{b}{2} J_0 (b Q_T)  
         \left( \ln \frac{b^2 Q^2}{b_0^2} \right)^n .
\label{eq:I}
\eea
This can be calculated by taking the $\eta \rightarrow 0$ limit of 
the similar Fourier transformation of $(b^2Q^2/b_0^2)^\eta$ 
[see (\ref{eq:Inew}) and (\ref{eq:Cnew1}) below].
For the present case, we need the results
\bea
  \mI_0 = \delta (Q_T^2) \ , \quad \mI_1 = - \frac{1}{\left( Q_T^2 \right)_+} ,
    \quad \mI_2 = - 2 \left( \frac{\ln Q^2 / Q_T^2}{Q_T^2} \right)_+ ,
\label{eq:I0-I2}
\eea
where the + distributions are defined as (\ref{eq:Q_T-plus}), and we get
\bea
\frac{d \sigma^{\rm (res.)}}{d Q^2 d Q_T^2 d y d \phi}&=&
\left. \frac{1}{S} \sum_{j}
\frac{d \sigma^{j \bar{j}}_{\rm (LO)}(Q^2)}{d \phi}\right(
\delta (Q_T^2)\ f_{j / h_1}\left( x_1^0 , \mu_F^2 \right)  f_{\bar{j} / h_2}
           \left( x_2^0 , \mu_F^2 \right) 
\nonumber\\
+ \frac{\alpha_s(\mu_R^2)}{2\pi} && \!\!\!\!\!\!\!\!\!
 \left. \left \{\left[ A^{(1)}_j 
\left( \frac{\ln (Q^2 / Q_T^2 )}{Q_T^2} \right)_+ 
+B^{(1)}_j \frac{1}{(Q_T^2)_+} \right]
f_{j / h_1}
           \left( x_1^0 , \mu_F^2 \right) 
          f_{\bar{j} / h_2}
           \left( x_2^0 , \mu_F^2 \right) \right.\right. \nonumber\\
&+& \delta (Q_T^2) \left[ \sum_{i} ( C_{j i}^{(1)} \otimes f_{i / h_1})
           \left( x_1^0 , \mu_F^2 \right) f_{\bar{j} / h_2} 
\left( x_2^0 , \mu_F^2 \right) \right.
\nonumber\\
&&\left. +
\sum_{k} f_{j / h_1}
           \left( x_1^0 , \mu_F^2 \right) ( C^{(1)}_{\bar{j} k} \otimes f_{k / h_2})
\left( x_2^0 , \mu_F^2 \right) 
 \right] \nonumber\\
+&& \!\!\!\!\!\!\!\!
\left(  \frac{1}{(Q_T^2)_+} 
        + \delta (Q_T^2) \ln \frac{Q^2}{\mu_F^2} \right)
\left[ \sum_{i} (  \mathscr{P}_{j i}^{(0)} \otimes f_{i / h_1})
\left( x_1^0 , \mu_F^2 \right) f_{\bar{j} / h_2} \left( x_2^0 , \mu_F^2 \right)\right.
\nonumber\\
&&\left. \left.\left.   +
\sum_{k} f_{j / h_1}
  \left( x_1^0 , \mu_F^2 \right) ( \mathscr{P}_{\bar{j} k}^{(0)}  \otimes f_{k / h_2})
\left( x_2^0 , \mu_F^2 \right) 
 \right] 
\right\} \right) + {\cal O}(\alpha_s^2 ) .
\label{eq:order1}
\eea
Here, we identify the structure characteristic of the singular component of 
the fixed-order cross section, like that observed in $\Delta_{T}X$ 
of (\ref{eq:X00})-(\ref{eq:X}).
In particular, the matching of (\ref{eq:order1}) with the corresponding 
fixed-order cross section completely determines the coefficients 
$A^{(1)}$, $B^{(1)}$, and $C^{(1)}$, and the LO cross section 
$\sigma^{j\bar{j}}_{\rm (LO)}(Q^2)$ for the parton-level process 
$j + \bar{j} \to F (Q^2)$;
for example, for the case of the spin-dependent cross section of tDY, the matching
with (\ref{eq:X00})-(\ref{eq:X}) gives results identical to (\ref{coeff_NLL}) 
for $A^{(1)}_q$ and $B^{(1)}_q$, and also yields
\be
C_{qq}^{(1)}(z)=C_{\bar{q}\bar{q}}^{(1)}(z)
     =  C_F \left( \frac{\pi^2}{2} - 4 \right) \, \delta(1 - z),\;\;\;\;\;
C_{q\bar{q}}^{(1)}(z)=C_{\bar{q}q}^{(1)}(z)=0
\label{eq:coeff.10}
\ee
in the $\overline{\rm MS}$ scheme and
\be
\frac{1}{S} 
\frac{d \sigma^{q \bar{q}}_{\rm (LO)}(Q^2)}{d \phi}=\frac{1}{S} 
\frac{d \sigma^{\bar{q}q}_{\rm (LO)}(Q^2)}{d \phi}
=\mN e_q^2 \cos\, (2 \phi_{k_1} - \phi_1 - \phi_2) ,
\label{eq:nom}
\ee
while all other quantities associated with the gluon index $j=g$ vanish.

The expansion (\ref{eq:order1}) can be extended to include higher orders 
in $\alpha_s$. The corresponding expansion up to ${\cal O}(\alpha_s^2)$ 
was carried out for $Q_T >0$ in unpolarized DY~\cite{DWS}, 
and the resulting ${\cal O}(\alpha_s^2)$ term involves the coefficients 
$A^{(2)}$ and $B^{(2)}$.\footnote{
The ${\cal O}(\alpha_s^2)$ term involving $C^{(2)}$ is proportional to $\delta(Q_T^2 )$,
similarly to the ${\cal O}(\alpha_s)$ term involving $C^{(1)}$ in (\ref{eq:order1}),
and it vanishes for $Q_T >0$.}
The matching of that result with the corresponding fixed-order cross section 
confirms the universal value of $A^{(2)}$ in (\ref{coeff_NLL}) 
and also determines $B^{(2)}$ for unpolarized DY [see also Ref.~\citen{dG} 
for results in other unpolarized processes of the type (\ref{eq:genpro})].
This fact also implies 
that when we use the resummed component (\ref{resum00}) up to NLL accuracy,
$d\sigma^{\rm (fin.)}/ d Q^2 d Q_T^2 d y d \phi$ in (\ref{cross section})
should be taken as the finite component of the fixed-order $\alpha_s$ cross section,
and the sum of those two components gives the cross section 
$d\sigma^{\rm (CSS)}/ d Q^2 d Q_T^2 d y d \phi$,
which is exact up to ${\cal O}(\alpha_s)$ when expanded in powers of $\alpha_s$.
Similarly, when we use the resummed component up to NNLL accuracy,
$d\sigma^{\rm (fin.)}/ d Q^2 d Q_T^2 d y d \phi$
should be taken as the finite component of the fixed-order cross section
to ${\cal O}(\alpha_s^2)$, so that we obtain 
$d\sigma^{\rm (CSS)}/ d Q^2 d Q_T^2 d y d \phi$, 
which is exact up to ${\cal O}(\alpha_s^2)$.
Generalizing these considerations, the second term, 
$d\sigma^{\rm (fin.)}/ d Q^2 d Q_T^2 d y d \phi$, 
in (\ref{cross section}) should be determined by  
\bea
\frac{d \sigma^{\rm (fin.)}}{d Q^2 d Q_T^2 d y d \phi} = 
\left. \frac{d \sigma}{d Q^2 d Q_T^2 d y d \phi}\right|_{\rm fo}-
\left.\frac{d \sigma^{\rm (res.)}}{d Q^2 d Q_T^2 d y d \phi}\right|_{\rm fo}  ,  
\label{finite}
\eea
where the notation $(\cdots)|_{\rm fo}$ represents the expansion of 
the quantity $(\cdots)$ in powers of $\alpha_s(\mu_R^2)$ up to 
a given fixed order; i.e.,  the first term on the RHS, 
$\left. d \sigma/d Q^2 d Q_T^2 d y d \phi \right|_{\rm fo}$,
is the cross section that is computed by truncating 
the customary perturbative expansion at a given fixed order in $\alpha_s(\mu_R^2)$. 
As noted above, when we consider the resummation at the NLL level, 
$(\cdots)|_{\rm fo}$ is the expansion up to ${\cal O}(\alpha_s(\mu_R^2))$. 
In this case, the second term of (\ref{finite}), 
$\left. d \sigma^{\rm (res.)}/d Q^2 d Q_T^2 d y d \phi \right|_{\rm fo}$,
is given by (\ref{eq:order1}), where the coefficients $A^{(1)}$ and $B^{(1)}$ 
in (\ref{expand}) and $C^{(1)}$,  $\sigma^{j\bar{j}}_{\rm (LO)}(Q^2)$, 
determined similarly to (\ref{eq:coeff.10}) and (\ref{eq:nom}), are substituted.
When we consider the resummation at the NNLL level, $(\cdots)|_{\rm fo}$
is the expansion up to ${\cal O}((\alpha_s(\mu_R^2)^2)$.

The matching procedure for the finite component, represented by
(\ref{finite}), guarantees that
$d\sigma^{\rm (CSS)}/ d Q^2 d Q_T^2 d y d \phi$ in (\ref{cross section})
retains the complete information of the perturbative calculation up to
the specified fixed order and, at the same time,
incorporates the resummation of the logarithmically enhanced contributions to all orders.
The fixed-order truncation of $d\sigma^{\rm (CSS)}/ d Q^2 d Q_T^2 d y d \phi$ 
exactly reproduces the fixed-order cross section 
obtained in the customary 
QCD perturbation theory.
In this sense, using the matching of (\ref{finite}),
the resummed and fixed-order components are consistently combined in 
$d\sigma^{\rm (CSS)}/ d Q^2 d Q_T^2 d y d \phi$ of (\ref{cross section})
without any double counting.

\section{Systematizing the $b$-space resummation formula}

In the $b$ integration in the resummed component (\ref{resum00}), 
we can distinguish three regions~\cite{PP,CSS}:
(i) $ 0 \le b \lesssim 1/Q$, (ii) $1/Q \lesssim b \ll 1/\Lambda_{\rm QCD}$,
and (iii) $b \gtrsim 1/\Lambda_{\rm QCD}$.
These are relevant to the kinematical regions 
$Q_T \gtrsim Q$, $\Lambda_{\rm QCD} \ll Q_T \lesssim Q$,
and $Q_T \lesssim \Lambda_{\rm QCD}$, respectively.
We now go into further detail concerning the behavior and the physical content
of the resummation formula in each of these three regions.

\subsection{Region (i): $0\le b \lesssim 1/Q$}

In the region (i), $0\le b \lesssim 1/Q$, 
the resummation is irrelevant and can be truncated at low order in $\alpha_s$. 
Thus here, the customary fixed-order perturbation theory
can be used. The resummed component associated with this region
should be given by (\ref{eq:order1}) with $Q_T \gtrsim Q$ to good accuracy.
Apparently, in general, if the functions $A, B$ and $C$ of
(\ref{expand}) and the QCD $\beta$ function are evaluated up to order
$\alpha_s^n$, then the corresponding 
resummed component is correct to order $\alpha_s^n$.

\subsection{Region (ii): $1/Q \lesssim b \ll 1/\Lambda_{\rm QCD}$}

In the region (ii), $1/Q \lesssim b \ll 1/\Lambda_{\rm QCD}$,
all-order resummed perturbation theory can be used 
[see the discussion below (\ref{eq:W2})]. 
In this region, $L$ in (\ref{eq:lambda}) can be large, 
but still we have, taking $\mu_R \simeq Q$ as usual,
\be 
0< \lambda \ll \beta_0\alpha_s(\mu^2_R)\ln \frac{Q^2}{\Lambda_{\rm QCD}^2}\sim 1,
\label{eq:ordercout}
\ee
so that the Sudakov factor $e^{\mS (b,Q)}$ with (\ref{sudakov:11})-(\ref{eq:h1})
can be expanded in a power series in $\lambda$. 
In principle, the $b$ dependence of the coefficient functions $C_{ji}$ and 
the parton distributions $f_{i/h}$ in (\ref{resum0}) 
can be expanded similarly in powers of $\lambda$ (see \S7).

It is instructive to apply the corresponding expansion to
(\ref{resum00}) and (\ref{resum0}) at LL accuracy,
which means that we retain only the first term with $h^{(0)}$ 
in (\ref{sudakov:11}) for the Sudakov factor $e^{\mS (b,Q)}$,
and also the terms at the same level in the other factors.
The coefficient function $C^{(1)}$ in (\ref{expand}), as well as
the scale dependence of the parton distributions $f_{i/h}$,
appears at the same logarithmic level as $B^{(1)}$, i.e., at the NLL level, 
as seen in (\ref{eq:order1}) (also compare (\ref{eq:h1}) and (\ref{eq:freplace}) below). 
Thus, at the LL level, those effects can be ignored as
$( C_{j i} \otimes f_{i / h}) ( x, b_0^2/b^2 ) \rightarrow f_{j / h} ( x, \mu_F^2)$ 
in (\ref{resum0}).  
As a result, the $b$ dependence of $\mW(b; Q, y,\phi)$ in the integrand 
of (\ref{resum00}) remains only in the Sudakov factor at the LL level 
[see (\ref{eq:h0})], 
\bea
&&\hspace{-0.5cm}
e^{\mS(b, Q)}=e^{h^{(0)}(\lambda)/\alpha_s}
= \sum_{n=0}^\infty \frac{1}{n!}\left(-\frac{A^{(1)}}{2\pi\beta_0^2 \alpha_s} 
\sum_{m=2}^{\infty}\frac{\lambda^m}{m}\right)^n  \non \\
&&\hspace{0.5cm}
= \sum_{n=0}^\infty \frac{\alpha_s^{n}}{n!}
\left( \varpi_{n}^{(0)} L^{2n} + \varpi_{n}^{(1)} L^{2n-1} +
\varpi_{n}^{(2)} L^{2n-2} +\cdots + \varpi_{n}^{(n-1)} L^{n+1} \right) ,
\label{eq:LLexp}
\eea
where $\alpha_s = \alpha_s(\mu_R^2 )$. In the second line,
(\ref{eq:lambda}) has been substituted, and the series are rearranged 
according to their powers in the large logarithm $L$ for each order in $\alpha_s$,
with $\varpi_{n}^{(0)} =( -A^{(1)}/4\pi)^n$, and other expansion coefficients,
$\varpi_{n}^{(1)}, \varpi_{n}^{(2)}, \ldots$, which can be expressed
similarly in terms of $A^{(1)}$ and $\beta_0$. 
Using the expansion (\ref{eq:LLexp}), the Fourier transformation 
relevant to (\ref{resum00}) at the LL level is given by
\bea
\int_0^{\infty} d b \, \frac{b}{2}\,J_0 (b Q_T) 
e^{\mS(b, Q)}&\equiv& \mathcal{T}^{\rm LL}(Q_T^2, Q^2) \nonumber\\
&=& \sum_{n=0}^\infty \int_0^{\infty} d b\ \frac{b}{2}\,J_0 (b Q_T) 
\frac{\alpha_s^{n}}{n!}
\left( \varpi_{n}^{(0)} L^{2n} + \varpi_{n}^{(1)} L^{2n-1} + \cdots\right) ,
\non\\
\label{eq:LLFou}
\eea
and it can be completely expressed in terms of
$\mI_m$ ($m=2n, 2n-1, \ldots$) in (\ref{eq:I}).
Generalizing (\ref{eq:I0-I2}) for an arbitrary integer $n\ (\ge 0)$,
we get [see (\ref{eq:Cnew1})],
\bea
\mI_n &=& \left.\frac{d^n}{d\eta^n}\left\{\left[\delta (Q_T^2) - \sum_{m=0}^\infty
\frac{\eta^{m+1}}{m!}\left( \frac{\ln^m (Q^2 / Q_T^2)}{Q_T^2} \right)_+ \right]
\exp\left[-2 \sum_{r=1}^{\infty}\frac{\zeta(2r+1)}{2r+1}\eta^{2r+1}\right]
\right\}\right|_{\eta =0}
\nonumber\\
&=&
\varrho_n \delta(Q_T^2)
-n \left( \frac{\ln^{n-1} (Q^2 / Q_T^2)}{Q_T^2} \right)_+ 
\nonumber\\
&&+ \frac{2}{3}\zeta(3)n(n-1)(n-2)(n-3)
\left( \frac{\ln^{n-4} (Q^2 / Q_T^2)}{Q_T^2} \right)_+ +\cdots ,
\label{eq:genIn}
\eea
where $\varrho_n \equiv \left. (d^n /d\eta^n)
\exp\left[-2 \sum_{r=1}^{\infty}\zeta(2r+1)\eta^{2r+1} /(2r+1) \right]\right|_{\eta =0}$,
$\zeta (n)$ is the Riemann zeta-function, 
and the ellipses denote the terms involving $(\ln^{m} (Q^2 / Q_T^2) /Q_T^2)_+$
with $m\le n-6$.
Thus, for $\Lambda_{\rm QCD} \ll Q_T \lesssim Q$, corresponding to the
region (ii) $1/Q \lesssim b \ll 1/\Lambda_{\rm QCD}$, (\ref{eq:LLFou}) yields
\be
\mathcal{T}^{\rm LL}(Q_T^2, Q^2) =\sum_{n=1}^\infty \frac{\alpha_s^n}{n!}\left[
-2n  \left(-\frac{A^{(1)}}{4\pi}\right)^n  \frac{\ln^{2n-1} (Q^2 / Q_T^2)}{Q_T^2}
+ \cdots 
\right] ,
\label{eq:resLL}
\ee
where the ellipses denote the terms involving $\ln^{m} (Q^2 / Q_T^2) /Q_T^2$
with $m=2n-2, 2n-3, \cdots$. This result demonstrates that (\ref{resum00}) 
indeed resums the towers of logarithms (\ref{eq:remnant}) 
to all orders in $\alpha_s$.
Summing up the leading tower in (\ref{eq:resLL}) over all $n$, we get
\be
\mathcal{T}^{\rm LL}(Q_T^2, Q^2) =\frac{A^{(1)}\alpha_s}{2\pi} 
\frac{\ln (Q^2 / Q_T^2)}{Q_T^2} 
\exp\left[-\frac{A^{(1)}\alpha_s}{4\pi} \ln^{2} (Q^2 / Q_T^2)\right]+\cdots .
\label{eq:resLL2}
\ee
Substituting $A^{(1)}$ of (\ref{coeff_NLL}),
it is seen that this result reproduces the Sudakov quark form factor in the so-called
``double leading logarithmic approximation (DLLA).''~\cite{DDT}

It is worth noting that, introducing the Fourier transform [see (\ref{eq:I0-I2})],
\be
\nu(k_T)\equiv \int \frac{d^2 b}{4\pi} 
e^{-i \boldsymbol{b}\cdot \boldsymbol{k}_T}
\left[ -\frac{A^{(1)}\alpha_s}{4\pi} L^2 \right]
= \frac{A^{(1)}\alpha_s}{2\pi} \left(\frac{\ln (Q^2 / k_T^2)}{k_T^2} \right)_+ ,
\label{eq:nuK}
\ee
the contribution from the leading tower of logarithms in (\ref{eq:LLFou}) 
can be expressed as
\bea
\sum_{n=1}^\infty && \frac{1}{n!}
\int \frac{d^2 b}{4\pi} e^{i \boldsymbol{b}\cdot \boldsymbol{Q}_T}
\left[ -\frac{A^{(1)}\alpha_s}{4\pi} L^2 \right]^n
\nonumber\\
=&& \pi \sum_{n=1}^\infty \frac{1}{n!}
\int d^2 k_{1,T}\cdots d^2 k_{n,T}\
\delta^{(2)}\left( \boldsymbol{Q}_T+ \sum_{l=1}^n \boldsymbol{k}_{l,T} \right) 
\frac{\nu(k_{1,T})}{\pi}\cdots \frac{\nu(k_{n,T})}{\pi}.
\label{eq:momencon}
\eea
Here, $\nu(k_T)$ in (\ref{eq:nuK}) represents the probability distribution 
for the emission of one soft gluon with transverse momentum $k_T$ in the final state, as 
$(d\sigma^{\mbox{\scriptsize 1-gluon}}/dk_T^2)/ \sigma^{\rm total}$ $= \nu(k_T)$, with
$\sigma^{\rm total}$ the total cross section~\cite{PP}. 
Indeed, terms with the same structure as (\ref{eq:nuK}) appear in
(\ref{eq:X}) and (\ref{eq:order1}).
Therefore, the $n$-th order term in (\ref{eq:momencon}) represents the probability 
for the emission of $n$ soft gluons in the final state as the product of 
the probabilities for $n$ independent gluons~\cite{ESRW}.
Thus, the exponentiation of the LL term in the Sudakov factor $e^{\mS(b, Q)}$
in fact sums up the multiple soft-gluon emission probabilities to all orders.

Note that in each term of the expansion on the RHS of (\ref{eq:order1}),
transverse-momentum conservation is ensured by the delta function,
$\delta^{(2)}\left( \boldsymbol{Q}_T+ \sum_{l=1}^n \boldsymbol{k}_{l,T} \right)$,
and thus $\boldsymbol{Q}_T$ of the dilepton in the final state
is provided by the recoil from the radiation of soft gluons.
This automatic and proper treatment of transverse-momentum conservation 
is one of the most important features of the impact parameter $b$-space 
approach~\cite{PP}.
In fact, transverse-momentum conservation is
particularly important when we approach the small $Q_T$ region. 
For example, the contribution for $n =2$ in (\ref{eq:momencon}) gives, 
using (\ref{eq:genIn}),
\bea
\frac{\pi}{2!}
\int&& d^2 k_{1,T}d^2 k_{2,T} 
\delta^{(2)}\left( \boldsymbol{Q}_T+  \boldsymbol{k}_{1,T} 
+ \boldsymbol{k}_{2,T}  \right) 
\frac{\nu(k_{1,T})}{\pi}\frac{\nu(k_{2,T})}{\pi}
\nonumber\\
&&=\frac{1}{2} \left(\frac{A^{(1)}\alpha_s}{2\pi} \right)^2
\left[ -\left( \frac{\ln^{3} (Q^2 / Q_T^2)}{Q_{T}^2} \right)_+
+ \frac{4\zeta(3)}{\left(Q_T^2 \right)_+} \right] ,
\label{eq:momencon2}
\eea
where the first term corresponds to the $n=2$ term given in (\ref{eq:resLL}). 
Inspecting the integration over the transverse momenta, $k_{1,T}$ and $k_{2,T}$, 
on the LHS of (\ref{eq:momencon2}), we confirm that the first term 
on the RHS comes from the ``strongly-ordered'' phase space as
$\delta^{(2)}(\boldsymbol{Q}_T+  \boldsymbol{k}_{1,T} )\theta(k_{1,T}^2 -k_{2,T}^2)
+(1 \leftrightarrow 2)$. 
Then, the second term is the non-strongly-ordered contribution corresponding 
to the emission of soft gluons whose transverse momenta, 
$k_{1,T}$ and $k_{2,T}$ ($\gtrsim Q_T$), add vectorially to balance $\boldsymbol{Q}_T$.
Although the second term is formally subleading in (\ref{eq:momencon2}),
contributions of this type, called ``kinematic logarithms,''~\cite{ESRW,EV}
dominate the cross section in the asymptotic limit, $Q\gg Q_T\gg \Lambda_{\rm QCD}$, 
where the DLLA Sudakov factor of (\ref{eq:resLL2}) leads to suppression of 
the (formally) leading contributions.
Here, the probability of having a parton-antiparton annihilation into
a lepton pair with no emission of gluons 
of transverse momenta greater than a fixed value decreases
asymptotically faster than any power of $Q^2$, and
events at small $Q_T$ may be obtained asymptotically only by the emission 
of at least two gluons whose transverse momenta are not small and add 
to a small value of $Q_T$. All the non-leading ``kinematic logarithms'' 
are correctly taken into account by imposing transverse-momentum
conservation as in (\ref{eq:momencon}) and (\ref{eq:momencon2}),
and this is naturally realized in the $b$-space
resummation approach with (\ref{resum00}).\footnote{
There have been efforts to formulate a ``$Q_T$-space'' resummation approach,
which is basically organized according to the strong ordering of the phase space.
Thus this approach requires a hard and sophisticated task of 
including effects of the non-leading ``kinematic logarithms''
(see Ref.~\citen{EV}).}
Also, apparently, 
transverse-momentum conservation with the $b$-space resummation approach is crucial 
in the case of smaller $Q_T$ corresponding to the region (iii).

Calculations similar to those in (\ref{eq:LLexp})-(\ref{eq:momencon2}) can be 
performed including the NLL terms and higher logarithmic terms.
It is straightforward to show that the NLL terms associated with $h^{(1)}$
of the Sudakov exponent (\ref{sudakov:11}) give rise to a series of contributions
$\alpha_s^n L^{2n-1}, \alpha_s^n L^{2n-2}, \ldots$ to (\ref{eq:LLFou}), 
and thus yield contributions $\alpha_s^n \ln^{m} (Q^2 / Q_T^2) /Q_T^2$
with $m=2n-2, 2n-3, \ldots$ to (\ref{eq:resLL}); similarly, the NNLL terms
give rise to the contributions $\alpha_s^n \ln^{m} (Q^2 / Q_T^2) /Q_T^2$
with $m=2n-4, 2n-5, \ldots$ to (\ref{eq:resLL}).
Therefore, the resummation formula (\ref{resum00}) 
at the LL level corresponds to summing up only the first tower of logarithms,  
$\alpha_s^n \ln^{2n-1}(Q^2/Q_T^2)/Q_T^2$ exactly, 
while the resummation formula up to NLL accuracy allows us to fully sum up 
the first three towers, i.e., $\alpha_s^n \ln^m(Q^2/Q_T^2)/Q_T^2$ 
with $m=2n-1$, $2n-2$, and $2n-3$.

The convergence of the resummed perturbation series for (\ref{resum00}) 
in the $Q_T$ space is governed by $\alpha_s \ln^{2} (Q^2 / Q_T^2)$ 
rather than $\alpha_s$, and when $Q_T$ is small, 
$\alpha_s \ln^{2} (Q^2/ Q_T^2)$ will be large even for $\alpha_s \ll 1$.
As $Q_T$ approaches $Q$, we have $\ln(Q^2/Q_T^2) \ll 1$,
and the all-order resummed perturbation theory of (\ref{resum00}) is smoothly
extrapolated to the region (i).

\subsection{Region (iii): $b \gtrsim 1/\Lambda_{\rm QCD}$}

Next we turn to the region (iii), $b \gtrsim 1/\Lambda_{\rm QCD}$.
In this region, $\lambda$ in (\ref{eq:lambda}) can be as large as 1,
and hence the Sudakov factor $e^{\mS (b,Q)}$ cannot be expanded in terms of $\lambda$,
and the functional form exponentiating (\ref{sudakov:11})-(\ref{eq:h1})
has to be used as it is. Moreover, in this long-distance region, 
nonperturbative effects become relevant, which could modify 
the resummation formula, (\ref{resum0}) and (\ref{resum}).
In fact, when $b$ becomes extremely large, so that $\lambda \rightarrow 1$,
the resummation formula in the form of (\ref{resum0}) and (\ref{resum})
breaks down. In this case, the functions (\ref{eq:h0}) and (\ref{eq:h1}) 
in the Sudakov exponent (\ref{sudakov:11}) are singular at 
$\lambda = 1$, and this singular behavior is related to the presence of the Landau pole 
in the perturbative running coupling $\alpha_s (\kappa^2)$ in QCD. 

Therefore, to properly define the $b$ integration in the resummation
formula (\ref{resum00}) for the corresponding long-distance region, 
relevant to $Q_T \lesssim \Lambda_{\rm QCD}$,
a prescription to deal with this singularity is required.
Such a prescription may be accompanied by the modification of (\ref{resum})
to complement the relevant nonperturbative effects.
For this purpose, two types of approaches are commonly used 
in the literature.\footnote{See Ref.~\citen{QZ}, for an attempt to 
incorporate the nonperturbative effects as power corrections to 
the RG equations that control the large logarithms arising in 
the integral on the RHS of (\ref{eq:W}).} 
One is the so-called $b_*$ prescription, which was proposed by CSS~\cite{CSS}.
In this approach,  $W_j (b;\, Q, x_1, x_2)$ in (\ref{resum}) is replaced as
\bea
W_j (b;\, Q, x_1, x_2) \to W_j (b_* ;\, Q, x_1, x_2)\, F^{NP} (b ;\,  Q, x_1, x_2)  ,
\label{replace}
\eea
where
\bea  
b_* = \frac{b}{\sqrt{1 + b^2 / b_{\rm lim}^2}}  ,
      \quad \quad  b_{\rm lim} \sim 0.5 \, {\rm GeV}^{-1}  .
\label{eq:blim}
\eea
Through this replacement, $W_j (b;\, Q, x_1, x_2)$ 
in (\ref{resum}), which is accurate for $b \ll 1/\Lambda_{\rm QCD}$, 
is smoothly extrapolated to the extremely large $b$ region, while
the singularity is avoided, since $b_* \le b_{\rm lim}$ in $W_j(b_* ;\, Q ,x_1, x_2)$. 
This corresponds to effectively taking into account
the ``freezing'' of the running coupling in the long-distance region
due to the nonperturbative effects. 
The function $F^{NP}$ represents possible nonperturbative effects, 
which may be interpreted as the contributions associated with
the intrinsic transverse momentum of partons inside the colliding hadrons.
$F^{NP}$ is normally taken as a smearing factor of Gaussian type in $b$~\cite{CSS}.
Thus, in the large $b$ region of the integration over $b$ in (\ref{resum00}),
$W_j(b_* ;\, Q ,x_1, x_2)$ approaches a constant, $W_j(b_{\rm lim} ;\, Q ,x_1, x_2)$, 
and $F^{NP}$ acts as a damping factor.
Many different parameterizations of $F^{NP}$ have been proposed 
(see Ref.~\citen{KS} and references therein).

Another approach~\cite{KLSV} 
consists of an extension of the minimal prescription~\cite{CMNT}
proposed in the context of the so-called threshold resummation.
This approach allows us to avoid the Landau singularity in a purely 
perturbative framework:
Decomposing the Bessel function in (\ref{resum00}) into two Hankel
functions as 
\bea
J_0(bQ_T)=\frac{1}{2}[H_0^{(1)}(bQ_T)+H_0^{(2)}(bQ_T)] \ , 
\label{bessel}
\eea
we deform the $b$-integration contour associated with $H_0^{(1)}(bQ_T)$ 
and $H_0^{(2)}(bQ_T)$ into the upper and lower half planes in the complex
$b$-space, respectively, and obtain the following two convergent 
integrals as $|b|\ra \infty$:
\bea
  \frac{d \sigma^{\rm (res.)}}{d Q^2 d Q_T^2 d y d \phi}
    &=& 
\int_{\cal C_+} d b \, \frac{b}{4}\,H_0^{(1)} (b Q_T) \mW (b;\, Q\,,\,y, \phi)
\non\\&& 
 + \int_{\cal C_-} d b \, \frac{b}{4}\,H_0^{(2)} (b Q_T) \mW (b;\, Q\,,\,y, \phi) .
  \label{resum:2}
\eea
The new contour ${\cal C_\pm}$ is taken to be
from 0 to $b_c$ on the real axis, followed by the two branches, 
$b=b_c+e^{\pm i\theta}t$ with $t \in \{0, \infty \}$ and 
$0<\theta<\pi/4$.
The constant $b_c$ here is arbitrary in the interval $0 \le b_c < b_L$, 
where $b_L=(b_0 /Q)e^{1/(2\beta_0\alpha_s(\mu_R^2))}$ corresponds to 
the position of the singularity on the real axis as the solution of $\lambda=1$.
Equation (\ref{resum:2}) provides us with a (formally) consistent definition 
of a finite $b$-integral for the resummation formula
within a perturbative framework with which,
unlike in the above case of the $b_*$-prescription, no extra cut-off
parameter $b_{\rm lim}$ is required in this contour deformation prescription. 
In this paper, we employ (\ref{resum:2}) to deal with the Landau singularity.
(Also, in this case, possible nonperturbative effects in the large $b$ region 
can be included, as seen in (\ref{eq:np}) below.\footnote{
Actually, we have chosen $0<\theta<\pi/4$ in the above contour 
${\cal C_\pm}$, anticipating the Gaussian smearing factor (\ref{eq:np})
as the corresponding nonperturbative effects.})

We note that the contour deformation used to obtain (\ref{resum:2})
can be performed ``safely'' order-by-order in $\alpha_s(\mu_R^2 )$, 
by expanding $\mW (b;\, Q\,,\,y, \phi)$ in the integrand 
in powers of $\alpha_s$.
The above discussion concerning the regions (i) and (ii) implies that
for (\ref{resum00}) in the region $Q_T \gg \Lambda_{\rm QCD}$ 
we can define $\mW (b;\, Q\,,\,y, \phi)$ in the integrand 
as a power series in $\alpha_s(\mu_R^2 )$,
which can be organized similarly as the integrand of (\ref{eq:LLFou}). 
In this case, (\ref{resum00}) can be completely expressed by $\mI_n$ 
in (\ref{eq:I}), whose $b$ integration is well-defined and yields distributions 
like (\ref{eq:I0-I2}) and (\ref{eq:genIn}).
It is straightforward to show that we can perform the above contour 
deformation into ${\cal C_\pm}$ with (\ref{bessel}) for the $b$ integration in $\mI_n$; 
in fact, on the LHS of the relation
\be 
\int_0^{\infty} db\, \frac{b}{2} J_0 (b Q_T)  
         \left(  \frac{b^2 Q^2}{b_0^2} \right)^{\eta}=
\sum_{m=0}^{\infty}\frac{\eta^m}{m!}\mI_m ,
\label{eq:Inew}
\ee
the corresponding contour deformation can be performed using Cauchy's theorem;
i.e., the contributions to the integral from the contour 
in the limits $b \rightarrow 0$ and $|b| \rightarrow \infty$ are found to vanish using 
an appropriate analytic continuation for $\eta$.
After such a contour deformation is performed for each term 
in the series of (\ref{resum00}), the corresponding series can be summed up
under the integrand of the deformed $b$-integration. 
This yields $\mW (b;\, Q\,,\,y, \phi)$, with the Sudakov factor
exponentiating (\ref{sudakov:11}). 
This final form coincides with (\ref{resum:2}) exactly,
and is applicable to the region $Q_T \lesssim \Lambda_{\rm QCD}$ as well as 
the region $Q_T \gg \Lambda_{\rm QCD}$.
Therefore, the choice of contours in (\ref{resum:2}) is equivalent to 
the original contour in (\ref{resum00}), order-by-order in $\alpha_s(\mu_R^2 )$,
and also extends the applicability of the formula even into the low $Q_T$ region.
This also implies that all the above results 
obtained for the regions (i) and (ii), as well as those 
obtained in \S5, are unchanged using the new form (\ref{resum:2}).
Note that this is a consequence of the fact that (\ref{resum:2}) does not
involve any cut-off parameter, like $b_{\rm lim}$ in (\ref{eq:blim}),
and it is an advantage of our approach over the $b_{*}$ prescription.

Now we can use (\ref{resum:2}) as the precise form of 
the resummation formula to 
investigate its behavior for $Q_T \lesssim \Lambda_{\rm QCD}$.
Here we mention in particular a remarkable point regarding the case $Q_T \approx 0$:
In this case, the behavior of (\ref{resum:2}) is controlled by a saddle point
in the $b$ integration. 
As shown in \S8, the corresponding saddle point is on the real-$b$ axis 
with $\lambda \sim 1$. The important role of the saddle point for $Q_T \approx 0$ 
was pointed out in Refs.~\citen{PP,CS} and \citen{CSS}, 
using the ``old'' form, (\ref{resum00}). 
Around the saddle point, we have 
$L \sim 1/\alpha_s(\mu_R^2 )\sim \ln(Q^2/\Lambda_{\rm QCD}^2 )$. 
Because this implies that all logarithms, $L$ and 
$\ln (Q^2 /\Lambda^2_{\rm QCD})\sim 1/\alpha_s$, 
are regarded as equally large for $Q \gg \Lambda_{\rm QCD}$,
the resulting contributions to the resummation formula (\ref{resum:2}) 
are organized in terms of the single small parameter $\alpha_s$,
but with a classification of the contributions with respect to the order 
of $\alpha_s$ that differs from that of the
customary perturbation theory that can be used in the region (i), corresponding to 
$Q_T\gtrsim Q$. In the Sudakov exponent (\ref{sudakov:1}),
one such logarithm, $\ln(Q^2/\kappa^2 )$, explicitly appears,
and another large logarithm is implicit in the integration over $\kappa^2$.
Suppose that one wants to evaluate the resummation formula (\ref{resum:2}) 
in an approximation of ``degree $N$,''~\cite{CSS} meaning
that any corrections are suppressed by a factor of 
$[\ln(Q^2 /\Lambda_{\rm QCD}^2 )]^{-(N+1)}$.
Then, since two large logarithms multiply the function $A$ in (\ref{sudakov:1}),
one must evaluate $A$ in (\ref{expand}) to order $\alpha_s^{N+2}$. 
Similarly, one needs $B$ to order $\alpha_s^{N+1}$, $C$ to order $\alpha_s^N$, and,
for the running of $\alpha_s$, the $\beta$ function to order $\alpha_s^{N+2}$.
In particular, 
if one wants an approximate result for (\ref{resum:2}) that will converge 
to the exact $Q_T \approx 0$ cross section as $Q\rightarrow \infty$,
one needs a degree 0 approximation: $A$ to order $\alpha_s^2$, $B$ to order $\alpha_s^1$,
$C$ to order $\alpha_s^0$, and $\beta$ to order $\alpha_s^2$~\cite{CSS}.
Note that this classification controlling (\ref{resum:2}) for $Q_T \approx  0$
is different from that in the 
customary perturbation theory for $Q_T\gtrsim Q$, and also from that in 
the all-order resummed perturbation theory,
which controls the towers of logarithms (\ref{eq:remnant}) 
in the kinematical region, $\Lambda_{\rm QCD} \ll Q_T \lesssim Q$.
This point is discussed in detail in \S8.

Combining the above considerations for the three regions in the resummation formula
(\ref{resum:2}) with (\ref{resum0}) and (\ref{resum}),
we see that this formula represents a complicated object involving 
contributions of various orders and from various distance scales, 
due to the ubiquitous gluons.
The impact parameter $b$-space formulation allows us to organize 
these complicated contributions into the universal form 
of (\ref{resum0}), and this form directly embodies a simple physical
picture behind the process (\ref{eq:genpro}):
When the transverse momentum $Q_T$ of the final-state system $F$ is small,
the emission of the accompanying radiation is strongly inhibited,
so that only soft and collinear partons, which have low transverse momenta $k_T$,
are radiated into the final state. The quantity
$d \sigma^{j \bar{j}}_{\rm (LO)}(Q^2)/d \phi$ represents the hard annihilation of
partons, and the associated virtual corrections at transverse-momentum 
scales $k_T \sim Q$ are included in the coefficient functions $C_{ji}$.
The form factor $e^{\mS_j (b,Q)}$ contains real and virtual contributions due to 
soft [the function $A_j(\alpha_s)$ in (\ref{sudakov:1})] and 
flavor-conserving collinear [the function $B_j(\alpha_s)$ in (\ref{sudakov:1})] 
radiation at scales $Q \gtrsim k_T \gtrsim 1/b$.
At very low momentum scales, $k_T\lesssim 1/b$, real and virtual 
soft-gluon corrections cancel, because the cross section is IR safe, 
and only real and virtual contributions due to collinear radiation remain,
inducing the contributions to the coefficient functions 
$C_{ji} (z, \alpha_s(b_0^2/b^2) )$ and the RG evolution of 
the parton distributions $f_{i/h}(x, b_0^2/b^2)$.

\section{The NLL+LO cross section for the tDY}

The $Q_T$-differential cross section (\ref{cross section}) is given by the sum of the 
resummed component (\ref{resum:2}) and the finite component determined by (\ref{finite}).
For $Q_T\ll Q$, (\ref{cross section}) is dominated by the resummed part, 
$d\sigma^{\rm (res.)} /dQ^2 dQ_T^2 dy d\phi$, which behaves as $\sim 1/Q_T^2$ 
[see (\ref{eq:order1}), (\ref{eq:resLL}) and (\ref{eq:resLL2})]. 
On the other hand, for $Q_T\sim Q$, it approaches the fixed-order result, 
$\left. d\sigma /dQ^2 dQ_T^2 dy d\phi \right|_{\rm fo}$, 
since there is no logarithmic enhancement and 
$d\sigma^{\rm (res.)}/dQ^2 dQ_T^2 dy d\phi$ and 
$\left. d\sigma^{\rm (res.)}/dQ^2 dQ_T^2 dy d\phi \right|_{\rm fo}$ 
almost cancel.\footnote{
This cancellation is often very subtle in a numerical sense.
See, for example, Ref.\citen{ERV}. }
Therefore, in (\ref{cross section}) with (\ref{finite}), 
the resummed cross section, which is dominant at small $Q_T$, and 
the fixed-order cross section controlling the large $Q_T$ region 
are most naturally matched at intermediate $Q_T$, without double counting.  
For example, as discussed in \S5.2, the resummed component 
$d\sigma^{\rm (res.)} /dQ^2 dQ_T^2 dy d\phi$ at NLL accuracy 
is exact up to ${\cal O}(\alpha_s (\mu_R^2 ))$ when expanded in powers of 
$\alpha_s (\mu_R^2 )$.
Therefore, the NLL resummed cross section should be matched with 
the fixed-order cross section at LO, which is of order $\alpha_s$ for $Q_T>0$
[see the discussion below (\ref{eq:nom})].
We refer to the resulting cross section (\ref{cross section}) 
as the ``NLL+LO'' cross section, $d\sigma^{\rm NLL+LO} /dQ^2 dQ_T^2 dy d\phi$.
Similarly, the NNLL resummed cross section is matched with the NLO cross
section, giving the result for the ``NNLL+NLO'' cross section, and so forth. 
Note also that, in order to evaluate the NLL+LO cross section with consistent accuracy,
the NLO parton distributions have to be used as the input for $f_{i/h}(x, \mu^2)$.
The evaluation of the NNLL+NLO cross section requires the NNLO
parton distributions.

\subsection{The spin-dependent cross section}

In what follows, we calculate the NLL+LO spin-dependent cross section of the tDY, 
performing the matching with the fixed-order result (\ref{owari}) obtained in \S4.
We also ``reorganize'' 
the resummed component given in (\ref{resum:2}) and (\ref{resum0}) beyond the CSS form.
This is necessary for the consistent evaluation of the $b$ integral in
(\ref{resum:2}) to the required accuracy.

Following the general formula described in \S\S5, 6, we perform the $Q_T$
resummation for tDY at the NLL level. 
At this level, we must know the coefficients
$\{ A^{(1)}, \, A^{(2)}, \, B^{(1)}, \, C^{(1)}\}$ in (\ref{expand}). 
Then, substituting (\ref{coeff_NLL}), (\ref{eq:coeff.10}) and (\ref{eq:nom}) 
into (\ref{resum:2}) and (\ref{resum0}),
the NLL-level resummed component of the cross section is given by
$\mN \cos\, (2 \phi_{k_1} - \phi_1 - \phi_2) \Delta_T X^{\rm NLL}$ 
[see (\ref{owari})], with
\bea
\Delta_T X^{\rm NLL}(Q_T^2,Q^2,y)=&& 
\int_{\cal C} db \frac{b}{2} J_0(bQ_T) 
e^{S(b,Q)}\int_{x_1^0}^1\frac{dz}{z}\int_{x_2^0}^1\frac{dz'}{z'}
\Delta_T C \left(z, \alpha_s(b_0^2/b^2)\right)
\nonumber\\
&&\times
\Delta_T C \left(z', \alpha_s(b_0^2/b^2)\right)
\delta H\left(\frac{x_1^0}{z}, \frac{x_2^0}{z'};\frac{b_0^2}{b^2}\right) .
\label{resum:pol} 
\eea 
Here, the relevant coefficient function is given by
\be
\Delta_T C \left(z, \alpha_s \right) 
= \delta(1-z) + \frac{\alpha_s}{2\pi}\Delta_TC^{(1)} (z) \ ,
\label{eq:coeff.100}
\ee
with 
\be
\Delta_TC^{(1)} (z) = C_F \left( \frac{\pi^2}{2} - 4 \right) \, 
\delta(1 - z) \ ,
\label{eq:coeff.1}
\ee
and
\be
S(b ,\,Q ) =
    \frac{1}{\alpha_s (\mu_R^2)}\  h^{(0)} (\lambda) + h^{(1)} (\lambda)
\label{sudakov:2}
\ee
with (\ref{eq:h0}), (\ref{eq:h1}) and (\ref{coeff_NLL})
is the exponent (\ref{sudakov:11}) for the Sudakov quark form factor up to NLL accuracy.
Above, we have also introduced the shorthand notation for the contour
integration of (\ref{resum:2}):
\bea
\int_{\cal C} db \frac{b}{2} J_0(bQ_T) \mW (b;\, Q\,,\,y, \phi)
&&\equiv
\int_{\cal C_+} d b \, \frac{b}{4}\,
     H_0^{(1)} (b Q_T)\mW (b;\, Q\,,\,y, \phi)
\nonumber\\ 
&& 
    + \int_{\cal C_-} d b \, \frac{b}{4}\,H_0^{(2)} (b Q_T)\mW (b;\, Q\,,\,y, \phi) .
\label{eq:shorti}
\eea

In (\ref{resum:pol}),
the parton distributions, as well as the coupling constant $\alpha_s(b_0^2 /b^2)$
associated with the coefficient function $\Delta_TC^{(1)}$, 
depend on $b$, in addition to the Sudakov factor $e^{S(b,Q)}$.
Similarly to the Sudakov exponent $S(b,Q)$ given in (\ref{sudakov:2}), 
we can reorganize this $b$ dependence in terms of a systematic large logarithmic 
expansion, using $\lambda$ of (\ref{eq:lambda}).
In order to extract this $b$ dependence explicitly, 
we take the double Mellin moments of (\ref{resum:pol})
with respect to the DY scaling variables $x_{1,2}^0$ at fixed $Q$,
\bea
\Delta_T X_{N_1,N_2}^{\rm NLL}(Q^2,Q_T^2)&&\equiv
\int_0^1 dx_1^0(x_1^0)^{N_1-1}\int_0^1dx_2^0(x_2^0)^{N_2-1}
\Delta_T X^{\rm NLL}(Q_T^2,Q^2,y)
\nonumber\\
=&& 
\int_{\cal C} db \frac{b}{2} J_0(bQ_T) 
e^{S(b,Q)}
\Delta_T C_{N_1}\left(\alpha_s( b_0^2 /b^2)\right)  
\Delta_T C_{N_2} \left(\alpha_s(b_0^2 /b^2)\right) 
\nonumber\\
&& \times
\delta H_{N_1,N_2}(b_0^2 /b^2) \ ,
\label{eq:dMel}
\eea 
where [see (\ref{tPDF})]
\bea
\delta H_{N_1,N_2}(\mu^2)&=&\int_0^1 dx_1 x_1^{N_1-1}\int_0^1dx_2 x_2^{N_2-1}
\delta H(x_1,x_2;\mu^2)
\nonumber \\
&=&\sum_{j=q,\bar{q}} e_j^2 f_{j/h_1,N_1}(\mu^2) f_{\bar{j}/h_2,N_2}(\mu^2) ,
\nonumber\\
f_{j/h,N}(\mu^2) &\equiv& \int_0^1dxx^{N-1}f_{j/h}(x,\mu^2) , 
\label{eq:Mellin000}
\eea
with $f_{q/h}(x,\mu^2)\equiv \delta q_h(x,\mu^2)$,
$f_{\bar{q}/h}(x,\mu^2)\equiv \delta \bar{q}_h(x,\mu^2)$, and\footnote{
In the present case, actually, the moment of the coefficient function, 
$C_{ji,N}$, is diagonal in $(j,i)$ and is independent of $N$, as seen from
(\ref{eq:Mellin0}) and (\ref{eq:coeff.1}).
However, even when the $C_{ji,N}$ have non-zero components 
for $j\neq i$ and depend on $N$, the results given in
(\ref{eq:Creplace0}) and (\ref{eq:Creplace}) below hold 
for each $(j,i)$-component and each moment $N$ of $C_{ji,N}$.}
\be
\Delta_T C_{N}\left(\alpha_s \right)  \equiv \int_0^1dxx^{N-1} 
\Delta_T C \left(x, \alpha_s \right) = 1
+ \frac{\alpha_s}{2\pi} \Delta_T C_{N}^{(1)}.
\label{eq:Mellin0}
\ee
We rewrite the moment of the coefficient function, 
$\Delta_T C_{N}\left(\alpha_s(b_0^2/b^2) \right)$, in (\ref{eq:dMel}) formally as
\be
\Delta_T C_{N}\left(\alpha_s(b_0^2/b^2) \right) 
=\Delta_T C_{N}\left(\alpha_s(Q^2) \right) 
\exp\left\{ \ln 
\frac{\Delta_T C_{N}\left(\alpha_s(b_0^2/b^2) \right)}
{\Delta_T C_{N}\left(\alpha_s(Q^2) \right)}\right\}.
\label{eq:RGid}
\ee
Also, note that we obtain the following large logarithmic expansion 
of the running coupling constant (\ref{eq:run}) using (\ref{eq:lambda})~\cite{KLSV}:
\bea
&&
\ln\left(\frac{\alpha_s(Q^2 )}{\alpha_s(b_0^2 /b^2 )}\right)
\non\\
&&\hspace{1cm}
=\ln(1-\lambda)+ \alpha_s(\mu_R^2 )\left[ \frac{\beta_1}{\beta_0}
\frac{\ln(1-\lambda)}{1-\lambda}
+\beta_0 \ln \left(\frac{Q^2}{\mu_R^2}\right) \frac{\lambda}{1-\lambda} \right]
+ {\cal O}\left(\alpha_s(\mu_R^2 )^2\right) \ .
\non\\
\label{eq:largel}
\eea
Substituting (\ref{eq:Mellin0}) into the exponent of (\ref{eq:RGid}), and using 
(\ref{eq:largel}), we find
\be
\ln \frac{\Delta_T C_{N} (\alpha_s(b_0^2/b^2))}{\Delta_T C_{N}(\alpha_s(Q^2) )} =
\frac{\alpha_s(\mu_R^2)}{2\pi}\Delta_T C_{N}^{(1)} \frac{\lambda}{1-\lambda}
+{\cal O}\left(\alpha_s(\mu_R^2)^2\right) .
\label{eq:Creplace0}
\ee
Therefore, the exponent on the RHS of (\ref{eq:RGid}) corresponds to 
the NNLL level, according to the order counting of the contributions 
in the Sudakov exponent discussed below (\ref{sudakov:11}).
Because the NNLL-level term was ignored in the Sudakov exponent (\ref{sudakov:2}),   
we can make the replacement 
\be
\Delta_T C_{N}\left(\alpha_s(b_0^2/b^2) \right) \rightarrow 
\Delta_T C_{N}\left(\alpha_s(Q^2) \right) 
\label{eq:Creplace}
\ee
for the coefficient function in (\ref{eq:dMel}) at NLL accuracy.

The $b$ dependence of the parton distributions in (\ref{eq:dMel})
can be reorganized similarly. 
In this case, the moments of the DGLAP kernel of (\ref{eq:DGLAPk}) 
give the corresponding anomalous dimensions $\gamma_{ij,N}(\alpha_s)$, 
and, for the NLO parton distributions relevant to the NLL+LO cross section, 
we have
\be
\gamma_{ij,N}(\alpha_s) \equiv\int_0^1 dz z^{N-1} \frac{\alpha_s}{2\pi}
\mathscr{P}_{ij} (z,\alpha_s) 
= \frac{\alpha_s}{2\pi} \gamma_{ij,N}^{(0)} + \left( \frac{\alpha_s}{2\pi}\right)^{2}
\gamma_{ij,N}^{(1)} , 
\label{eq:Mellin}
\ee
where the RHS constitutes the perturbative expansion up to the two-loop term,
and $\gamma_{ij,N}^{(0)}$ and $\gamma_{ij,N}^{(1)}$ are given by (\ref{eq:P0N})
and (\ref{P1N}) for the transversity distributions.
Then, the DGLAP equation (\ref{eq:DGLAP}) yields the RG evolution,
\be
f_{i/h,N}(b_0^2/b^2)=\sum_{j=q,\bar{q}} \mU_{ij,N}(b_0^2/b^2,Q^2)f_{j/h,N}(Q^2)  ,
\label{eq:mom}
\ee
where $\mU_{ij,N} (b_0^2/b^2,Q^2) \equiv [ \boldsymbol{\mU}_N(b_0^2/b^2,Q^2) ]_{ij}$ 
denotes the evolution operator, where 
\be
\boldsymbol{\mU}_N(b_0^2/b^2,Q^2) 
={\cal P} \exp\left\{\int_{Q^2}^{b_0^2/b^2}\frac{d\kappa^2}{\kappa^2}
\boldsymbol{\gamma}_{N}(\alpha_s(\kappa^2))\right\} ,
\label{eq:RGevol}
\ee
with $\gamma_{ij,N}(\alpha_s) \equiv [ \boldsymbol{\gamma}_N(\alpha_s) ]_{ij}$,
and the symbol ``${\cal P}$'' denotes the ``path ordering'' expansion 
of the exponential matrix.
Substituting (\ref{eq:Mellin}) into (\ref{eq:RGevol}), the result possesses 
the well-known structure,\cite{FP,BV}
\be
\boldsymbol{\mU}_N(b_0^2/b^2,Q^2) 
= \left[1 + {\cal O}\left(\alpha_s(b_0^2 /b^2 ) \right)\right] 
\exp \left\{ \frac{ \boldsymbol{\gamma}^{(0)}_{N}}{2\pi\beta_0}
\ln \left(\frac{\alpha_s(Q^2)}{\alpha_s(b_0^2 /b^2 )}\right) \right\} 
\left[1 + {\cal O}\left(\alpha_s(Q^2 )\right) \right],
\label{eq:RGevol2}
\ee
where the ${\cal O}\left(\alpha_s(b_0^2 /b^2 ) \right)$ term 
and the ${\cal O}\left(\alpha_s(Q^2 )\right)$ term
involve the two-loop anomalous dimension matrix
$\boldsymbol{\gamma}^{(1)}_{N}$, and their explicit forms for 
the transversity distributions can be found in (\ref{DGLAP}). 
Using (\ref{eq:mom}), (\ref{eq:RGevol2}) and (\ref{eq:largel}) 
for the parton distributions in (\ref{eq:dMel}), we find that 
we can replace the parton distributions in (\ref{eq:dMel}) as~\cite{BCFG}
\be
f_{i/h,N}(b_0^2/b^2) \rightarrow \sum_{j=q,\bar{q}} U_{ij,N}(b_0^2/b^2,Q^2)
f_{j/h,N}(Q^2 ) ,
\label{eq:freplace}
\ee 
with
\be
U_{ij,N}(b_0^2/b^2,Q^2)=\delta_{ij}e^{R_N(\lambda)},~~~~
R_N(\lambda)\equiv
\frac{\gamma^{(0)}_{qq,N}}{2\pi\beta_0}\ln(1-\lambda) ,
\label{eq:RN}
\ee
up to irrelevant corrections which are down by $\alpha_s(\mu_R^2 )$ and 
vanish for $\lambda=0$. These corrections correspond to the NNLL-level
term omitted from the Sudakov exponent (\ref{sudakov:2}).
Now, (\ref{eq:dMel}) reads, to NLL accuracy,
\bea
&&\hspace{-1cm}
\Delta_T X_{N_1,N_2}^{\rm NLL}(Q^2,Q_T^2)
= \left[1+\frac{\alpha_s(Q^2)}{2\pi}C_F(\pi^2-8)\right]
\delta H_{N_1,N_2}(Q^2) I_{N_1,N_2}(Q_T^2,Q^2) \ ,
\label{eq:X_Mellin}\\&&\hspace{-1cm}
I_{N_1,N_2}(Q_T^2,Q^2)\equiv 
\int_{\cal C} db\frac{b}{2}J_0(bQ_T)
e^{S(b,Q)+R_{N_1}(\lambda)+R_{N_2}(\lambda)} \ .
\label{eq:I_Mellin}
\eea 
In (\ref{eq:X_Mellin}), the large logarithms $L=\ln(Q^2b^2/b_0^2)$ and the associated 
$b$-integral is factorized into $I_{N_1,N_2}(Q_T^2,Q^2)$,
and therefore, the all-order resummation of the large logarithmic 
contributions is now realized at the partonic level.

If we expand $e^{S(b,Q)+R_{N_1}(\lambda)+R_{N_2}(\lambda)}$
in powers of $\lambda$ in the integrand of (\ref{eq:I_Mellin}), 
similarly to (\ref{eq:LLexp}) and (\ref{eq:LLFou}),
the most enhanced logarithmic contributions coincide with the
leading tower $\alpha_s^nL^{2n}$ in (\ref{eq:LLFou}), which comes from
the expansion of $e^{-(A^{(1)}/2\pi) \alpha_sL^2}$
associated with the LL term, $h^{(0)}/\alpha_s$, 
of the Sudakov exponent (\ref{sudakov:2}), and result in the contributions
$\alpha_s^n \ln^{2n-1}(Q^2/Q_T^2)/Q_T^2$ in the cross section,
as shown in (\ref{eq:resLL}).
Similarly, noting that [see (\ref{eq:RN})]
\be 
S(b,Q)+R_{N_1}(\lambda)+R_{N_2}(\lambda) 
= -\frac{A^{(1)}}{2\pi}\alpha_sL^2- \frac{B^{(1)}+ \gamma^{(0)}_{qq,N_1}
+\gamma^{(0)}_{qq,N_2}}{2\pi}\alpha_s L + {\cal O}(\alpha_s^2) ,
\ee
the next-to-leading tower $\alpha_s^nL^{2n-1}$, which leads to 
$\alpha_s^n\ln^{2n-2}(Q^2/Q_T^2)/Q_T^2$,  comes from, e.g., 
the ``cross terms'' $\propto (A^{(1)}\alpha_sL^{2})^{n-1} \times
[B^{(1)}+ \gamma^{(0)}_{qq,N_1}+\gamma^{(0)}_{qq,N_2}]\alpha_s L$.
For the terms corresponding to the next-to-next-to-leading tower,
$\alpha_s^nL^{2n-2}$, which lead to $\alpha_s^n\ln^{2n-3}(Q^2/Q_T^2)/Q_T^2$, 
the coefficients $A^{(2)}$ and $\Delta_TC^{(1)}_N$ participate, 
in addition to $A^{(1)}$, $B^{(1)}$, $\gamma^{(0)}_{qq,N_1}$ and 
$\gamma^{(0)}_{qq,N_2}$. 
As mentioned below (\ref{eq:momencon2}), to completely determine 
the next tower, $\alpha_s^nL^{2n-4}$, 
the NNLL coefficients $A^{(3)}$ and $B^{(2)}$ associated with $h^{(2)}(\lambda)$
of (\ref{sudakov:11}) and also the coefficient functions and 
anomalous dimensions at the two-loop level, 
$\Delta_TC_{ij,N}^{(2)}$ and $\gamma_{ij,N}^{(1)}$, are necessary~\cite{DWS}.
The NLL resummation formula given in (\ref{eq:X_Mellin}) and (\ref{eq:I_Mellin})
fully sums up the first three towers of logarithms contributing 
to the cross section, i.e., $\alpha_s^n\ln^m(Q^2/Q_T^2)$ with $m=2n-1,2n-2$ 
and $2n-3$, for all $n$. 

As discussed in \S6.1, for $Q_T \gtrsim Q$, corresponding to the region (i), 
the resummation is irrelevant, and the cross section can be accurately 
obtained with fixed-order perturbation theory. We can
switch between the resummed and fixed-order calculations at a certain 
value $Q_T \sim Q$\cite{AK}, but such procedure, introducing ad-hoc boundaries
between the large-$Q_T$ and small $Q_T$-regions, is not convenient for 
the actual evaluation of the cross section. 
For this reason, we use (\ref{eq:X_Mellin}) and (\ref{eq:I_Mellin})
over the entire region of $Q_T$. 
The formula (\ref{eq:I_Mellin}) exponentiates the functions of $\lambda$ 
in the integrand, as a result of the large logarithmic expansion, 
which is valid in the regions (ii) and (iii) of \S6.
Thus, this exponentiated form is effective when $Q_T \lesssim Q$, 
corresponding to (ii) and (iii), but its behavior cannot be trusted 
for small $b\lesssim 1/Q$ ($Q_T \gtrsim Q$); indeed, $L$ in
(\ref{eq:lambda}) is large for small $b$ as well as for large $b$, and thus
(\ref{eq:I_Mellin}) also exponentiates the 
unjustified large logarithmic contributions at large $Q_T$.~\footnote{When we expand 
the integrand of (\ref{eq:I_Mellin}) in powers of $\alpha_s$, 
the enhanced behavior of $L$ for small $b$ is harmless,
order-by-order in $\alpha_s$ [see (\ref{eq:order1}), (\ref{eq:resLL})]. 
This is because the $b$ integration in $\mI_n$ of (\ref{eq:I}) 
converges as $b\rightarrow 0$.}
This problem can be avoided by making the replacement~\cite{BCFG,CTTW}
\bea   L = \ln \left( Q^2 b^2 / b_0^2 \right)
       \ \to \ \tilde{L} = \ln \left(Q^2 b^2 / b_0^2 + 1 \right)  
\label{tilde-L}
\eea
in the definition (\ref{eq:lambda}) of $\lambda$.
Through this replacement, the 
unjustified logarithmic contributions are suppressed: as $b \rightarrow 0$, 
we get $L \rightarrow 0$, and $e^{S(b,Q)+R_{N_1}(\lambda)+R_{N_2}(\lambda)}\ra 1$.
By contrast, the resummation of the large logarithmic contributions in 
(\ref{eq:X_Mellin}) and (\ref{eq:I_Mellin}) for $b\gtrsim 1/Q$ is not 
affected by (\ref{tilde-L}), since $L$ and $\tilde{L}$ are equivalent 
for organizing the large logarithmic expansion, 
as we have $\tilde{L}  = L+{\cal O}(1/(Q^2b^2))$.
After this replacement, (\ref{eq:X_Mellin}) and (\ref{eq:I_Mellin}) read
\bea
\Delta_T\tilde{X}_{N_1,N_2}^{\rm NLL}(Q^2,Q_T^2)
&=& \left[1+\frac{\alpha_s(Q^2)}{2\pi}C_F(\pi^2-8)\right]
\delta H_{N_1,N_2}(Q^2) \tilde{I}_{N_1,N_2}(Q_T^2,Q^2) \ ,\non\\
\label{tilde-X_Mellin}\\
\tilde{I}_{N_1,N_2}(Q_T^2,Q^2)
&=& \int_{\cal C} db\frac{b}{2}J_0(bQ_T)
\left[e^{S(b,Q)+R_{N_1}(\lambda)+R_{N_2}(\lambda)}\right]_{L\rightarrow \tilde{L}},
\label{eq:tilde-I_Mellin}
\eea
and we denote the double inverse Mellin transform of 
$\Delta_T \tilde{X}^{\rm NLL}_{N_1, N_2}  (Q_T^2 , Q^2)$ from $(N_1 ,N_2 )$ 
space to $(x_1^0 , x_2^0 )$ space as $\Delta_T \tilde{X}^{\rm NLL}  (Q_T^2 , Q^2, y)$,
which represents the new resummed component to NLL accuracy.

The NLL resummed component obtained in this way is now combined with 
the fixed order cross section (\ref{owari}). 
Specifically, the finite component in the NLL+LO cross section is determined by
(\ref{finite}), where we use (\ref{owari}) as the first term on the RHS.
The result represents the NLL+LO cross section of the tDY
in the $\overline{\rm MS}$ scheme, 
\bea
&&   
\frac{\Delta_T d \sigma^{\rm NLL+LO}}{d Q^2 d Q_T^2 d y d \phi_{k_1}}
\non\\
&&\hspace{2cm}   
= \mN \cos\, (2 \phi_{k_1} - \phi_1 - \phi_2)
      \left[ \Delta_T\tilde{X}^{\rm NLL}\, (Q_T^2, Q^2,y)
             + \Delta_T\tilde{Y}\, (Q_T^2, Q^2, y) \right]
   ,\nonumber\\
\label{eq:NLL+LO}
\eea
where
\bea
\Delta_T\tilde{Y}(Q_T^2,Q^2,y)
&\equiv&
\Delta_TX(Q_T^2,Q^2,y)+ \Delta_TY(Q_T^2,Q^2,y)
-\Delta_T\tilde{X}^{\rm NLL}(Q_T^2,Q^2,y)
\biggl|_{\rm fo}  .
\nonumber\\
\label{eq:tilde-Y}
\eea
Here, $(\cdots)|_{\rm fo}$ now denotes the expansion up to 
${\cal O}\left(\alpha_s(\mu_R^2)\right)$.
Note that $\Delta_T\tilde{Y}(Q_T^2,Q^2,y)$ would coincide with 
$\Delta_TY(Q_T^2,Q^2,y)$, if we did not perform the replacement
(\ref{tilde-L}) for $\Delta_T \tilde{X}^{\rm NLL}  (Q_T^2 , Q^2, y)$.
We calculate the ``subtraction term'' 
$\Delta_T\tilde{X}^{\rm NLL}|_{\rm fo}$ in (\ref{eq:tilde-Y})
explicitly [compare with (\ref{eq:order1})] and obtain
\bea
\left. \Delta_T \tilde{X}^{\rm NLL}(Q_T^2,Q^2,y)\right|_{\rm fo}
&&= \left[1+\frac{\alpha_s(\mu_R^2)}{2\pi}C_F(\pi^2-8)\right] \tilde{\mI}_0 \
\delta H(x_1^0,x_2^0;\mu_F^2  )\nonumber\\
+&&
\frac{\alpha_s(\mu_R^2)}{2\pi}C_F \ \Biggl\{ \delta H(x_1^0,x_2^0;\mu_F^2  )
(-\tilde{\mI}_2+3\tilde{\mI}_1)
\non\\
+ &&\left(\tilde{\mI}_0  \ln \frac{Q^2}{\mu_F^2} -\tilde{\mI}_1\right)
\left[ \int^1_{x_1^0} \frac{d z}{z}
\Delta_T \, P_{qq} (z)
        \delta H 
     \left( \frac{x_1^0}{z}, x_2^0 ;\ \mu_F^2 \right)\right. \nonumber\\
&&\left.  +  \int^1_{x_2^0} \frac{d z}{z}\,  \Delta_T \, P_{qq} (z)\, 
         \delta H \left( x_1^0\,,\,\frac{x_2^0}{z};\ \mu_F^2  \right)
 \right]\Biggr\} ,
\label{subtract}
\eea
where
\bea
\tilde{\mI}_n&\equiv& \int_{\cal C} db\frac{b}{2} J_0(bQ_T)
\ln^n \left(Q^2b^2/b_0^2+1\right)  .
\label{eq:tilde-I}
\eea
Similarly to the case of $\mI_n$, (\ref{eq:Inew}), we can relate this to 
$\int_{\cal C} db(b/2) J_0(bQ_T) (Q^2b^2/b_0^2+1)^{\eta}$, where 
the contour ${\cal C}$ can be deformed into that from $b=0$ to $b=\infty$ 
along the real $b$-axis using analytic continuation for $\eta$ and 
Cauchy's theorem. The resulting integral yields [see (\ref{eq:Cnew2})]
\bea
\tilde{\mI}_0&=& \delta(Q_T^2) ,
\nonumber\\
\tilde{\mI}_1&=& - \frac{b_0}{Q}
\left[ \frac{1}{Q_T}
K_1\left(\frac{b_0Q_T}{Q}\right)\right]_{+\infty}, \label{eq:tilde-I1}\\
\tilde{\mI}_2&=& -\frac{2b_0}{Q} \left[
K_1\left(\frac{b_0Q_T}{Q}\right)\frac{\ln\left(Q/Q_T\right)}{Q_T}
+\frac{Q}{b_0Q_T^2}K_0\left(\frac{b_0Q_T}{Q}\right)\right]_{+\infty}
\label{eq:tilde-I2} ,
\eea
where $K_{n}(x)$ denotes the modified Bessel functions of the second kind,
and we have introduced the notation ``$[ \cdots ]_{+\infty}$'' to indicate
the ``generalized + distribution'' defined between $0\leq Q_T^2<\infty$, 
as $\int_0^\infty dQ_T^2 [ \cdots ]_{+\infty} =0$ (see Appendix~B of Ref.~\citen{BCFG}).
The quantity $\Delta_TX$ given in (\ref{eq:X0}) and (\ref{eq:X}) is 
identical to (\ref{subtract}) with $\tilde{\mI}_n$ replaced by $\mI_n$,
and $\tilde{\mI}_n$ is given by $\mI_n$ with the replacement
$L\ra\tilde{L}$ of (\ref{tilde-L}).
Because the replacement $L\ra\tilde{L}$ does not affect the large $b$ behavior of the
corresponding integrand,
the small $Q_T$ behavior of $\tilde{\mI}_n$ is the same as that of $\mI_n$.
In fact, we can show~\cite{BCFG} that for $Q_T \ll Q$, we have 
$\tilde{\mI}_n= \mI_n\times(1+{\cal O}(Q_T^2/Q^2))$, i.e., 
$\tilde{\mI}_n -\mI_n = {\cal O}(\ln^{n-1}(Q_T^2/Q^2)/Q^2)$ [see (\ref{eq:genIn})],
and the difference $\Delta_TX - \Delta_T\tilde{X}^{\rm NLL}|_{\rm fo}$ 
is given by contributions less singular than
$1/Q_T^2$ or $\delta(Q_T^2)$ as $Q_T^2 \rightarrow 0$.\footnote{
The fact that the terms proportional to $\delta(Q_T^2)$ 
cancel out in the difference $\Delta_TX-\Delta_T\tilde{X}^{\rm NLL}|_{\rm fo}$
is actually a nontrivial point.
In the present NLL+LO case, they do indeed cancel, as explicitly shown in Appendix~C.}
Therefore, $\Delta_T\tilde{Y}$ differs from $\Delta_TY$
only by the ${\cal O}(\alpha_s)$ terms, which are less singular than
$1/Q_T^2$ or $\delta(Q_T^2)$ as $Q_T^2 \rightarrow 0$;
the entire effect of the replacement $L\ra\tilde{L}$ is to move a portion of 
the ``finite'' component to the ``singular'' component in the cross section.
We can rewrite (\ref{eq:tilde-Y}) as
\bea
\Delta_T\tilde{Y}(Q_T^2,Q^2,y)
&=&\Delta_TX^{(1)}(Q_T^2,Q^2,y)\biggr|_{Q_T^2>0}+\Delta_TY(Q_T^2,Q^2,y)
\non\\&&\hspace{3cm}
-\Delta_T\tilde{X}^{\rm NLL}(Q_T^2,Q^2,y)\biggl|_{{\cal O}(\alpha_s),~Q_T^2>0} \ ,
\non
\label{eq:tilde-Y:2}\\
\eea
where $(\cdots)|_{{\cal O}(\alpha_s)}$ denotes the order 
$\alpha_s(\mu_R^2)$ term in the expansion of ($\cdots$).
Note that $\Delta_T\tilde{Y}$ is regular over the entire region of $Q_T$, 
except for a ``weak'' singularity $\propto\ln(Q^2/Q_T^2)/Q^2$, which 
also exists in $\Delta_TY$.\footnote{This weak singularity 
does not matter in $d\sigma/dQ_T$ instead of $d\sigma/dQ_T^2$.} 
The contributions of the first and second terms in (\ref{eq:tilde-Y:2}) 
correspond to the LO cross section (\ref{cross section2}), 
and hence, the NLL+LO cross section (\ref{eq:NLL+LO}) is actually 
the NLL resummed cross section with the ${\cal O}(\alpha_s)$ terms of 
its expansion subtracted, plus the LO cross section.   

One of the advantages of making the replacement $L\ra\tilde{L}$ is that 
the NLL+LO cross section defined above satisfies the ``unitarity
constraint,''\cite{BCFG} which means that (\ref{eq:NLL+LO})
integrated over $Q_T$ exactly reproduces the $Q_T$-integrated cross section 
at NLO. Noting that $e^{S(b,Q)+R_{N_1}(\lambda)+R_{N_2}(\lambda)}=1$ for $b=0$ 
[see (\ref{eq:tilde-I_Mellin})],
we find that the $Q_T$-integral of $\Delta_T\tilde{X}^{\rm NLL}$ is given by
\bea
\int_0^\infty dQ_T^2\ \Delta_T\tilde{X}^{\rm NLL}(Q_T^2,Q^2,y)
&=&\left[1+\frac{\alpha_s(Q^2)}{2\pi}(\pi^2-8)\right] \delta
H(x_1^0,x_2^0;Q^2)
\label{eq:int-tilde-X}\\
&=&\int_0^{Q^2} dQ_T^2\ \Delta_TX(Q_T^2,Q^2,y)  ,
\nonumber
\eea
with $\Delta_TX(Q_T^2,Q^2,y)$ given by (\ref{eq:X0}) and (\ref{eq:X}).
In the second equality, we have used the relations,
\be
\int_{0}^{Q^2}dQ_T^2\ \mI_{n}= \varrho_n ,
\;\;\;\;\;\;\;\;\;\;\;\;\;\;\;\;\;
\int_{0}^\infty d Q_T^2\ \tilde{\mI}_{n} = \delta_{n,0} ,
\label{eq:int-I}
\ee
with $\varrho_0=1$ and $\varrho_1=\varrho_2 = 0$,
which follow from (\ref{eq:genIn}) and (\ref{eq:tilde-I}), respectively.\footnote{
We note that $\varrho_3 = -4 \zeta(3), \varrho_4 =0, \varrho_5=-48 \zeta(5), \cdots$,
and the nonzero contributions in (\ref{eq:int-I}) for $n\ge 3$ 
come from the ``contact term'' proportional to $\delta(Q_T^2)$ in (\ref{eq:genIn}).
Therefore, in general cases at NNLL+NLO or higher, in principle,
we have to take into account those contact-term contributions
to satisfy the corresponding unitarity constraint.
To avoid this complication, one convenient method is to perform 
the matching between the resummed 
and the fixed-order cross sections at $Q_T > 0$ from the beginning~\cite{BCFG}.}
For the $\Delta_T\tilde{Y}$-term, using (\ref{eq:tilde-Y}) and
(\ref{eq:int-I}), we find
\bea
&&
\int_0^\infty dQ_T^2 \Delta_T\tilde{Y}(Q_T^2,Q^2,y)
\non\\&&\hspace{2cm}
=\int_{0}^{Q_{T,max}^2} dQ_T^2 
\left[\Delta_TX(Q_T^2,Q^2,y)+\Delta_TY(Q_T^2,Q^2,y) \right]
\non\\&&\hspace{8cm}
-\int_{0}^\infty dQ_T^2 \Delta_T\tilde{X}^{\rm NLL}(Q_T^2,Q^2,y)\biggr|_{\rm fo}
\non\\&&\hspace{2cm}
=\int_{Q^2}^{Q_{T,max}^2} dQ_T^2 \Delta_TX(Q_T^2,Q^2,y)
+\int_0^{Q_{T,max}^2} dQ_T^2 \Delta_TY(Q_T^2,Q^2,y) ,\non\\
\label{eq:int-tilde-Y}
\eea
where $Q_{T,max}$ denotes the maximum value of $Q_T$ determined 
from the partonic kinematics,
\bea
Q_{T,max}=Q\frac{\sqrt{\left[1-(x_1^0)^2\right]\left[1-(x_2^0)^2\right]}}{x_1^0+x_2^0}  .
\eea
Adding (\ref{eq:int-tilde-X}) and (\ref{eq:int-tilde-Y}), 
we find that the $Q_T$-integral of the NLL+LO cross section (\ref{eq:NLL+LO}) 
is identical to that of the fixed-order $Q_T$ differential cross section (\ref{owari}), 
and hence
\bea
\int_0^\infty d Q_T^2 \frac{\Delta_T d \sigma^{\rm NLL+LO}}
{d Q^2 d Q_T^2 d yd \phi_{k_1}}
=\frac{\Delta_T d \sigma}{d Q^2d y d \phi_{k_1}} .
\label{eq:unitarity}
\eea
Here, we note that the RHS, i.e. the $Q_T$-integral of (\ref{owari}), is counted as 
the NLO cross section, because the $Q_T$-integrated cross section at LO is 
of ${\cal O}(\alpha_s^0)$, and its partonic subprocess is expressed 
by the tree diagram in Fig.~1.

\subsection{The unpolarized cross section}

The NLL+LO $Q_T$-differential cross section for the unpolarized DY process 
is obtained in the same way as the spin-dependent cross section in \S7.1, 
and the results are given in Appendix~A of Ref.~\citen{KKT07}
(see also Refs.\cite{AEGM,CSS}). 
The NLL+LO $Q_T$-differential cross section for unpolarized DY reads
\bea
   \frac{d \sigma^{\rm NLL+LO}}{d Q^2 d Q_T^2 d y d \phi_{k_1}}
   = 2 \, \mN 
      \left[ \tilde{X}^{\rm NLL}\, (Q_T^2, Q^2,y) 
    + \tilde{Y}\, (Q_T^2, Q^2, y) \right] \ ,
\label{eq:NLL+LO:unp}
\eea
where the NLL resummed component $\tilde{X}^{\rm NLL}$ is given by
\bea
&&
\tilde{X}^{\rm NLL}(Q_T^2,Q^2,y)= 
\int_{\cal C} db \frac{b}{2} J_0(bQ_T)\left( e^{S(b,Q)}
\left[ H\left(x_1^0,x_2^0;\frac{b_0^2}{b^2}\right)
\right.\right.
\non\\&&
+\frac{\alpha_s(Q^2)}{2\pi}\left\{
\int_{x_1^0}^1\frac{dz}{z} C_{qq}^{(1)}(z)
H\left(\frac{x_1^0}{z},x_2^0;\frac{b_0^2}{b^2}\right)+
\int_{x_2^0}^1\frac{dz}{z} C_{qq}^{(1)}(z)
H\left(x_1^0,\frac{x_2^0}{z};\frac{b_0^2}{b^2}\right)
\right.\nonumber\\&&
\left.\left.\left.
+\int_{x_1^0}^1\frac{dz}{z} C_{qg}^{(1)}(z)
K_2\left(\frac{x_1^0}{z},x_2^0;\frac{b_0^2}{b^2}\right)+
\int_{x_2^0}^1\frac{dz}{z} C_{qg}^{(1)}(z)
K_1\left(x_1^0,\frac{x_2^0}{z};\frac{b_0^2}{b^2}\right)
\right\}
\right]\right)_{L\rightarrow \tilde{L}} \ ,\non\\
\label{resum:unpol}
\eea
where the coefficient functions in the $\overline{\rm MS}$ scheme are given by
\bea
C_{qq}^{(1)}(z)=C_F(1-z)+C_F\left(\frac{\pi^2}{2}-4\right) \ , ~~~
C_{qg}^{(1)}(z)=2T_Rz(1-z) \ ,
\label{eq:coeff:unpol}
\eea
with $T_R=1/2$, and $H(x_1,x_2; \mu^2)$ and $K_{1,2}(x_1,x_2; \mu^2)$ denote 
the products of the NLO
unpolarized parton distributions,
\bea
H (x_1,x_2; \mu^2) &=& \sum_qe_q^2
\left[q_{h_1}(x_1,\mu^2)\bar{q}_{h_2}(x_2,\mu^2)
+\bar{q}_{h_1}(x_1,\mu^2)q_{h_2}(x_2,\mu^2)\right]  , \label{eq:739}\\
K_1(x_1,x_2; \mu^2) &=& \sum_qe_q^2
\left[q_{h_1}(x_1,\mu^2)+\bar{q}_{h_1}(x_1,\mu^2)\right] g_{h_2}(x_2,\mu^2)  , \\
K_2(x_1,x_2; \mu^2) &=& \sum_qe_q^2 \,
g_{h_1}(x_1,\mu^2)
\left[q_{h_2}(x_2,\mu^2)+\bar{q}_{h_2}(x_2,\mu^2)\right]  ,\label{eq:742}
\eea
where $q_h(x,\mu^2)$ and $g_h(x,\mu^2)$ are the quark and gluon density
distributions inside the hadron $h$. In (\ref{resum:unpol}), 
the same manipulations as for $\Delta_T\tilde{X}^{\rm NLL}$, 
such as the replacement (\ref{tilde-L}) and the contour deformation of 
(\ref{eq:shorti}),
are applied. Also, the reorganization of the $b$ dependence of 
the parton distributions according to (\ref{eq:freplace}) and (\ref{eq:RN})
should be understood in terms of the corresponding anomalous dimensions.
Note that the Sudakov exponent $S(b,Q)$ in $(\ref{resum:unpol})$ 
for the unpolarized DY process is same as that 
in (\ref{resum:pol}) for the polarized one, given by (\ref{sudakov:2}).   
Similarly to $\Delta_T\tilde{Y}$ in (\ref{eq:tilde-Y}), 
$\tilde{Y}$ in (\ref{resum:unpol}) is defined using the finite component 
$Y$ of the fixed-order $\alpha_s$ cross section.
(The explicit expression for $Y$ is given in Appendix~A of Ref.~\citen{KKT07}.)
Using logic similar to that for (\ref{eq:int-tilde-X})-(\ref{eq:unitarity}),
we can show that (\ref{eq:NLL+LO:unp}) satisfies the unitarity condition
\be
\int_0^\infty d Q_T^2 \frac{d \sigma^{\rm NLL+LO}}
{d Q^2 d Q_T^2 d yd \phi_{k_1}}
=\frac{d \sigma}{d Q^2d y d \phi_{k_1}} ,
\label{eq:unitarity2}
\ee
where the RHS denotes the NLO $Q_T$-integrated cross section for unpolarized DY.

\section{Asymptotic behavior of the resummed cross section at $Q_T=0$}

In \S6, we discussed how the resummation formula given in 
(\ref{resum00}) and (\ref{resum:2})
is controlled in each of three regions (i), (ii), and (iii)
of the impact parameter $b$, corresponding to the kinematical regions, 
$Q_T\grtsim Q$, $\Lambda_{\rm QCD}\ll Q_T \lsim~ Q$ and 
$Q_T\lsim\Lambda_{\rm QCD}$, respectively. 
In the region (iii), the Sudakov form factor $e^{\mS(b,Q)}$ 
with (\ref{sudakov:11}) cannot be
expanded in terms of $\lambda$, since $|\lambda|= {\cal O}(1)$.
In this section we show that the behavior of (\ref{resum:2}) in the region (iii), 
in particular its $Q_T\rightarrow 0$ limit, is controlled by the saddle point 
in the $b$ integration. Such analysis has been done 
in Refs.~\citen{PP} and \citen{CSS} for the LL resummation formula. 
Here we extend the analysis to the case of the NLL-level resummation formula with 
(\ref{tilde-X_Mellin}) and (\ref{eq:tilde-I_Mellin}).
It is possible to carry out the corresponding extension 
on the basis of the present formalism, which realizes the
resummation at the partonic level, as discussed in \S7.

Let us start from the NLL resummation formula (\ref{eq:tilde-I_Mellin}) with 
the exponentiation of the functions of $\lambda$
in the integrand.
For simplicity, we fix the renormalization scale as $\mu_R=Q$ in the following. 
When $Q$ is large enough that $\alpha_s (Q^2) \ll 1$, 
the $b$ integral in (\ref{eq:tilde-I_Mellin}) at small $Q_T$
is dominated by a saddle point, which is determined mainly by  
the LL term in the Sudakov exponent (\ref{sudakov:2}), 
$h^{(0)}(\lambda)/\alpha_s (Q^2) \rightarrow \infty$.
In this case, the contributions to the $b$ integration from very short 
($|b| \ll 1/Q$) as well as very long distances ($|b| \gg 1/\Lambda_{\rm QCD}$)
along the integration contour ${\cal C_\pm}$ are exponentially suppressed 
[see (\ref{resum:2}), (\ref{eq:shorti})].
This allows us to give up the replacement (\ref{tilde-L}) in (\ref{eq:tilde-I_Mellin}).
Also, we may ignore the integration along the two branches, 
$b=b_c + e^{\pm i\theta}t$ with $t \in \{0, \infty \}$ in  ${\cal C_\pm}$, 
when $b_c$ is sufficiently large, but less than the position of the singularity 
in the Sudakov exponent, $b_L=(b_0 /Q)e^{1/(2\beta_0\alpha_s(Q^2))}$.
(In fact, the relevant integrand has a nice saddle point well below 
$b_L$ (above $0$) for $Q \gg \Lambda_{\rm QCD}$,
as seen from (\ref{eq:lsp000}) and (\ref{eq:bspbsp}) below.)
Then (\ref{eq:tilde-I_Mellin}) reads, up to the exponentially suppressed corrections,
\be
\tilde{I}_{N_1,N_2}(Q_T^2,Q^2)
= \int_0^{b_c} db\ \frac{b}{2}J_0(bQ_T)\
\exp\left[\frac{h^{(0)}(\lambda)}{\alpha_s (Q^2)}
+h^{(1)}(\lambda) +R_{N_1}(\lambda)+R_{N_2}(\lambda)\right],
\label{eq:tilde-I_Mellin2}
\ee
with $b_c < b_L$. 
To evaluate the integral in (\ref{eq:tilde-I_Mellin2})
with the saddle-point method, we consider the simplest case, with $Q_T=0$, 
and change the integration variable from $b$ to $\lambda$ of (\ref{eq:lambda}), as
\begin{equation}
\tilde{I}_{N_1 , N_2}(Q_T^2 = 0, Q^2)=
\frac{b_0^2}{4Q^2 \beta_0 \alpha_s(Q^2)}\int_{-\infty}^{\lambda_c}  d\lambda 
e^{-\zeta^{(0)}(\lambda)
+ h^{(1)}(\lambda) + R_{N_1}(\lambda) + R_{N_2}(\lambda)} ,
\label{eq:sp0}
\end{equation} 
where $\lambda_c = \beta_0 \alpha_{s}(Q^2) \ln(Q^2 b_c^2/b_0^2 )$ and 
\bea
\zeta^{(0)}(\lambda)&& \equiv - \frac{\lambda}{\beta_0 \alpha_s(Q^2)}
-\frac{h^{(0)}(\lambda)}{\alpha_s (Q^2)} . 
\label{eq:fxi0}
\eea
Here, $\zeta^{(0)}(\lambda)$ denotes the leading term $\propto 1/\alpha_s(Q^2)$ 
in the exponent of (\ref{eq:sp0}), 
while the remaining terms,  
$h^{(1)}(\lambda) + R_{N_1}(\lambda) + R_{N_2}(\lambda)$, 
collect the contributions of the NLL level, which are down by $\alpha_s(Q^2)$ 
from the leading term in the relevant region, $\lambda\sim 1$.
The precise position of the saddle point in the integral of (\ref{eq:sp0}) 
is determined by the condition
\bea
-{{\zeta}^{(0)}}'(\lambda)
+{h^{(1)}}'(\lambda ) + {R_{N_1}}'(\lambda) 
+ {R_{N_2}}' (\lambda) = 0  .
\label{eq:nlsp}
\eea
Now we express its solution as
$\lambda = \lambda_{SP} + \Delta \lambda_{SP}$, where $\lambda_{SP}$ 
is the solution of ${\zeta^{(0)}}'(\lambda)=0$; i.e.,
$\lambda_{SP}$ satisfies
\begin{equation}
1-\frac{A_q^{(1)}}{2\pi \beta_0}\frac{\lambda_{SP}}{1-\lambda_{SP}} = 0 ,
\label{eq:lsp0}
\end{equation}
and $\Delta \lambda_{SP}$ denotes the shift of the saddle point 
due to the subleading terms in (\ref{eq:nlsp}), 
${h^{(1)}}'(\lambda ) + {R_{N_1}}'(\lambda) + {R_{N_2}}' (\lambda)$.
From (\ref{eq:lsp0}), we get
\be
\lambda_{SP} =\frac{2\pi \beta_0}{2\pi \beta_0+A_q^{(1)}}
\label{eq:lsp000}
\ee
which is independent of $Q$ and coincides with the saddle point at the LL level
discussed in Refs.~\citen{PP,CSS} and \citen{EV}.
Substituting $A_q^{(1)}=2C_F$ from (\ref{coeff_NLL}), 
we obtain $\lambda_{SP}\simeq 0.6$, which is ${\cal O}(1)$ as it should be. 
Contrastingly, the shift, $\Delta \lambda_{SP}=
[{h^{(1)}}'(\lambda_{SP} ) + {R_{N_1}}'(\lambda_{SP}) 
+ {R_{N_2}}' (\lambda_{SP})]/{{\zeta}^{(0)}}''( \lambda_{SP}) +\cdots$, 
is ${\cal O}(\alpha_s(Q^2))$ or higher.
Up to the exponentially suppressed corrections,
the saddle point evaluation of (\ref{eq:sp0}) around
$\lambda=\lambda_{SP} + \Delta \lambda_{SP}\equiv \omega$ can be 
performed in the usual way. 
Defining $\xi(\lambda) \equiv \zeta^{(0)}(\lambda) 
- h^{(1)}(\lambda) - R_{N_1}(\lambda) - R_{N_2}(\lambda)$, we obtain
\bea
&&
\tilde{I}_{N_1 , N_2}(0, Q^2)
\non\\&&\hspace{1cm}
= \frac{b_0^2}{4Q^2 \beta_0 \alpha_s(Q^2)} e^{-\xi(\omega)}
\int_{-\infty}^{\infty}d\lambda 
e^{-[\xi''(\omega)/2](\lambda-\omega)^2}
\nonumber\\&&\hspace{3cm}
\times 
\left(1 
- \frac{1}{24} \xi''''(\omega) (\lambda-\omega)^4
+\frac{1}{72} [\xi'''(\omega)]^2(\lambda-\omega)^6
+\cdots
\right) 
\nonumber\\&&\hspace{1cm}
=\frac{b_0^2}{4Q^2 \beta_0 \alpha_s(Q^2)} e^{-\xi(\omega)} 
\sqrt{\frac{2\pi}{\xi''(\omega)}}
\left(1-\frac{1}{8} \frac{\xi''''(\omega)}{[\xi''(\omega)]^2}+
\frac{5}{24} \frac{[\xi'''(\omega)]^2}{[\xi''(\omega)]^3} + \cdots \right) .
\non\\\label{eq:speval2speval2}
\eea
Because $\xi(\omega) \sim 1/\alpha_s(Q^2)$ for $\alpha_s(Q^2) \ll 1$, 
the second and third terms in the parentheses on the RHS of (\ref{eq:speval2speval2})
are of ${\cal O}(\alpha_s(Q^2))$,
and similarly it is straightforward to show that the subsequent terms 
denoted by the ellipses are of ${\cal O}\left(\alpha_s(Q^2)^2\right)$ or higher.
We ignore the second and following terms in the parentheses, 
because they correspond to the NNLL or higher-level contributions
that were omitted from our starting point (\ref{eq:tilde-I_Mellin}).
As discussed below (\ref{sudakov:11}), the NNLL contributions are associated with
the terms $\alpha_s^{n+1}L^n \propto \alpha_s \lambda^n$ ($n \ge 1$) in the exponent
of the resummation formula. 
More precisely, in the regions (i) and (ii) of \S6, those NNLL contributions 
behave as $\sim \alpha_s^{2}L,  \alpha_s^{3}L^2, \ldots$,
while at the saddle point $\lambda =\omega=\lambda_{SP} + \Delta \lambda_{SP} \sim 1$,
they would behave as $\sim \alpha_s$. 
A similar situation occurs for the NNNLL and higher-level contributions. 
As mentioned in \S6.3, this indicates that, for $Q_T\approx 0$ in the region (iii),
all contributions are completely controlled by the single small
parameter $\alpha_s(Q^2)$, and we have to use a classification 
with respect to the order counting of the terms
that is different from the customary and all-order resummed perturbation theory
used in the regions (i) and (ii).
In particular, when we start with the NLL-level resummation formula and go to the 
$Q_T\approx 0$ region, 
we should ignore the contributions that are of the same order as 
the ${\cal O}(\alpha_s)$ terms in the exponent of 
the corresponding formula~\cite{CSS,KKT07}.

Substituting $\omega= \lambda_{SP} + \Delta \lambda_{SP}$ and
$\xi=\zeta^{(0)}- h^{(1)} - R_{N_1} - R_{N_2}$ into (\ref{eq:speval2speval2}),
and expanding the result in terms of $\Delta \lambda_{SP}$,
we find 
\begin{equation}
\tilde{I}_{N_1 , N_2}(0, Q^2)=
\frac{b_0^2}{4Q^2 \beta_0 \alpha_s(Q^2)} 
\sqrt{\frac{2\pi}{{\zeta^{(0)}}''(\lambda_{SP})}} 
e^{-[\zeta^{(0)}(\lambda_{SP})+h^{(1)}(\lambda_{SP})
+ R_{N_1}(\lambda_{SP}) + R_{N_2}(\lambda_{SP})]} ,
\label{eq:speval}
\end{equation}
to NLL accuracy. 
Note that using ${\zeta^{(0)}}'(\lambda_{SP})=0$, 
the shift $\Delta \lambda_{SP}$ yields only NNLL-level corrections.
We substitute this into (\ref{tilde-X_Mellin}) and combine the result 
with the coefficient function and the parton distributions. 
Then, using the same logic as above, 
the ${\cal O}(\alpha_s(Q^2))$ term in the coefficient function
in (\ref{tilde-X_Mellin}) is also ignored as  
$[1+ \alpha_s(Q^2) C_F (\pi^2 -8 )/2\pi] \rightarrow 1$, and thus we have
\begin{equation}
\Delta_T \tilde{X}^{\rm NLL}_{N_1, N_2}  (0 , Q^2)
=\delta H_{N_1,N_2}(Q^2) \tilde{I}_{N_1,N_2}(0, Q^2) \ .
\label{resum:3} 
\end{equation}
Then, performing the double inverse Mellin transformation 
to the ($x_1^0$, $x_2^0$) space, we obtain~\cite{KKT07}
\bea
&&
\Delta_T \tilde{X}^{\rm NLL} (0 , Q^2, y)
\non\\
&&\hspace{1cm}
=\left(\frac{b_0^2}{4Q^2 \beta_0 \alpha_s(Q^2)} 
\sqrt{\frac{2\pi}{{\zeta^{(0)}}''(\lambda_{SP})}} 
e^{-\zeta^{(0)}(\lambda_{SP})+h^{(1)}(\lambda_{SP})} \right) 
\delta H \left(x_1^0,x_2^0; \frac{b_0^2}{b_{SP}^2}\right)  ,
\non\\\label{resum:4} 
\eea
with
\be
b_{SP}= \frac{b_0}{Q}e^{\lambda_{SP}/[2\beta_0
\alpha_s(Q^2)]} ,
\label{eq:bspbsp}
\ee
which is the value of $b$ corresponding to $\lambda_{SP}$ in (\ref{eq:lsp000}).
Here, we have used the fact that 
$e^{R_{N_1}(\lambda_{SP})}$ and $e^{R_{N_2}(\lambda_{SP})}$ in (\ref{eq:speval}) 
can be identified with the NLO evolution operators from the scale $Q$ 
to $b_0/b_{SP}$, to the present accuracy. This yields  
$\delta H(x_1^0,x_2^0; Q^2)$ with the scale shifted as
$Q \rightarrow b_0/b_{SP}$ [see (\ref{eq:freplace}), (\ref{eq:RN})].

The saddle-point evaluation of the NLL resummation formula (\ref{resum:unpol}) for 
the unpolarized case can be performed similarly, and 
the result is given by~\cite{KKT07}
\bea
&&
\tilde{X}^{\rm NLL} (0 , Q^2, y)
\non\\
&&\hspace{1cm}
=\left(\frac{b_0^2}{4Q^2 \beta_0 \alpha_s(Q^2)} 
\sqrt{\frac{2\pi}{{\zeta^{(0)}}''(\lambda_{SP})}} 
e^{-\zeta^{(0)}(\lambda_{SP})+h^{(1)}(\lambda_{SP})} \right) 
H \left(x_1^0,x_2^0; \frac{b_0^2}{b_{SP}^2}\right)  ,
\non\\\label{resumunpol:4} 
\eea
up to NLL accuracy, which corresponds to the above result (\ref{resum:4}) 
with the replacement 
$\delta H ( x_1^0, x_2^0;\ b_0^2 /b_{SP}^2 ) \rightarrow 
H ( x_1^0, x_2^0;\ b_0^2 /b_{SP}^2 )$. 
At the present accuracy, the gluon distribution decouples 
for $Q_T \approx 0$, since the corresponding coefficient function $C_{qg}^{(1)}$
in (\ref{eq:coeff:unpol}) is of order $\alpha_s(Q^2)$.   

We emphasize that the results (\ref{resum:4})-(\ref{resumunpol:4}) are
exact, up to the NNLL corrections corresponding to 
the ${\cal O}(\alpha_s (Q^2) )$ effects. 
We also see that these results give the explicit realization of
the ``degree 0'' approximation discussed in \S6.3 for polarized and 
unpolarized DY processes.
The common factor for both the polarized and unpolarized results, 
given by the contribution in the parentheses of (\ref{resum:4}) 
and (\ref{resumunpol:4}), represents ``very large perturbative effects'' 
due to the universal Sudakov factor.
Substituting (\ref{eq:lsp000}) and the running coupling constant 
at two-loop level [see (\ref{eq:run})] into this common factor, 
we obtain, to the present accuracy, 
\bea
&&\frac{b_0^2}{4Q^2 \beta_0 \alpha_s(Q^2)} \sqrt{\frac{2\pi}
{{\zeta^{(0)}}''(\lambda_{SP})}} 
e^{-\zeta^{(0)}(\lambda_{SP})+h^{(1)}(\lambda_{SP})} 
\nonumber\\
&& \;\;\;\;\;\;\;\;\;\;
=
\frac{b_0^2 \sqrt{\pi a}}{2\sqrt{2} (a+1)} \frac{
   \sqrt{
\ln\left(Q^2 /\Lambda_{\rm QCD}^2\right)
}}{\Lambda_{\rm QCD}^2}
\left(\frac{\Lambda_{\rm QCD}^2}{Q^2}\right)^{a \ln\left(1+1/a \right)} 
\nonumber\\
&&\hspace{3cm}
\times
\left[\ln \left(Q^2/\Lambda_{\rm QCD}^2\right)\right]^{(
   \beta_1/\beta_0^2)\left[1-a \ln \left(1+1/a\right)\right]}
e^{h^{(1)}(\lambda_{SP})} ,
\label{eq:asmpasymp}
\eea
with $a\equiv A_q^{(1)}/(2\pi \beta_0)\simeq 0.6$.
Here, the factors in the second line reproduce the well-known 
asymptotic behavior for $Q \gg \Lambda_{\rm QCD}$, 
derived from the conventional LL-level saddle-point calculation~\cite{PP} 
(see also Refs.~\citen{EV} and \citen{QZ}),
while those in the third line represent the NLL-level effects.
We note that the subleading, NLL contributions 
do not disappear in the asymptotic limit $Q\ra \infty$;
e.g., $e^{h^{(1)}(\lambda_{SP})}$ provides the ${\cal O}(1)$ constant factor 
[see (\ref{eq:h1})].
In fact, according to the discussion of the ``degree $N$'' approximation in \S6.3,
our results (\ref{resum:4})-(\ref{eq:asmpasymp}),
representing the degree 0 approximation, 
are exact in the asymptotic limit, while the ``degree $-1$'' approximation, 
which corresponds to the LL resummation, is not~\cite{CSS}.  

In our saddle-point formulae (\ref{resum:4}) and (\ref{resumunpol:4}), 
the ``new scale'' $b_0/b_{SP}$ participating in the parton distributions
is another NLL-level effect, which does not disappear in the asymptotic limit 
$Q\ra \infty$. In fact, for the polarized case (\ref{resum:4}), 
the corresponding contribution is associated with the ${\cal O}(1)$ terms, 
$R_{N_{1}}(\lambda_{SP})+R_{N_{2}} (\lambda_{SP})$, in the exponent of (\ref{eq:speval}).
Substituting (\ref{eq:lsp000}), we 
can express (\ref{eq:bspbsp}) as 
\be
\frac{b_0}{b_{SP}}= 
\Lambda_{\rm QCD} \left(\frac{Q}{\Lambda_{\rm QCD}
   }\right)^{\frac{a}{a+1}}
\left[\ln \left(Q^2/\Lambda_{\rm QCD}^2 \right)\right]^{-\frac{\beta_1}
{2 \beta_0^2 (a+1)}} ,
\label{eq:814}
\ee
where the first two factors on the RHS coincide with the corresponding 
LL-level expression~\cite{CSS,EV} of $b_0/b_{SP}$,
while the third one, involving the logarithms of $Q^2/\Lambda_{\rm QCD}^2$, 
represents the NLL-level modification induced by the two-loop running of 
the coupling constant in (\ref{eq:bspbsp}).
Because $a\simeq 0.6$, this implies $b_0/b_{SP} \sim
\Lambda_{\rm QCD}\left(Q/\Lambda_{\rm QCD}\right)^{a/(a+1)}$
$\simeq \sqrt{\Lambda_{\rm QCD}Q}$, modulo logarithms.
Therefore, our saddle-point formulae (\ref{resum:4}) and (\ref{resumunpol:4}) are 
directly applicable to the production of very high mass DY pairs 
such that $b_0/b_{SP} \gg \Lambda_{\rm QCD}$.
In fact, in such cases, the integral over $b$ in our resummation 
formula (\ref{eq:tilde-I_Mellin}) is actually dominated by the region of 
$b$ near $b_{SP}$ ($\ll 1/\Lambda_{\rm QCD})$), 
where the form of the integrand in (\ref{eq:tilde-I_Mellin}) is accurate 
[see the discussion below (\ref{eq:W2})].
Practically, however, the polarized DY experiments at RHIC, J-PARC, GSI, etc.,
will measure dileptons with masses $Q \sim$ several GeV, where $b_0 /b_{SP}$ may
be of order 1 GeV, and our formulae (\ref{resum:4}) and (\ref{resumunpol:4}) are
not useful for analyzing those cases.
Apparently, a similar problem arises at the LL-level~\cite{PP,CSS,QZ}, 
and the present result indicates that
the higher-order perturbative corrections at NLL level cannot solve this problem.

The root of this frustrating conclusion is in the fact that 
the integrand of our resummation formula
(\ref{eq:tilde-I_Mellin})
is composed of purely perturbative quantities and is inaccurate 
in the long-distance region (iii), $|b| \gtrsim \Lambda_{\rm QCD}$.
We have to introduce nonperturbative effects relevant to such a large $b$ region
in order to treat the small $Q_T$ behavior in the tDY
for the kinematics of the relevant experiments.
We discuss this point and the application of the results to
phenomenology in the next section.

\section{Predictions for the transversely polarized Drell-Yan process 
at RHIC and J-PARC in the NLL+LO $Q_T$-resummation framework}

On the basis of the QCD factorization and the transverse-momentum resummation framework,
we have obtained all the necessary analytic formulae of the higher-order 
perturbative corrections for the QCD prediction of the dilepton $Q_T$ 
spectrum and the spin asymmetry in the transversely polarized DY process. 
In this section, we specify the nonperturbative inputs
necessary for estimating them and calculate those observables
numerically with the kinematics of an ongoing experiment at RHIC 
and a possible future experiment at J-PARC.
A detailed numerical study has already been given in Ref.~\citen{KKT07}, and
here we highlight
the important results to show how the soft-gluon resummation contributions  
control the behavior of the $Q_T$ spectrum and the double
transverse-spin asymmetry, especially in the small $Q_T$ region. 
We also analyze the double-spin asymmetries, utilizing the saddle-point 
evaluation of the resummation formula.

\subsection{The $Q_T$-differential cross sections}

First, we calculate the $Q_T$-differential cross sections (\ref{eq:NLL+LO}) and 
(\ref{eq:NLL+LO:unp}) for polarized and unpolarized Drell-Yan processes, respectively,
with the kinematics of $pp$ collisions at RHIC.
For simplicity, here and below we denote the parton distributions of 
the proton as [see (\ref{tPDF}), (\ref{pdf}), (\ref{eq:739})-(\ref{eq:742})]
\be
\delta q(x,\mu^2)\equiv \delta q_{h=p}(x,\mu^2), \;\;\;\;
q(x,\mu^2)\equiv q_{h=p}(x,\mu^2), \;\;\;\; g(x,\mu^2)\equiv g_{h=p}(x,\mu^2), 
\label{eq:qqgqqg}
\ee
omitting the subscript, and also we denote the combination of the azimuthal angles
appearing in (\ref{eq:NLL+LO}) as $2\phi_{k_1}-\phi_1 -\phi_2 \equiv 2 \phi$.

To calculate the unpolarized cross section (\ref{eq:NLL+LO:unp}),
we need nonperturbative inputs for the unpolarized quark and gluon 
distributions, $q(x,\mu^2)$ and $g(x,\mu^2)$; 
we use the NLO GRV98 distributions for them~\cite{GRV98}.
To calculate the polarized cross section (\ref{eq:NLL+LO}),
we need nonperturbative inputs for the transversity $\delta q(x,\mu^2)$.
Recently, Anselmino et al.~\cite{Anselmino} gave the first estimation 
of the transversity distributions from semi-inclusive DIS (SIDIS) data, 
combined with $e^+e^-$ data for the associated (Collins) fragmentation function. 
They performed a global fit of the SIDIS spin-asymmetry
data with the corresponding LO QCD formula,
assuming that the antiquark distributions vanish, i.e. $\delta \bar{q}(x, \mu^2)=0$,
and obtained the LO quark transversity distribution.
Thus, their results cannot be used for the present calculation of tDY 
at the NLL+LO accuracy, in which the {\it NLO} transversity
distributions for {\it both the quark and antiquark} are necessary.
At the moment, no other experimental information on the transversity 
exists\footnote{For a recent result of a lattice QCD simulation of 
the first and second moments of the $u$- and $d$-quark transversity distributions,
see Ref.~\citen{QCDSF}. 
The transversity distributions  fitted to the prediction of 
the chiral quark soliton model of nucleons are given in Ref.~\citen{Wakamatsu}.}  
Theoretically, a nontrivial relation among the transversity $\delta q(x)$,
unpolarized $q(x)$, and longitudinally polarized $\Delta q(x)$ distributions
is provided by the Soffer inequality~\cite{S},
\bea
2 \, |\delta q(x)| \leq q (x) + \Delta q(x) ,
\label{eq:Soffer}
\eea
for each flavor of quark and antiquark.
In this paper, we use a model of the NLO transversity distributions 
which satisfy the Soffer inequality, following Refs.\citen{MSSV} and \citen{MSV}.
In this model, we saturate the Soffer bound of (\ref{eq:Soffer}) at 
the low input scale $\mu_0 \simeq 0.6$ GeV,
as $\delta q(x , \mu_0^2 ) = [q (x , \mu_0^2 ) + \Delta q(x , \mu_0^2 )]/2$.
Here, we cannot determine the signs of the distributions from (\ref{eq:Soffer}), 
and in this calculation, we have chosen all the signs to be positive.
For the input functions on the RHS of this formula, $q(x, \mu_0^2)$ and 
$\Delta q(x, \mu_0^2)$,  we use the NLO GRV98\cite{GRV98} and 
GRSV2000 (``standard scenario'')\cite{GRSV00} distributions, respectively.
Then we evolve $\delta q(x , \mu_0^2 )$ to higher scales, according to 
the NLO DGLAP kernel for the transversity as (\ref{DGLAP}).

We also introduce another nonperturbative contribution to the $Q_T$-differential
cross sections. As mentioned in \S8, in the kinematical regions 
for most experiments, including those at RHIC and J-PARC, 
the small $Q_T$ behavior of our resummation formula (\ref{eq:tilde-I_Mellin}) 
is dominated by the contribution from $|b|\grtsim 1/\Lambda_{\rm QCD}$ 
in the $b$ integral, which corresponds to the region (iii) of \S6.3.
However, the integrand of (\ref{eq:tilde-I_Mellin}) involving
purely perturbative quantities is not accurate for such a large $|b|$ region, 
and the corresponding long-distance behavior has to be complemented 
by relevant nonperturbative effects.
Formally, those nonperturbative effects play a role of compensating for the ambiguity 
that the prescription for the $b$ integration in (\ref{eq:tilde-I_Mellin}) 
needed to avoid the singularity in the Sudakov exponent $S(b, Q)$ 
of (\ref{sudakov:2}) is not unique, such as the $b_*$ prescription of 
(\ref{replace}), (\ref{eq:blim}),
and the present prescription with (\ref{resum:2}).
Therefore, following Refs.~\citen{CSS,KLSV} and \citen{BCFG}, 
in (\ref{eq:tilde-I_Mellin}) we make the replacement
\begin{equation}
e^{S (b , Q)}\rightarrow e^{S (b , Q)-  g_{NP} b^2} ,
\label{eq:np}
\end{equation}
with a nonperturbative parameter $g_{NP}$.
Because exactly the same Sudakov factor, $e^{S(b, Q)}$, participates 
in the corresponding formula for the unpolarized case, (\ref{resum:unpol}), 
as noted in \S7.2, we perform the replacement (\ref{eq:np}) with 
the same nonperturbative parameter $g_{NP}$ in the NLL+LO unpolarized 
differential cross section.
This may be interpreted as assuming the same ``intrinsic transverse momentum'' 
of partons inside nucleons for both polarized and unpolarized cases, 
corresponding to the Gaussian smearing factor of (\ref{eq:np}).
For the present calculations, we use $g_{NP} = 0.5$ GeV$^2$, which is
consistent with a study of the $Q_T$ spectrum in the unpolarized case \cite{KS}.

Assuming $e^{-g_{NP} b^2} \rightarrow 1$ in customary perturbation theory,
it is straightforward to show that all results derived in \S7 are unchanged 
under the replacement (\ref{eq:np}).
In particular, the NLL+LO $Q_T$-differential cross sections (\ref{eq:NLL+LO}) and 
(\ref{eq:NLL+LO:unp}) with (\ref{eq:np}) satisfy the unitarity condition, 
given in (\ref{eq:unitarity}) and (\ref{eq:unitarity2}), exactly.

For all the following numerical calculations, we choose $\phi=0$ for 
the azimuthal angle of one of the leptons, $\mu_F=\mu_R=Q$ for the
factorization and renormalization scales, and $b_c=0$ and
$\theta=\frac{7}{32}\pi$ for the integration contour 
${\cal C}_{\pm}$ defined below (\ref{resum:2}).

\begin{figure}
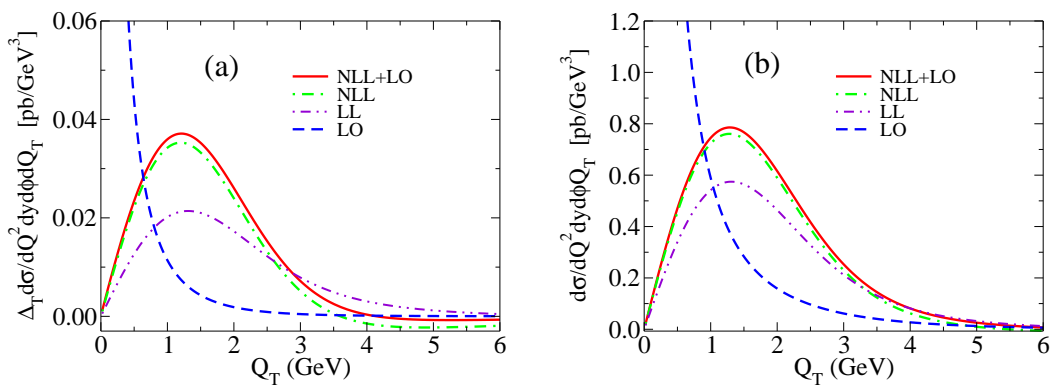

\vspace*{0.5cm}
\bc
\includegraphics[height=5cm]{RHIC_200_5_y2_pol_2.eps}~~~~
\includegraphics[height=5cm]{RHIC_200_5_y2_unpol_2.eps}
\ec
\caption{The spin-dependent and spin-averaged differential 
cross sections for tDY: (a) $\Delta_Td\sigma/dQ^2 dQ_T dy d\phi$ 
and 
(b) $d\sigma/dQ^2 dQ_T dyd\phi$, as functions of $Q_T$ 
with RHIC kinematics, $\sqrt{S}=200$ GeV, $Q=5$ GeV, $y=2$ and $\phi=0$,
with $g_{NP}=0.5$ GeV$^2$. }
\label{fig:2}
\end{figure}

The tDY $Q_T$-differential cross sections for the spin-dependent and spin-averaged cases
are shown in Fig.~\ref{fig:2}(a) and  Fig.~\ref{fig:2}(b), respectively, 
for $\sqrt{S}=200$ GeV, $Q= 5$ GeV and $y = 2$, 
corresponding to the detection of dileptons with the PHENIX detector at RHIC.
The solid curve in Fig.~\ref{fig:2}(a) shows the NLL+LO differential cross section 
(\ref{eq:NLL+LO}), multiplied by $2Q_T$.
The dot-dashed curve represents the contribution from the NLL resummed component 
$\Delta_T\tilde{X}^{\rm NLL}$ 
in (\ref{eq:NLL+LO}).\footnote{In principle, we should take 
the ``degree 0'' approximation for the numerical 
calculations when $Q_T \approx 0$, as discussed in \S6 and \S8. 
But we do not make the corresponding replacement for the coefficient
functions, $C_{ji}(z, \alpha_s) \rightarrow \delta_{ji}\delta(1-z)$, 
for $Q_T \approx 0$ in our numerical calculations,
because the effect of the replacement is invisible in 
the solid as well as dot-dashed curve for $Q_T \approx 0$ in 
Fig.~\ref{fig:2}(a), and also in Fig.~\ref{fig:2}(b) below.
For the spin asymmetry, which is defined as the ratio of the cross
sections in Figs.~\ref{fig:2}(a) and (b) and is discussed in \S9.2,
the replacement would increase the NLL+LO asymmetry (\ref{asym:NLL+LO}) 
and the NLL asymmetry (\ref{asym:NLL}) for $Q_T \approx 0$ 
in Figs.\ref{fig:3}-\ref{fig:5} by about 5\%.} 
The LO cross section (\ref{cross section2}) is depicted by the dashed curve,
which becomes large and diverges as $Q_T\ra 0$, due to the singular terms
$\alpha_s\ln(Q^2/Q_T^2)/Q_T^2$ and $\alpha_s/Q_T^2$ in $\Delta_TX^{(1)}$
of (\ref{eq:X}).    
This singular behavior disappears in the NLL resummed component 
$\Delta_T\tilde{X}^{\rm NLL}$
when we sum up the recoil logarithms to all orders in $\alpha_s$.
The NLL resummed component dominates the NLL+LO cross section in
the peak region of the solid curve, i.e., at intermediate 
as well as small values of $Q_T$.
As a result, the NLL+LO cross section (\ref{eq:NLL+LO}) is finite and
well behaved over all regions of $Q_T$.  
We also show the LL result, represented by the two-dot-dashed curve, 
obtained from the corresponding NLL result by omitting the contributions 
from the NLL terms, i.e., 
$h^{(1)}(\lambda)$, $R_{N_1}(\lambda)$, $R_{N_2}(\lambda)$ in 
(\ref{eq:tilde-I_Mellin}) and $\alpha_s(Q^2)C_F(\pi^2-8)/2\pi$ in (\ref{tilde-X_Mellin}).
The LL contributions are sufficient to obtain a finite cross section, 
suppressing the cross sections for $Q_T\approx 0$.

The curves in Fig.~\ref{fig:2}(b) are the same as those in Fig.~\ref{fig:2}(a), 
but for the unpolarized cross sections; e.g., the solid curve is obtained by
calculating (\ref{eq:NLL+LO:unp}), and the dot-dashed curve is the contribution from
the NLL resummed component $\tilde{X}^{\rm NLL}$ in (\ref{eq:NLL+LO:unp}). 
The behavior of each curve is similar to that in the polarized case.

Comparing the NLL results with the corresponding LL results in
Figs.~\ref{fig:2}(a) and (b), we find that the contributions from 
the NLL terms considerably enhance the cross sections in both polarized 
and unpolarized cases, while the enhancement is larger for the former case:
The ``universal'' term $h^{(1)}(\lambda)$ gives a similar (enhancement) 
effect for both channels, while the other NLL contributions associated 
with the evolution operators and the ${\cal O}(\alpha_s)$ terms 
in the coefficient functions give different effects in the polarized 
and unpolarized cases 
[see (\ref{eq:RN}), (\ref{eq:coeff.1}) and (\ref{eq:coeff:unpol})].

\begin{figure}
\vspace{0.5cm}
\bc
\includegraphics[height=5cm]{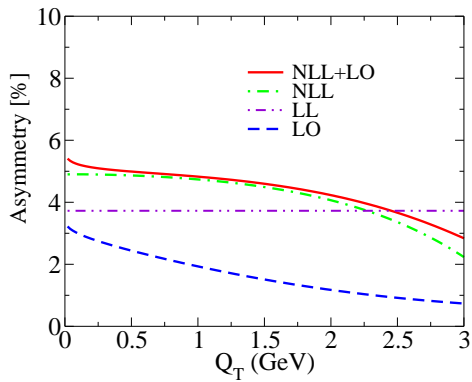}
\ec
\caption{The asymmetries for RHIC kinematics,
$\sqrt{S}=200$ GeV, $Q=5$ GeV, $y=2$ and $\phi=0$, 
obtained from each curve in Fig.~\ref{fig:2}.}
\label{fig:3}
\end{figure}

\subsection{The double transverse-spin asymmetries}

We consider the double transverse-spin asymmetry, 
defined by the ratio of the NLL+LO cross sections (\ref{eq:NLL+LO}) 
and (\ref{eq:NLL+LO:unp}),
\bea
\aqtnn =\frac{1}{2} \cos(2\phi) 
\frac{\Delta_T\tilde{X}^{\rm NLL}(Q_T^2 , Q^2, y)
+\Delta_T\tilde{Y}(Q_T^2, Q^2, y)}
{\tilde{X}^{\rm NLL}(Q_T^2, Q^2, y)+\tilde{Y}(Q_T^2, Q^2, y)} ,
\label{asym:NLL+LO}
\eea
as a function of $Q_T$. Figure~\ref{fig:3} plots the asymmetries for the tDY
at RHIC, given by the ratio of the value of each line in Fig.~\ref{fig:2}(a) 
to that of the corresponding line in Fig.~\ref{fig:2}(b);
that is, the solid curve gives the NLL+LO asymmetry $\aqtnn$ of (\ref{asym:NLL+LO}), 
and the dot-dashed curve gives the NLL result, defined as
\bea
\aqtn^{\rm NLL} (Q_T) = \frac{1}{2}\cos(2\phi)
\frac{\Delta_T \tilde{X}^{\rm NLL} (Q_T^2 , Q^2 ,y)}{\tilde{X}^{\rm NLL} 
(Q_T^2 , Q^2 ,y)} .
\label{asym:NLL}
\eea 
The NLL+LO asymmetry $\aqtnn$ in Fig.~\ref{fig:3} is almost flat in the
peak region of the NLL+LO cross sections in Fig.~\ref{fig:2}. 
This flat behavior is dominated by the NLL resummed components,
and it reflects the fact that the soft gluon effects resummed into the
Sudakov factor $e^{S(b,Q)}$ with (\ref{sudakov:2}) are universal to 
the NLL accuracy between the numerator and the denominator of 
$\aqtn^{\rm NLL}(Q_T)$ of (\ref{asym:NLL}).\footnote{
We note that the shape of the $Q_T$ spectra in Figs.~\ref{fig:2}(a) and (b) is actually 
sensitive to the value of the nonperturbative parameter
$g_{NP}$ of (\ref{eq:np}) (see Ref.~\citen{KKST06}). 
But the corresponding $g_{NP}$ dependence almost cancels between 
the numerator and denominator in the asymmetries of (\ref{asym:NLL+LO}) 
and (\ref{asym:NLL}) in the range $g_{NP}=0.3$-$0.8$ GeV$^2$~\cite{KKT07}. }
A slight increase of $\aqtnn$ as $Q_T\ra 0$ is due to the terms $\propto\ln(Q^2/Q_T^2)$
in the ``regular components'' $\Delta_T\tilde{Y}$ and $\tilde{Y}$ in (\ref{asym:NLL+LO}).
The dashed curve in Fig.~\ref{fig:3} shows the LO asymmetry $\aqtn^{\rm LO}(Q_T)$,
defined by the ratio of (\ref{cross section2}) to 
the corresponding LO cross section for unpolarized DY process,  
\bea
\aqtn^{\rm LO}(Q_T)=\frac{1}{2}\cos(2\phi)
\frac{\Delta_TX^{(1)}(Q_T^2,Q^2,y)|_{Q_T^2>0}+\Delta_TY(Q_T^2,Q^2,y)}
{X^{(1)}(Q_T^2,Q^2,y)|_{Q_T^2>0}+Y(Q_T^2,Q^2,y)} \ ,
\label{asym:LO}
\eea
where $X^{(1)}$ denotes the singular part of the ${\cal O}(\alpha_s)$ 
unpolarized cross section~\cite{AEGM},
defined similarly to $\Delta_T{X}^{(1)}$ in (\ref{eq:X}).
Although both the numerator and denominator of (\ref{asym:LO}) diverge as 
$Q_T \rightarrow 0$, as shown in Figs.~\ref{fig:2}(a) and (b), 
$\aqtn^{\rm LO}(Q_T)$, their ratio, appears to be finite. However,
this LO result decreases as $Q_T$ increases, and it is much smaller than 
the NLL+LO result, indicating that the soft gluon resummation is crucial
for the prediction of the asymmetries at small $Q_T$ and intermediate $Q_T$. 
The two-dot-dashed curve in Fig.~\ref{fig:3} shows the LL result
[see (\ref{tilde-X_Mellin}), (\ref{eq:tilde-I_Mellin}), (\ref{resum:unpol})],
\bea
\aqtn^{\rm LL}(Q_T)= \frac{1}{2}\cos(\phi)
\frac{\delta H(x_1^0,x_2^0;Q^2)}{H(x_1^0,x_2^0;Q^2)}  ,
\label{asym:LL}
\eea
which is constant in $Q_T$, because the soft-gluon corrections in the
numerator and denominator exactly cancel at the LL level [see also (\ref{eq:resLL2})]. 
As expected from Fig.~\ref{fig:2}, $\aqtn^{\rm NLL}(Q_T)$ is larger 
than $\aqtn^{\rm LL}(Q_T)$, due to the effects of the NLL contributions. 
In particular, among the NLL-level effects noted at the end of \S9.1,
the contributions associated with the evolution operators play an important 
role to amplify $\aqtn^{\rm NLL}(Q_T)$ in comparison with $\aqtn^{\rm LL}(Q_T)$:
In the resummation formula (\ref{eq:tilde-I_Mellin}), the large
logarithmic contributions due to the Sudakov factor in (\ref{sudakov:2})
and (\ref{sudakov:1}) strongly enhance the contributions from the region $b\sim 1/Q_T$ 
in the $b$ integration, where the evolution operator (\ref{eq:RN}) 
causes the scale of the parton distributions to be $b_0/b\sim Q_T$, instead of $Q$.
Therefore, the numerator of (\ref{asym:NLL}) is dominated by the transversity
distributions at the scale $\sim Q_T$. A similar mechanism arises
in the denominator, given by (\ref{resum:unpol}), so that, after 
the cancellation of the universal Sudakov factor between the numerator and denominator,
the dominant contribution to (\ref{asym:NLL}) is given by
$\aqtn^{\rm NLL}(Q_T) \sim \left[\cos(2\phi)/2\right]
\delta H(x_1^0,x_2^0;Q_T^2)/H(x_1^0,x_2^0;Q_T^2)$. 
Note that the transversity distributions in (\ref{tPDF})
obey QCD evolution of the non-singlet type, described by (\ref{DGLAP}).
Thus, as the scale $\mu$ increases, the antiquark transversity distribution 
$\delta \bar{q}(x, \mu^2)$ in the small $x$ region grows less rapidly 
than the unpolarized sea distribution $\bar{q}(x,\mu^2)$ in (\ref{eq:739}). 
The latter grows rapidly for small $x$ through mixing with the gluon distribution.
Therefore, the ratio of the sea distributions 
$\delta\bar{q}(x,\mu^2)/\bar{q}(x,\mu^2)$ at small $x$
becomes enhanced as the scale $\mu$ is reduced from $Q$ to $Q_T$.  
This explains why the NLL asymmetry $\aqtn^{\rm NLL}(Q_T) \sim \left[\cos(2\phi)/2\right]
\delta H(x_1^0,x_2^0;Q_T^2)/H(x_1^0,x_2^0;Q_T^2)$
is enhanced in comparison with the LL result $\aqtn^{\rm LL}(Q_T)$ 
appearing in (\ref{asym:LL}) in the present case, with $x_{1}^0=0.185$
and $x_{2}^0=0.003$ in (\ref{eq:hadronvar}) [see also the discussion in \S9.3 
using the saddle-point formula (\ref{eq:attnll}) below].

Now we compare these asymmetries in Fig.~\ref{fig:3} with the asymmetry for 
the $Q_T$-integrated cross sections $A_{TT}$, defined as
\bea
A_{TT}&&\equiv\frac{\int dQ_T^2 \left[ 
\Delta_T d\sigma^{\rm NLL+LO}/dQ^2dQ_T^2dyd\phi \right]}
{\int dQ_T^2 \left[ d\sigma^{\rm NLL+LO} /dQ^2dQ_T^2dyd\phi \right]}
\nonumber\\
&&=\frac{\left[\Delta_T d\sigma /dQ^2 dyd\phi\right]}
{\left[d\sigma  /dQ^2 dyd\phi\right]}
=\frac{1}{2}\cos(2 \phi) \frac{\delta H(x_1^0, x_2^0; Q^2)+\cdots}
{H(x_1^0, x_2^0; Q^2)+\cdots} ,
\label{eq:att}
\eea
where we have used the fact that our NLL+LO $Q_T$-differential cross sections
satisfy the unitarity conditions (\ref{eq:unitarity}) and (\ref{eq:unitarity2}) 
exactly, and the ellipses on the RHS stand for the NLO [${\cal O} (\alpha_s )$] 
correction terms in the $\overline{\rm MS}$ scheme. 
Therefore, $A_{TT}$ defined in this way coincides with the NLO asymmetry
calculated in Ref.~\citen{MSSV}.\footnote{
In Ref.~\citen{MSSV},
the corresponding asymmetry is defined through a certain integration over $\phi$,
and it is equal to (\ref{eq:att}) with the formal replacement 
$\cos(2\phi) \rightarrow 2/\pi$.
Also, the input parton distributions in Ref.~\citen{MSSV} are not completely 
the same as ours (see the discussion in Ref.~\citen{KKT07}).}
Using the present nonperturbative inputs, we get $A_{TT}= 4.0\%$. 
This is close to the value of $\aqtn^{\rm LL}(Q_T)$ and is smaller 
than that of $\aqtnn$ in the flat region, 
which is reasonable, because $A_{TT}$ in (\ref{eq:att}) coincides with 
$\aqtn^{\rm LL}(Q_T)$ in (\ref{asym:LL}), up to NLO QCD corrections.

\begin{figure}
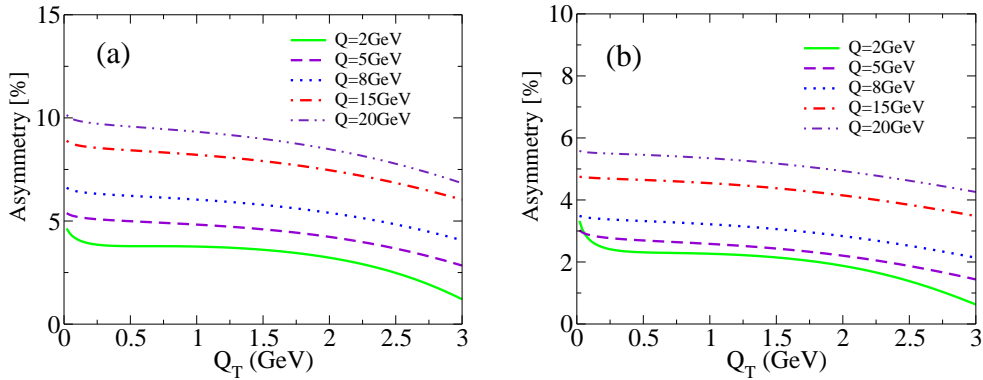

\vspace*{0.5cm}
\bc
\includegraphics[height=5cm]{RHIC_200_y2_asym.eps}~~~~
\includegraphics[height=5cm]{RHIC_500_y2_asym.eps}
\ec
\caption{The NLL+LO $\aqtnn$ 
in 
(\ref{asym:NLL+LO}) with (\ref{eq:np}) 
using $g_{NP}=0.5$GeV$^2$ with RHIC kinematics: 
(a) $\sqrt{S}=200$ GeV, $y=2$ and $\phi=0$; 
(b) $\sqrt{S}=500$ GeV, $y=2$ and $\phi=0$.}
\label{fig:4}
\end{figure} 

Figures~\ref{fig:4}(a) and (b) show the NLL+LO asymmetries $\aqtnn$ 
with RHIC kinematics, $\sqrt{S}=200$ GeV and $\sqrt{S}=500$ GeV, respectively, 
with various values of $Q$ and $y=2$.
The dashed curve in Fig.~\ref{fig:4}(a) is the same as the solid curve 
in Fig.~\ref{fig:3}, and similarly to this dashed curve, all other curves 
in Figs.~\ref{fig:4}(a) and (b) exhibit the characteristic flat behavior 
in the small $Q_T$ region.
We mention that the dependence of $\aqtnn$ on $Q$ comes from 
the small-$x$ behavior of the relevant parton distributions:
A smaller $Q$ corresponds to a smaller $x_{1,2}^0=e^{\pm y}Q/\sqrt{S}$, so
that the small-$x$ rise of the unpolarized sea distributions enhances 
the denominator of (\ref{asym:NLL+LO}).
A similar mechanism also explains why the results for $\sqrt{S}=500$ GeV are smaller than
those for $\sqrt{S}=200$ GeV.
Comparing with the NLO $A_{TT}$ in (\ref{eq:att}), we find~\cite{KKT07} that
the values of $\aqtnn$ in the flat region of Figs.~\ref{fig:4}(a) and (b) 
are larger by 20-30\% than the corresponding values of $A_{TT}$.

\begin{figure}
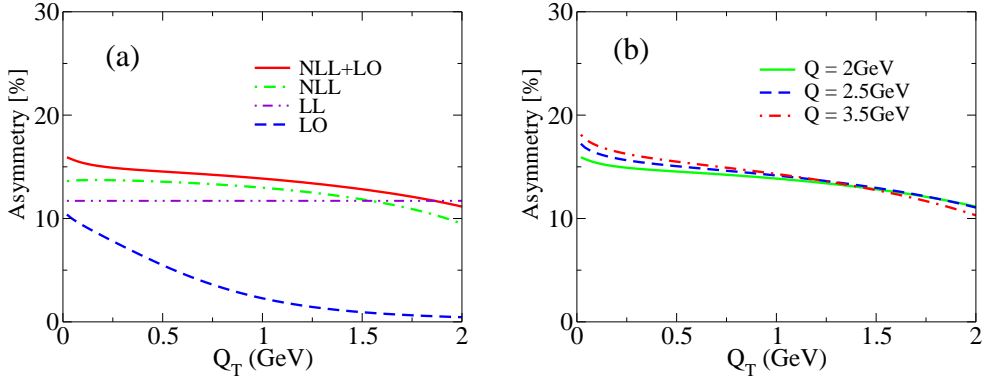

\vspace{0.5cm}
\bc
\includegraphics[height=5cm]{J-PARC_10_2_y0_asym.eps}~~~~
\includegraphics[height=5cm]{J-PARC_10_y0_asym.eps}
\ec
\caption{The asymmetries with J-PARC kinematics.
(a) The asymmetries obtained similarly to Fig.~\ref{fig:3}(a), with
    $\sqrt{S}=10$ GeV, $Q=2$ GeV, $y=0$ and $\phi=0$.
(b) The NLL+LO $\aqtnn$ with $\sqrt{S}=10$ GeV, $y=0$ and $\phi=0$.}
\label{fig:5}
\end{figure}

Next, we present the results for J-PARC kinematics, 
i.e., the results for the tDY
in $pp$ collisions at the moderate CM energy $\sqrt{S}=10$ GeV. 
Figure~\ref{fig:5}(a) shows the double transverse-spin asymmetries for
$\sqrt{S}=10$ GeV, $Q=2$ GeV and $y=0$, 
and the curves have the same meanings as the corresponding curves 
in Fig.~\ref{fig:3}.
The results for J-PARC have features similar to those in Fig.~\ref{fig:3} for RHIC: 
The NLL+LO asymmetry $\aqtnn$ is flat in the small $Q_T$ region, and it is dominated by 
the NLL asymmetry $\aqtn^{\rm NLL}(Q_T)$;
the NLL asymmetry is enhanced in comparison with the LL result,
and the fixed-order LO result is far below the others. 
We note that $\aqtnn\simeq 15\%$ in the flat region is again larger than 
the corresponding prediction using (\ref{eq:att}), $A_{TT}=12.8\%$.  
Figure~\ref{fig:5}(b) shows the NLL+LO asymmetries $\aqtnn$ with J-PARC kinematics, 
$\sqrt{S}=10$ GeV, $Q=2,~2.5,~3.5$ GeV, and $y=0$, 
and the solid curve is the same as the solid curve in Fig.~\ref{fig:5}(a).
We observe that the asymmetries $\aqtnn$ are flat and are approximately 15\%
for all $Q$; we note that this value of $\aqtnn$ is larger by 20-30\% than
the corresponding values of $A_{TT}$ in (\ref{eq:att})~\cite{KKT07}. 
Moreover, the asymmetries found at J-PARC are in general larger than
those at RHIC energies, because the parton distributions for larger
$x_{1,2}^0\grtsim 0.2$ 
are probed, where the enhancement mechanism of the denominator of (\ref{asym:NLL+LO})
by the unpolarized sea distributions, as noted in Figs.~\ref{fig:4}(a) and (b), 
is absent.\footnote{It is expected that even larger asymmetries 
will be observed in future $p\bar{p}$ experiments at GSI~\cite{SSVY:05,BCCGR:06}. 
The application of our $Q_T$ resummation formalism to $p\bar{p}$ collisions 
will be presented elsewhere.~\cite{KKT07-2} } 

Finally, we emphasize that in all cases considered in Figs.~\ref{fig:3}-\ref{fig:5}, 
we have observed that $\aqtnn$ in (\ref{asym:NLL+LO}) in the flat region 
is considerably larger than the corresponding $A_{TT}$ in (\ref{eq:att}).
Indeed, this is generally true for the tDY
in $pp$ collisions, regardless of the specific kinematics or the detailed behavior of
nonperturbative inputs, 
because this phenomenon is mainly governed by the partonic mechanism associated 
with the soft gluon resummation at the NLL level, 
as demonstrated in Fig.~\ref{fig:2}.
By contrast, apparently the absolute magnitudes of both $\aqtnn$ and $A_{TT}$
are influenced by the detailed behavior of the input parton distributions,
in particular, at RHIC energies, by their small-$x$ behavior.

\subsection{The saddle-point formula incorporating nonperturbative smearing}

In the numerical results presented in \S9.2,
we have observed the universal flat behavior of $\aqtnn$ at small $Q_T$,
where the bulk of the DY pairs is produced. 
We have also observed that in this flat region, the numerator and 
denominator of (\ref{asym:NLL+LO}) are dominated by the NLL resummed 
components, $\Delta_T\tilde{X}^{\rm NLL}$ and $\tilde{X}^{\rm NLL}$, respectively, 
and as a result, $\aqtnn$ is well approximated by $\aqtn^{\rm NLL}(Q_T)$. 
This suggests that the experimental value of the double transverse-spin asymmetry
obtained from the data in the flat region can
be compared with the theoretical value of (\ref{asym:NLL}) directly.
Noting that the flat behavior of the NLL asymmetry (\ref{asym:NLL}) 
can be extrapolated to $Q_T\ra 0$, as demonstrated in Figs.~\ref{fig:3} 
and \ref{fig:5}(a), we can use the relations
\bea 
\aqt\approx\aqtn^{\rm NLL}(Q_T)\approx\aqtn^{\rm NLL}(0) = \frac{1}{2}\cos(2\phi)
\frac{\Delta_T \tilde{X}^{\rm NLL} (0 , Q^2 ,y)}{\tilde{X}^{\rm NLL} 
(0 , Q^2 ,y)}  ,
\label{approx}
\eea
for $Q_T$ in the flat region, where $\aqt$ 
denotes the experimental value of the double transverse-spin asymmetry.
In \S8, the $Q_T\rightarrow 0$ limit of the NLL resummed components, 
$\Delta_T\tilde{X}^{\rm NLL} (0 , Q^2 ,y)$ and 
$\tilde{X}^{\rm NLL} (0 , Q^2 ,y)$, has been evaluated analytically 
using the saddle-point method for the case without the replacement (\ref{eq:np}).
In view of (\ref{approx}), it would be interesting to extend the results in \S8,
incorporating the nonperturbative smearing factor as in (\ref{eq:np})~\cite{KKT07}.

Starting from (\ref{eq:tilde-I_Mellin}) with the replacement (\ref{eq:np}), 
we can use logic similar to that used 
in \S8 in order to obtain (\ref{eq:tilde-I_Mellin2}), which now reads, at $Q_T=0$,
\be
\tilde{I}_{N_1,N_2}(0,Q^2)= \frac{1}{2} \int_0^{b_c} db b
\exp\left[\frac{h^{(0)}(\lambda)}{\alpha_s (Q^2)}
+h^{(1)}(\lambda) +R_{N_1}(\lambda)+R_{N_2}(\lambda)-g_{NP}b^2 \right].
\label{eq:tilde-I_Mellin3}
\ee
Because $b^2 = (b_0^2/Q^2)e^{\lambda/(\beta_0 \alpha_s (Q^2 ))}$ using (\ref{eq:lambda}),
the nonperturbative function produces a ``double exponential'' behavior, 
and the saddle-point evaluation with such an integrand is a nontrivial task.
Nevertheless, 
we can check numerically that the integrand of (\ref{eq:tilde-I_Mellin3}) has 
a nice saddle point well below $b_c$ (above $0$) for the kinematics of interest.
When $g_{NP} \rightarrow 0$, the integrand is controlled by the perturbative functions 
in (\ref{eq:tilde-I_Mellin3}), which was also the case in \S8. 
Contrastingly, when $g_{NP}\rightarrow \infty$, 
the integrand is controlled solely by the nonperturbative function, such that
the integral of (\ref{eq:tilde-I_Mellin3}) is dominated by the $b \approx 0$ region.
Our task is to find a formula for finite $g_{NP}$ which interpolates
between these two extreme cases, and, for this purpose, 
we perform the saddle-point evaluation of (\ref{eq:tilde-I_Mellin3}), 
assuming $- g_{NP} b^2=- g_{NP} (b_0^2/Q^2)e^{\lambda/(\beta_0 \alpha_s (Q^2 ))}$
as a leading term in the exponent.
As a result, we obtain (\ref{eq:sp0}), where $\zeta^{(0)}(\lambda)$ is
now replaced as
\be
\zeta^{(0)}(\lambda) \rightarrow  - \frac{\lambda}{\beta_0 \alpha_s(Q^2)}
-\frac{h^{(0)}(\lambda)}{\alpha_s (Q^2)} 
+ \frac{g_{NP}b_0^2}{Q^2}e^{\frac{\lambda}{\beta_0 \alpha_s (Q^2 )}} ,
\label{eq:fxi}
\ee
and correspondingly the equation (\ref{eq:lsp0}) to determine 
the saddle point is replaced by
\begin{equation}
1-\frac{A_q^{(1)}}{2\pi\beta_0}\frac{\lambda_{SP}}{1-\lambda_{SP}}=
\frac{g_{NP}b_0^2}{Q^2}e^{\frac{\lambda_{SP}}{\beta_0 \alpha_s (Q^2 )}} .
\label{eq:lsp}
\end{equation}

An important observation is that the ratio
$[ h^{(1)}(\lambda) + R_{N_1}(\lambda) + R_{N_2}
(\lambda)]/\zeta^{(0)}(\lambda)$ with (\ref{eq:fxi}) 
actually behaves as a quantity of order $\alpha_s(Q^2 )$ 
in the relevant region, $\lambda \approx \lambda_{SP}$, of the integration 
in (\ref{eq:sp0}), even for nonzero $g_{NP} \simeq 0.5$ GeV$^2$.
This fact supports the validity of the present approach, 
incorporating the nonperturbative function as (\ref{eq:fxi}). 
Also, this fact suggests that we may use an order counting that is
similar to that explained below (\ref{eq:speval2speval2})
for the contributions arising in the saddle-point evaluation. 
This is apparently correct when $g_{NP}$ is sufficiently small,
but in general it has to be checked. Here we assume the same order counting as in \S8,
and check the accuracy of the resulting formula below, comparing it 
with the result of the numerical $b$-integration of the original formula, 
(\ref{eq:tilde-I_Mellin}) with (\ref{eq:np}) (see Table~\ref{tab:3} below). 

Now we can perform the saddle-point evaluation of (\ref{eq:sp0}) at NLL accuracy
and also a similar calculation starting from (\ref{resum:unpol}) 
for the unpolarized case.
Immediately we find that the results (\ref{eq:speval})-(\ref{resumunpol:4})
with (\ref{eq:fxi}) and (\ref{eq:lsp}) hold, up to the NNLL corrections
corresponding to the ${\cal O}(\alpha_s)$ effects~\cite{KKT07}.
These results realize the ``degree 0'' approximation of \S6.3. 
Also, it is straightforward to show that these results 
reproduce the results in the two extreme limits $g_{NP}\rightarrow 0$ 
and $g_{NP}\rightarrow \infty$, as desired for the ``interpolation formula''
mentioned below (\ref{eq:tilde-I_Mellin3}).
Taking the ratio of the saddle point formulae (\ref{resum:4}) and (\ref{resumunpol:4}),
we obtain~\cite{KKT07}, for the RHS of (\ref{approx}),
\begin{equation}
\aqtn^{\rm NLL}(0)= \frac{1}{2}\cos(2\phi)
\frac{\Delta_T \tilde{X}^{\rm NLL} (0 , Q^2 ,y)}{\tilde{X}^{\rm NLL} 
(0 , Q^2 ,y)}  =\frac{1}{2}\cos(2 \phi ) 
\frac{\delta H \left( x_1^0, x_2^0;\ b_0^2 /b_{SP}^2 \right)}
{H \left( x_1^0, x_2^0;\ b_0^2 /b_{SP}^2 \right)},
\label{eq:attnll}
\end{equation}
up to NLL accuracy.
Here, the common factor in (\ref{resum:4}) and (\ref{resumunpol:4}),
associated with the ``very large perturbative effects'' due to the Sudakov factor,
as well as with the Gaussian smearing factor involving $g_{NP}$,
has canceled out.

The new scale, $b_0/b_{SP}$, must be determined by solving (\ref{eq:lsp})
numerically, substituting $A^{(1)}_q=2C_F$ from (\ref{coeff_NLL}) and 
the input values for $Q$ and $g_{NP}$, but it is useful to consider 
its general behavior: 
The LHS of (\ref{eq:lsp}) equals 1 at $\lambda_{SP}=0$, decreases as a concave
function as $\lambda_{SP}$ increases, and vanishes at 
$\lambda_{SP}=1/[1+A_q^{(1)}/(2 \pi \beta_0 )] \cong 0.6$, 
which corresponds to the solution (\ref{eq:lsp000}) in the $g_{NP}=0$ case, 
while the RHS is in general much smaller than 1 at $\lambda_{SP}=0$, 
increases as a convex function as $\lambda_{SP}$ increases,
and is larger than 1 at $\lambda_{SP} \simeq 1$.
Thus, the solution of (\ref{eq:lsp}) corresponds to the case with 
${\rm LHS}={\rm RHS}\simeq 1/2$,
more or less independently of the specific values of $Q$ and $g_{NP}$, so
that we get $b_0 /b_{SP} \simeq b_0 \sqrt{2g_{NP}}$. 
This result depends only weakly on the nonperturbative parameter $g_{NP}$,
and it suggests that one may always use $b_0 /b_{SP} \simeq 1$ GeV
for the cases of our interest, in which $Q$ is on the order of several GeV and 
$g_{NP} \simeq 0.5$ GeV$^2$, as in Figs.~\ref{fig:2}-\ref{fig:5}.
The actual numerical solution of (\ref{eq:lsp}) verifies these 
simple considerations at the level of 20\% accuracy.
This fact is particularly helpful in the first attempt to compare 
(\ref{eq:attnll}) with the experimental data using (\ref{approx}),
so as to extract the transversity distributions.

The above considerations also demonstrate that (\ref{eq:tilde-I_Mellin3}) is dominated 
by the region of $b$ near $b_{SP} \simeq 1$ GeV $\sim 1/\Lambda_{\rm QCD}$,
corresponding to the boundary of perturbative and nonperturbative physics.
We can ``safely'' treat such a long-distance region, 
owing to the fact that the nonperturbative smearing (\ref{eq:np}) 
suppresses the too long-distance region,
$b \gg 1/\Lambda_{\rm QCD}$ [see the discussion below (\ref{eq:814})].
Moreover, the model dependence due to the specific choice of $g_{NP}$
largely cancels in the final result (\ref{eq:attnll}).

\begin{table}
\bc
\caption{The $Q_T \rightarrow 0$ limit of $\aqtn^{\rm NLL}(Q_T)$ in (\ref{asym:NLL})
for RHIC and J-PARC kinematics. NB is obtained using
the numerical $b$-integration of 
(\ref{tilde-X_Mellin}), (\ref{eq:tilde-I_Mellin}) and
 (\ref{resum:unpol}) with (\ref{eq:np}) and $g_{NP}=0.5$ GeV$^2$;
SP-I and SP-II are the results of the saddle-point formula 
(\ref{eq:attnll}) with (\ref{eq:lsp}) using the evolution operators 
from $Q$ to $b_0 /b_{SP}$, to NLL accuracy and to the customary 
NLO accuracy, respectively. }
\renewcommand{\arraystretch}{1.2}
\begin{tabular}{|c|r|r|r|r|r||r|r|r|}
\hline
& \multicolumn{5}{|c||}{$\sqrt{S}=200$ GeV, \hspace{0.1cm} $y=2$}&
\multicolumn{3}{|c|}{$\sqrt{S}=10$ GeV,  \hspace{0.1cm} $y=0$}\\
\hline
 $Q$  &  2GeV  &  5GeV  & 8GeV & 15GeV & 20GeV 
& 2GeV & 2.5GeV  & 3.5GeV \\
\hline
NB
& 3.8\% & 4.9\%  &  6.1\% & 8.2\% & 9.4\% 
& 13.4\% & 14.0\% & 14.9\% \\
SP-I
& 4.3\% & 5.4\% & 6.6\% & 8.7\% & 9.8\% 
& 14.1\% & 14.5\% & 14.8\%  \\
SP-II
& 7.3\% & 8.7\% & 9.8\% & 11.8\% & 12.7\%  
& 14.7\% & 14.8\% & 14.2\% \\
\hline
\end{tabular}
\label{tab:3}
\ec
\end{table}

In Table~\ref{tab:3}, ``NB'' lists $\aqtn^{\rm NLL}(Q_T =0)$ 
obtained from (\ref{asym:NLL}) using the numerical $b$-integration of 
(\ref{tilde-X_Mellin}), (\ref{eq:tilde-I_Mellin}) and (\ref{resum:unpol}),
with the kinematics of Fig.~\ref{fig:4}(a) for RHIC and of
Fig.~\ref{fig:5}(b) for J-PARC. 
The result for $Q=5$ GeV at RHIC and that for $Q=2$ GeV 
at J-PARC coincide with the $Q_T \rightarrow 0$ limit of 
the dot-dashed curve in Figs.~\ref{fig:3} and \ref{fig:5}(a), respectively.
The row labeled ``SP-I'' lists the results obtained using the saddle-point 
formula (\ref{eq:attnll}):
We use $b_0 /b_{SP}$ obtained as the solution of (\ref{eq:lsp}) with 
$g_{NP}=0.5$ GeV$^2$, and, according to (\ref{eq:speval}), (\ref{resum:3}) 
and similar formulae for the unpolarized case,
the parton distributions participating in (\ref{eq:attnll}) are obtained
by evolving the customary NLO distributions at the scale $Q$ to $b_0 /b_{SP}$ with 
the NLO evolution operators up to NLL accuracy, like (\ref{eq:RN}).
Here the ``customary NLO distributions'' are constructed as described 
above (\ref{eq:np}).
Comparing NB and SP-I, we observe the remarkable accuracy 
of our simple analytic formula (\ref{eq:attnll}) for both RHIC and
J-PARC, reproducing the results of NB to 10\% accuracy,
i.e., to the canonical size of ${\cal O}(\alpha_s)$ corrections associated with 
the NLL accuracy.\footnote{If we use the fixed value
$b_0 /b_{SP} = 1$ GeV for all cases, instead of the solution of (\ref{eq:lsp}), 
the results in SP-I change by at most 5\% for both RHIC and J-PARC kinematics.}
This supports the assumption concerning the order counting in our 
saddle-point evaluation, discussed below (\ref{eq:lsp}).

Although the compact formula (\ref{eq:attnll}) is reminiscent of 
$\aqtn^{\rm LL}(Q_T)$ in (\ref{asym:LL}), which retains only the LL level resummation,
or the $Q_T$-independent asymmetry of (\ref{eq:att}),
it is different in the scale of the parton distributions from those 
leading-order results, and the only effect of this new scale $b_0/b_{SP}$
allows SP-I to reproduce NB in Table~\ref{tab:3}.
Combined with this fact, the above derivation of (\ref{eq:attnll}), 
using (\ref{resum:4}) and (\ref{resumunpol:4}),
clearly demonstrates the characteristic features 
of the NLL soft gluon resummation effects on the asymmetries, 
which lead to the mechanism to ``enhance'' the dot-dashed curve 
in Figs.~\ref{fig:3} and \ref{fig:5}(a):
In the $Q_T \rightarrow 0$ limit, the all-order soft-gluon-resummation effects on 
the asymmetry largely cancel between the numerator and the denominator 
of (\ref{eq:attnll}), but certain contributions at the NLL level survive 
the cancellation and are entirely absorbed into the unconventional scale
$b_0/b_{SP}\simeq 1$ GeV for the relevant distribution functions.
Combined with the properties of $\aqtn^{\rm NLL}(Q_T)$ in (\ref{approx}),
this explains why $\aqtn^{\rm NLL}(Q_T)$ is always larger than 
$\aqtn^{\rm LL}(Q_T)$ in (\ref{asym:LL}) or $A_{TT}$ in (\ref{eq:att})
as the NLL-level effects associated with the evolution 
[compare with the discussion below (\ref{asym:LL})].

Using (\ref{approx}), the saddle-point formula (\ref{eq:attnll}) 
can be directly compared with the experimental value of the asymmetries $\aqt$, 
observed around the peak of the $Q_T$ spectrum of the corresponding DY cross sections.
This approach is particularly useful, because it does not 
require as large a numerical computations as that in \S\S9.1, 9.2.
Furthermore, the approach with (\ref{eq:attnll}) reqires
one point of caution about the accuracy, associated with the mismatch 
in classifying the terms between NLL and NLO. 
We recall that the parton distributions appearing
in (\ref{eq:attnll}) are the NLO distributions up to the corrections 
at the NNLL level [see (\ref{eq:speval}), (\ref{resum:3})].
To obtain SP-I in Table~\ref{tab:3}, the transversity distributions
appearing in the numerator of (\ref{eq:attnll}) were obtained 
by evolving the customary NLO transversity $\delta q(x, Q^2)$ 
at the scale $Q$, to the scale $b_0 /b_{SP}$ using (\ref{eq:RN}), 
which consists of {\it the NLO evolution operators up to the NNLL corrections}.
Therefore, the formula (\ref{eq:attnll}) can be safely used
in the region where the NNLL corrections are small;
the NNLL corrections at $Q_T \approx 0$ correspond to 
${\cal O}(\alpha_s(Q^2 ))$ effects, applying the order counting of 
\S\S6.3 and 8, and should be negligible in general. 
However, such a straightforward estimate might fail in the edge region 
of the phase space, as in the small $x$ region.
This follows from the fact that 
the relevant evolution operators (\ref{eq:RN}) actually coincide with 
the leading contributions in the large-logarithmic expansion of the usual
LO DGLAP evolution,\footnote{This fact also suggests that one may use the fixed value
$b_0 /b_{SP} \simeq 1$ GeV in (\ref{eq:attnll}) for all $Q$ (and $g_{NP}$)
rather than solving (\ref{eq:lsp}) numerically for each different 
input value of $Q$ and $g_{NP}$, because the sensitivity of the LO evolution 
with respect to a small change of the scale is weak.} 
(\ref{eq:attnll}) would not be accurate when the NLO corrections
in the usual DGLAP evolution are large compared with 
the contributions of (\ref{eq:RN}).
Such a situation would typically occur in a region with small $x_{1,2}^0$,
corresponding to the case with large $\sqrt{S}$.
In Table~\ref{tab:3}, we show other saddle-point results, ``SP-II'',
using the customary NLO distributions at the scale $b_0 /b_{SP}$
for the parton distributions participating in (\ref{eq:attnll}).
Thus, SP-I and SP-II differ by contributions at the NNLL level.
The results for SP-II indicate that the NNLL corrections are moderate 
for large $\sqrt{S}$ at RHIC, while they are expected to be small 
for small $\sqrt{S}$ at J-PARC.

We propose that our simple formula (\ref{eq:attnll}) is applicable to
analysis of low-energy experiments at J-PARC in order to extract 
the NLO transversity distributions directly from the data.  
On the other hand, (\ref{eq:attnll}) will not be very accurate for analyzing 
the data obtained from RHIC, but it will still be useful for obtaining 
the first estimate of the transversities. 
We emphasize that such a (moderate) uncertainty in applying our formula
(\ref{eq:attnll}) to the RHIC case is not caused by the saddle-point evaluation, 
nor by considering the $Q_T \rightarrow 0$ limit, but rather
is inherent in the general $Q_T$ resummation framework which, at the NLL
level, implies the use of the evolution operators
(\ref{eq:RN}) with the LO anomalous dimension.
A more accurate treatment of the small-$x$ region of the parton distributions 
relevant to the RHIC case would require the resummation formula 
to NNLL accuracy, where the NLO anomalous dimensions participate
in the evolution operators (\ref{eq:RN}) from $Q$ to $b_0 /b$ 
(see e.g. Ref.~\citen{BCFG}).

\section{Conclusions}

The era of \lq\lq testing QCD\rq\rq\ is clearly finished and today 
the improvement of the precision of predictions is one of the most 
important subjects in QCD.
We have reported our analysis of the transversely polarized DY process, 
which can be measured in ongoing experiments at RHIC and in future experiments 
at J-PARC, GSI, etc.
We have demonstrated that the QCD corrections relevant for the tDY
at a measured $Q_T$ are now under control over the entire range of $Q_T$, 
so that the dilepton $Q_T$ spectrum as well as the $Q_T$ dependence 
of the double-spin asymmetry for 
the tDY is predictable in a systematic QCD factorization framework.
In particular, in the small $Q_T$ region, where most of the 
dileptons are produced in experiments,
these observables are dominated by universal logarithmically-enhanced contributions, 
which we have taken into account by an all-order 
resummation of multiple soft-gluon emission.
Based on this framework, we have performed a detailed numerical study 
of the observables in the tDY 
at RHIC and J-PARC, using a model of the transversity which is constructed 
to satisfy a known theoretical constraint.
We find that the logarithmically-enhanced higher-order QCD corrections 
associated with soft gluon resummation drive a mechanism that amplifies 
the double transverse-spin asymmetries in the small $Q_T$ region.
Quantitatively, our results indicate that the double transverse-spin 
asymmetries seem to be small at RHIC, while they will be large at J-PARC.
This would imply that the measurement of the tDY double transverse-spin
asymmetries at RHIC might require considerable effort.
Still, we expect the data from RHIC, as well as that from J-PARC, GSI, etc.,
that are comparable with our predictions.
In particular, we have proposed a remarkably simple formula for the double 
transverse-spin asymmetries, which provides us with a new direct approach 
to extracting the transversity distributions from experimental data.
We believe that the results presented in this paper
will be useful not only for experiments at RHIC and J-PARC 
but also for future precision measurements.

We have obtained all our results concerning the QCD corrections for the DY 
in the $\overline{\rm MS}$ scheme, for the first time,
directly in the dimensional regularization with transverse polarization,
and our calculation technique can be straightforwardly applied to
other polarized and unpolarized processes.
As described in this paper,
the formalism of transverse-momentum resummation we employed is 
the same for many other polarized and unpolarized processes, and has a wide range
of application, in particular, to the processes observed at LHC, 
where rigorous treatments of multi-scale problems in QCD are mandatory and 
require resumming ubiquitous soft gluons.
QCD is now facing a new challenging era regarding its precision and applicability. 
We hope that various kinds of new experiments and theoretical investigations 
will be able to clarify not only perturbative and nonperturbative
aspects of QCD but also the full structure of all interactions in Nature.
%
\section*{Acknowledgments}
We thank Hirotaka Shimizu for collaboration in the early stage of 
this work and Werner Vogelsang, Hiroshi Yokoya and Stefano Catani 
for valuable discussions and comments. We also express our thanks to 
Masao Ninomiya for his encouragement to complete the paper.
We are sincerely grateful to Mieko Kodaira for her constant support 
and encouragement during more than ten years of our collaboration. 
This work was supported by the Grant-in-Aid for Scientific Research 
Nos. C-16540255, C-16540266 and B-19340063. 
\appendix
\section{The Transversity Distribution}

The transversity distribution of a quark inside a spin-$1/2$ 
hadron $h$ is defined in terms of a matrix element of a non-local 
light-cone operator as\cite{BDR:02}
\bea
\delta q_h(x,\mu^2)
=\int \frac{d\lambda}{4\pi} e^{i\lambda x}
\langle h(P,S)| \, \bar{\psi}(0) [0,\lambda n]\slashm{n}
\gamma_5\slashm{S} \psi(\lambda n)\, | h(P,S) \rangle \ ,       
\label{pdf}
\eea 
where $P$ and $S$ are the momentum and transverse spin of 
the hadron $h$, respectively, and $n^\mu$
is a light-like vector which satisfies $n^2 =1$ and $P\cdot n=1$; for example,
when we ignore the nucleon mass, setting $P^2 =0$, and work 
in a frame where $P^\mu =(P^+/\sqrt{2})(1,\boldsymbol{0},1)$, 
we have $S^\mu=(0,\boldsymbol{S},0)$ and $n^\mu= (1/\sqrt{2}P^+)(1,\boldsymbol{0},-1)$.
The path-ordered link operator (Wilson line)
\bea
[0,\lambda n]={\cal P}\exp\left(-ig\int_0^{\lambda}d\lambda^\prime
n \cdot A(\lambda^\prime n)\right)  ,
\eea
which connects the two quark operators $\bar{\psi}(0)$ and $\psi(\lambda n)$,
is required to ensure the gauge invariance of the non-local operator.
In the definition (\ref{pdf}), the scale $\mu^2$ of the distribution 
corresponds to the renormalization scale of the non-local operator on the RHS. 

The ``quark transversity-distribution inside a quark'' can be obtained by evaluating 
the matrix element (\ref{pdf}), with $h= $ quark $q'$, in perturbation theory
with an appropriate regularization of the IR divergence. 
Employing the dimensional regularization in $D=4-2\epsilon$ dimensions 
for the IR and UV divergences and evaluating the matrix element up to one-loop accuracy,
(\ref{pdf}) for $h=q'$ yields\footnote{A similar calculation for the unpolarized 
quark distribution $q_{q'}(x,\mu^2)$ is described in Ref.~\citen{CSS:89}.}
\bea
\delta q_{q'}(x,\mu^2)
=\delta_{q q'}\left[\delta(1-x)
-\frac{\alpha_s(\mu^2)}{2\pi}\frac{1}{\hat{\epsilon}} \Delta_TP_{qq}(x) \right],
\label{pdf_1-loop}
\eea
in the $\overline{\rm MS}$ scheme for renormalization of the non-local 
operator of (\ref{pdf}), where 
$1/\hat{\epsilon}\equiv 1/\epsilon -\gamma_{E} +\ln(4\pi )$,
and $\Delta_T P_{qq}(x)$ is the LO splitting function for the transversity:  
\be
    \Delta_T P_{qq} (x) = C_F \left\{ \frac{2 \,x}{(1-x)_+} +
           \frac{3}{2}\ \delta (1-x) \right\}. 
\label{eq:LOAP}
\ee
We note that we can derive the DGLAP evolution equation at the LO level
by differentiating (\ref{pdf_1-loop}) with respect to $\mu^2$. 
The result is [compare with (\ref{eq:DGLAP})]
\be
\mu^2 \frac{\partial}{\partial \mu^2}\delta q_h(x, \mu^2)=\frac{\alpha_s(\mu^2)}{2\pi}
 \int_{x}^1\frac{dz}{z} \Delta_TP_{qq}\left(\frac{x}{z}\right)
\delta q_h(z, \mu^2) .
\label{eq:DGLAPtr}
\ee

Together with the density distributions ($q_h(x,\mu^2)$, $g_h(x,\mu^2)$) 
and the helicity distributions ($\Delta q_h(x,\mu^2)$, $\Delta g_h(x,\mu^2)$), 
the transversity distribution $\delta q_h(x,\mu^2)$ forms a complete set of 
twist-2 parton distributions for a spin-$1/2$ hadron $h$. 
Here the quark distributions $q_h(x,\mu^2)$ and $\Delta q_h(x,\mu^2)$
are defined by the matrix element (\ref{pdf}) for the unpolarized and 
longitudinally-polarized hadron states, with the chiral-odd Dirac structure 
$\slashm{n}\gamma_5\slashm{S}$ replaced by the corresponding chiral-even 
structures $\slashm{n}$ and $\slashm{n}\gamma_5$, respectively.
Thus, unlike these two chiral-even quark distributions, the transversity is chiral-odd.
Because all gluon distributions at twist-2 are chiral-even, and the QCD interaction 
conserves chirality, the gluon distributions do not mix into the DGLAP evolution 
(\ref{eq:DGLAPtr}) for the transversity, even at NLO or higher-order level.
As a result, the QCD evolution of the transversity beyond the LO is of 
the non-singlet type; i.e., it involves only the mixing between the quark 
and antiquark distributions associated with the same flavor, and 
it is much simpler than the evolution for the other two quark distributions.  
For example, in the Mellin $N$-space, 
the NLO evolution of the transversity distribution is written as
\bea
\delta q^{\pm}_{N}(Q^2)&=&\left[\frac{\alpha_s(Q^2)}
{\alpha_s(Q_0^2)}\right]^{-\gamma_{qq,N}^{(0)}/(2\pi\beta_0)}
 \non\\
&&\times\left[1+\frac{\alpha_s(Q_0^2)-\alpha_s(Q^2)}{4\pi^2 \beta_0}
\left(\gamma_{qq,N}^{(1)\pm}
-\frac{2\pi\beta_1}{\beta_0} \gamma_{qq,N}^{(0)}\right)\right]
\delta q^{\pm}_N(Q_0^2) \ ,
\label{DGLAP}
\eea
where $\delta q^\pm_N(\mu^2)\equiv \int_0^1 dxx^{N-1}\delta q^\pm(x,\mu^2)$ with
$\delta q^{\pm}(x,\mu^2)\equiv \delta q_h(x,\mu^2)\pm \delta \bar{q}_h(x,\mu^2)$.
The LO anomalous dimension is given by~\cite{AM}
\bea
\gamma_{qq,N}^{(0)}= \int_0^1 dxx^{N-1}\Delta_T P_{qq} (x) 
= C_F\left[-2S_1(N)+\frac{3}{2}\right]\ , \label{eq:P0N}
\eea
and the NLO anomalous dimensions in the $\overline{\rm MS}$ scheme are~\cite{HKKKM,V}
\bea
\gamma_{qq,N}^{(1)\eta}&&
=C_F^2\left[\frac{3}{8}+\frac{1-\eta}{N(N+1)}-3S_2(N)
-4S_1(N)\left(S_2(N)-S^\prime_2\left(\frac{N}{2}\right)\right)
\right.\nonumber\\
&& \;\;\;\;\; \;\;\;\;\;  
\left.-8\widetilde{S}(N)+S^\prime_3\left(\frac{N}{2}\right)\right]
\nonumber\\
&&+\frac{1}{2}C_FN_c\left[\frac{17}{12}-\frac{1-\eta}{N(N+1)}-\frac{134}{9}S_1(N)
+\frac{22}{3}S_2(N)\right.\nonumber\\&&\left.\hspace{3cm}
+4S_1(N)\left(2S_2(N)-S^\prime_2\left(\frac{N}{2}\right)\right)
+8\widetilde{S}(N)-S^\prime_3\left(\frac{N}{2}\right)  \right]
\nonumber\\
&&+\frac{2}{3}C_FN_fT_R\left[-\frac{1}{4}+\frac{10}{3}S_1(N)-2S_2(N)\right]  ,
\label{P1N} 
\eea
where $\eta=\pm$, and the harmonic sums are defined by
\bea
S_k(N)&=&\sum_{j=1}^N\frac{1}{j^k} \ ,\\
S^\prime_k\left(\frac{N}{2}\right)&=&2^{k-1}\sum_{j=1}^N\frac{1+(-1)^j}{j^k}
\ ,\\
\widetilde{S}(N)&=&\sum_{j=1}^N\frac{(-1)^j}{j^2}S_1(j) \ .
\eea
The first moment of the LO splitting function is negative
($\gamma_{qq,N=1}^{(0)}=-\frac{2}{3}$), so that
the transverse polarization in total (the ``tensor charge'') decreases 
as the scale increases.\footnote{A pedagogical 
explanation of the transverse-spin flip by collinear radiation 
is given in the appendix of Ref.~\citen{DS04}.} 

We note that the dominant large-$z$ behavior (\ref{eq:z1}) of 
the DGLAP splitting functions (\ref{eq:DGLAPk}) reflects the dominant 
large $N$ behavior of the corresponding anomalous dimensions (\ref{eq:Mellin}). 
For the present case, the anomalous dimensions given in 
(\ref{eq:P0N}) and (\ref{P1N}) at large $N$ behave as
\be
\gamma_{qq,N}^{(0)} 
= -A_q^{(1)}\left (\ln N+\gamma_E \right) -\frac{B_q^{(1)}}{2} +\cdots ,
\label{eq:A12A11}
\ee
where the ellipses stand for the terms that vanish as 
$N\rightarrow \infty$,  and 
\be
\gamma_{qq,N}^{(1)\eta} = -A_q^{(2)}\left (\ln N+\gamma_E \right) 
+\cdots ,
\label{eq:A12A12}
\ee
where the ellipses stand for the corrections of ${\cal O}(1)$.
The coefficients on the RHS of (\ref{eq:A12A11}) and (\ref{eq:A12A12}) 
coincide with (\ref{coeff_NLL}).
The same asymptotic behavior arises in the anomalous dimensions for 
the density and helicity distributions $q_h(x,\mu^2)$
and $\Delta q_h(x,\mu^2)$ (see e.g. Ref.~\citen{Korch}).

\section{The Cross Section Integrated over the Rapidity $y$}

Integrating (\ref{owari}) over the rapidity $y$ of the dileption in the final state,
we obtain the rapidity-integrated cross section in the $\overline{\rm MS}$ 
factorization scheme, 
\be
   \frac{\Delta_T d \sigma}{d Q^2 d Q_T^2 d \phi_{k_1}}
    = \mN \cos (2 \phi_{k_1} - \phi_1 - \phi_2)
     \int dy  \left[ \Delta_T X (Q_T^2 , Q^2 , y)
             + \Delta_TY (Q_T^2 , Q^2 , y) \right] , \label{owariowari} 
\ee
including the ${\cal O}(\alpha_s)$ QCD mechanism.
We use the following integration formulae:
\bea
   \int d y \, \delta H (x_1^0 , x_2^0)
    &=& \int d x_1 d x_2 \, \delta H (x_1 , x_2) \, 
   \delta (x_1 x_2 - \tau)  , \label{eq:B2}\\
   \int d y\, \int_{x_1^0}^1\, \frac{d z}{z} \Delta_T P_{qq} (z)\, 
        \delta H \left( \frac{x_1^0}{z}\,,\,x_2^0 \right)
      &=& \int \frac{d x_1 d x_2}{x_1 x_2} \, \Delta_T P_{qq} (\tau_{12})
         \, \delta H (x_1 , x_2) \, \theta (x_1 x_2 - \tau) ,\nonumber\\
\\
   \int d y \, \int_{\sqrt{\tau_+} e^y}^{1}\,
         \frac{d x_1}{x_1 - x_1^+}\, 
           \frac{\tau \, \delta H (x_1 , x_2^*)}{x_1 x_2^*}
      &=& \int \frac{d x_1 d x_2}{x_1 x_2}\,
        \frac{\tau_{12}\, \delta H (x_1 , x_2)}
                  {\sqrt{(1 - \tau_{12})^2 - 4 \rho_{12}}}\,
           \theta (x_1 x_2 - \tau_+) ,\nonumber\\
\eea
and
\bea
   \lefteqn{\int d y \, 
      \left( \int d x_1\, \delta H_1 + \delta H (x_1^0 , x_2^0 ) 
           \, \ln \frac{1 - x_1^+}{1 - x_1^0} \right)}\nonumber\\
      && \quad = \,  
       \int_{\tau_+}^1\, \frac{d x_1}{x_1}\, \delta H (x_1 , \tau / x_1)\,
      \ln \frac{\sqrt{(x_1 - \tau_+)(x_1 - \tau_-)} + x_1 - \tau -  2 \rho} 
          {2\, \sqrt{\rho ( \tau + \rho)}} \non\\
      && \hspace{5cm}
           - \, \int_{\tau}^1\, \frac{d x_1}{x_1}\,
    \delta H (x_1 , \tau / x_1)\, \ln \frac{x_1 - \tau}{\sqrt{\rho \tau}}\nonumber\\
      && \quad\quad\quad + \, \int \, \frac{d x_1 d x_2}{x_1 x_2}\,
        \frac{\tau_{12}\, \delta H (x_1 , x_2) - \delta H (x_1 , \tau / x_1)}
               {\sqrt{( 1 - \tau_{12})^2 - 4 \rho_{12}}}\,
           \theta ( x_1 x_2 - \tau_+ )\nonumber\\
      && \hspace{3.5cm}
          - \, \int \, \frac{d x_1 d x_2}{x_1 x_2}\,
        \frac{\tau_{12}\, \delta H (x_1 , x_2) - \delta H (x_1 , \tau / x_1)}
               {1 - \tau_{12}}\,
           \theta ( x_1 x_2 - \tau )  ,
\non\\\label{eq:B5}
\eea
where $\delta H (x_1 , x_2)$ denotes the product of the transversity distributions, 
(\ref{tPDF}), suppressing the scale dependence, $\delta H_1$ is defined 
in (\ref{eq:H1H1}), $\tau$ is given in (\ref{eq:hadronvar}), and we have
\bea
\rho = \frac{Q_T^2}{S}\ , \;\; \;\;\; \;
      \sqrt{\tau_{\pm}} = \sqrt{\tau + \rho} \pm \sqrt{\rho}\ ,\; \;\;\;\; \;
      \tau_{12} = \frac{\tau}{x_1 x_2}\ ,\; \;\;\;\; \;
      \rho_{12} = \frac{\rho}{x_1 x_2}  .
\eea
We thus get
\bea
\int dy  \Delta_T&& X (Q_T^2 , Q^2 , y)
=\int \frac{dx_1}{x_1}\delta H(x_1,\tau/x_1;\mu_F^2)
\label{eq:B7}\\
&&\;\;\;\;\;\;\;\; \times
	 \left\{\delta(Q_T^2)+
           \frac{\alpha_s}{2\pi}C_F\left[(\pi^2-8)\delta(Q_T^2)
         +2\left(\frac{\ln(Q^2/Q_T^2)}{Q_T^2}\right)_+-\frac{3}{(Q_T^2)_+}\right]\right\}
       \nonumber \\
   &&  +\frac{\alpha_s}{\pi}\left(\frac{1}{(Q_T)_+}
                                    +\delta(Q_T^2)\ln{\frac{Q^2}{\mu_F^2}}\right)
      \int_\tau^1\frac{dx_1}{x_1}\int_{\tau/x_1}^1\frac{dx_2}{x_2}
             \Delta_TP_{qq}(\tau_{12})\delta H(x_1,x_2;\mu_F^2) 
\nonumber
\eea
and
\bea
\int dy && \Delta_T Y (Q_T^2 , Q^2 , y)
\nonumber\\
=&&\frac{\alpha_s}{\pi}C_F\left\{
\left[\frac{4}{Q^2}-\frac{6}{Q_T^2}\ln\left(1+\frac{Q_T^2}{Q^2}\right)\right]\right.
\int_{\tau_+}^1\frac{dx_1}{x_1}\int_{\tau_+/x_1}^1\frac{dx_2}{x_2}
\frac{\tau_{12}\delta H(x_1,x_2;\mu_F^2)}{\sqrt{(1-\tau_{12})^2-4\rho_{12}}}\nonumber\\
&&+\frac{2}{Q_T^2}\left[
\int_{\tau_+}^1\frac{dx_1}{x_1}\int_{\tau_+/x_1}^1\frac{dx_2}{x_2}
\frac{\tau_{12}\delta H(x_1,x_2;\mu_F^2)-\delta H(x_1,\tau/x_1;\mu_F^2)}
{\sqrt{(1-\tau_{12})^2-4\rho_{12}}}\right. \label{eq:B8}\\
&&\hspace{2cm}
-\int_{\tau}^1\frac{dx_1}{x_1}\int_{\tau/x_1}^1\frac{dx_2}{x_2}
\frac{\tau_{12}\delta H(x_1,x_2;\mu_F^2)-\delta H(x_1,\tau/x_1;\mu_F^2)}
{1-\tau_{12}}\nonumber\\
&&\hspace{0cm}
+\int_{\tau_+}^1\frac{dx_1}{x_1}\delta H(x_1,\tau/x_1;\mu_F^2)
\ln{\frac{x_1-\tau-2\rho+\sqrt{(x_1-\tau_+)(x_1-\tau_-)}}
{2\sqrt{\rho(\tau+\rho)}}}\nonumber\\
&&\hspace{4cm}\left.
-\int_{\tau}^1\frac{dx_1}{x_1}\delta H(x_1,\tau/x_1;\mu_F^2)
\ln{\frac{x_1-\tau}
{\sqrt{\rho\tau}}}
\right\}  .
\nonumber						   
\eea
For $Q_T^2 >0$, the delta function $\delta(Q_T^2 )$ involved in (\ref{eq:B7})
vanishes, and, after some calculation, (\ref{owariowari}) with
(\ref{eq:B7}) and (\ref{eq:B8}) reduces to
\bea
  \frac{\Delta_T d \sigma}{d Q^2 d Q_T^2 d \phi_{k_1}}
   && = \mN \cos (2 \phi_{k_1} - \phi_1 - \phi_2)\frac{\alpha_s}{\pi}C_F
    \left[\frac{2}{Q_T^2}+\frac{4}{Q^2}-  \frac{6}{Q_T^2} 
       \ln \left( 1 + \frac{Q_T^2}{Q^2} \right)  \right] \nonumber\\
&& \;\;\;\;\;\;\;\; \times
\int_{\tau_+}^1\frac{dx_1}{x_1}\int_{\tau_+/x_1}^1\frac{dx_2}{x_2}
\frac{\tau_{12}\delta H(x_1,x_2;\mu_F^2)}
{\sqrt{(1-\tau_{12})^2-4\rho_{12}}}.
\label{owariowari2} 
\eea
This coincides with the result obtained by imposing the restriction 
$Q_T^2 >0$ in Ref.~\citen{VW},
and it provides the LO QCD prediction in the large $Q_T$ region.

To obtain the QCD prediction for the rapidity-integrated tDY cross section 
over the entire region of $Q_T$, we integrate (\ref{eq:NLL+LO}) over $y$. 
This gives the rapidity-integrated NLL+LO cross section in 
the $\overline{\rm MS}$ scheme.
The $y$-integral of $\Delta_T\tilde{X}^{\rm NLL} (Q_T^2, Q^2,y)$, as well as 
$\Delta_T\tilde{Y} (Q_T^2, Q^2, y)$, can be performed straightforwardly 
using (\ref{eq:B2})-(\ref{eq:B5}), similarly to the above fixed-order case. 
For $\Delta_T\tilde{X}^{\rm NLL}\, (Q_T^2, Q^2,y)$, the result can be expressed 
in the Mellin moment space with respect to $\tau=x_1^0 x_2^0$ at fixed $Q$:
Changing the integration variables from $x_1^0$ and $x_2^0$ to $\tau$ and $y$
in the definition of the double Mellin moment (\ref{tilde-X_Mellin})
[see (\ref{eq:dMel})],
we obtain 
$\int dx_1^0  dx_2^0(x_1^0)^{N_1-1} (x_2^0)^{N_2-1}
\Delta_T \tilde{X}^{\rm NLL}(Q_T^2,Q^2,y)
=\int d\tau dy\ \tau^{(N_1+N_2)/2-1}  e^{(N_1 - N_2)y}
\Delta_T \tilde{X}^{\rm NLL}(Q_T^2,Q^2,y)$.
Setting $N_1=N_2$ in this relation, we get the Mellin-space representation of
the corresponding $y$-integral as
\bea
\int_0^1 d\tau \tau^{N_1-1}&& 
\left[\int dy \Delta_T\tilde{X}^{\rm NLL}\, (Q_T^2, Q^2,y)\right]
\nonumber\\
&& \;\;\;
=\left[1+\frac{\alpha_s(Q^2)}{2\pi}C_F(\pi^2-8)\right]
\delta H_{N_1,N_1}(Q^2) \tilde{I}_{N_1,N_1}(Q_T^2,Q^2) ,
\eea
with $\tilde{I}_{N_1,N_1}(Q_T^2,Q^2)$ of (\ref{eq:tilde-I_Mellin}).

\section{Cancellation at $Q_T=0$ in Eq.~(\ref{eq:tilde-Y})}

In this appendix, we show that the terms proportional to the delta 
function $\delta(Q_T^2)$ cancel out in the difference 
$\Delta_TX-\tilde{\Delta_TX^{\rm NLL}}|_{\rm fo}$ in (\ref{eq:tilde-Y}),
and as a result, $\Delta_T\tilde{Y}$ differs from $\Delta_TY$
only by the terms that are less singular than $1/Q_T^2$ or $\delta(Q_T^2)$ 
as $Q_T^2 \rightarrow 0$.
This shows that the NLL+LO cross section (\ref{eq:NLL+LO}) is smooth and
well-defined in the entire region of $Q_T$, including $Q_T=0$. 
For this purpose, we explicitly show that [see (\ref{subtract})]
the relation
\bea
\lim_{\epsilon\ra 0}\int_0^{\epsilon^2}dQ_T^2 (\mI_n-\tilde{\mI}_n)=
\lim_{\epsilon\ra 0}\left(
-\int_{\epsilon^2}^{Q^2}dQ_T \mI_n + \int_{\epsilon^2}^\infty dQ_T
\tilde{\mI}_n \right) =0 
\label{eq:int-I:2}
\eea
holds for $n=1$ and $2$, which means that the delta function 
$\delta(Q_T^2)$ implicit in the + distributions in $\mI_{1,2}$ cancels with 
that in $\tilde{\mI}_{1,2}$.

The first equality in (\ref{eq:int-I:2}) follows from (\ref{eq:int-I}).
To show the second equality, 
it is convenient to consider the generating functions of $\mI_n$ and $\tilde{\mI}_n$.
The explicit form of the former is already given in (\ref{eq:genIn}) as
\bea
\mI(\eta)&\equiv&\sum_{n=0}^{\infty}\frac{\mI_n}{n!}\eta^n
\non\\\hspace{1cm}
&=& \left[\delta (Q_T^2) - \sum_{m=0}^\infty
\frac{\eta^{m+1}}{m!}\left( \frac{\ln^m (Q^2 / Q_T^2)}{Q_T^2} \right)_+ \right]
\exp\left[-2
\sum_{r=1}^{\infty}\frac{\zeta(2r+1)}{2r+1}\eta^{2r+1}\right] ,
\non\\
\label{eq:Cnew0}
\eea 
which can be obtained from the relations
\be
\mI(\eta)=
\frac{1}{2}\int_0^\infty db b J_{0}(bQ_T) \left[\frac{b^2 Q^2}{b_0^2}\right]^\eta
=\frac{\Gamma(1+\eta)}{\Gamma(-\eta)}
\frac{1}{Q_T^2}\left[\frac{4 Q^2}{b_0^2Q_T^2}\right]^\eta 
\label{eq:Cnew1}
\ee
and $\Gamma(1+\eta)/\Gamma(1-\eta) = (b_0^2/4)^\eta\exp\left[-2
\sum_{r=1}^{\infty}\frac{\zeta(2r+1)}{2r+1}\eta^{2r+1}\right]$.
The $Q_T$ integration of (\ref{eq:Cnew0}) reads
\bea
\int_{\epsilon^2}^{Q^2} dQ_T^2 ~\mI(\eta) = 
\left\{1-\left(\frac{Q^2}{\epsilon^2}\right)^\eta \right\}
\exp\left[-2
\sum_{r=1}^{\infty}\frac{\zeta(2r+1)}{2r+1}\eta^{2r+1}\right] \ .
\label{eq:int-Igen}
\eea
Next, the generating function of $\tilde{\mI}_n$, 
\be
\tilde{\mI}(\eta)\equiv \sum_{n=0}^{\infty}\frac{\tilde{\mI}_n}{n!}\eta^n ,
\ee
is given by~\cite{BCFG} 
\bea
\tilde{\mI}(\eta)=
\frac{1}{2}\int_0^\infty db b J_{0}(bQ_T) \left[\frac{b^2 Q^2}{b_0^2}+1\right]^\eta
=
\frac{ b_0^2}{Q^2}\frac{2^\eta}{\Gamma(-\eta)}
\frac{K_{1+\eta}(b_0Q_T/Q)}{(b_0Q_T/Q)^{1+\eta}} \ ,
\label{eq:Cnew2}
\eea
and the $Q_T$ integration of this function is obtained as 
\bea
\int_{\epsilon^2}^{\infty} dQ_T^2 \tilde{\mI}(\eta) &=& 
\frac{2^{1+\eta}}{\Gamma(-\eta)}
\frac{K_{\eta}(b_0\epsilon/Q)}{(b_0\epsilon/Q)^{\eta}} \ .
\eea
Expanding $K_{\eta}(b_0\epsilon/Q)$ about $\epsilon=0$, we can derive
\bea
\int_{\epsilon^2}^{\infty} dQ_T^2 \tilde{\mI}(\eta) &=&
\left\{
1-\left(\frac{Q^2}{\epsilon^2}\right)^\eta 
\exp\left[-2
\sum_{r=1}^{\infty}\frac{\zeta(2r+1)}{2r+1}\eta^{2r+1}\right] 
\right\}\left[1
+ {\cal O}(\epsilon^2)\right] .
\label{eq:int-tilde-Igen}
\eea
Taylor expanding (\ref{eq:int-Igen}) and (\ref{eq:int-tilde-Igen}) about
$\eta=0$ and comparing the results, we see that (\ref{eq:int-I:2}) is valid
for $n=1$ and $2$. For $n\geq 3$, (\ref{eq:int-I:2}) is not valid,
because terms including the zeta function $\zeta(n)$ appear differently 
in (\ref{eq:int-Igen}) and (\ref{eq:int-tilde-Igen}).



\begin{thebibliography}{99}
\bibitem{CSS:89}
    For a general review of QCD factorization see, for example,
    J.~C.~Collins, D.~Soper and G.~Sterman, 
    Factorization of hard processes 
    in QCD, in {\it Perturbative Quantum Chromodynamics} (World Scientific,
    Singapore, 1989) ed. A.~H.~Mueller, p.~1.\\
    G.~Sterman, Partons, Factorization and Resummation,
    in {\it QCD \& Beyond} (World Scientific,
    Singapore, 1996) ed. D.~E.~Soper, p.~327.
\bibitem{Alt94}
    G. Altarelli, {\it The Development of Perturbative QCD} 
       (World Scientific, Singapore, 1994).\\
    R.~Brock et al.  (CTEQ Collaboration),
    Rev.\ Mod.\ Phys.\ \andvol{67,1995,157}.
\bibitem{Bass}
    S.~D.~Bass, Rev.~Mod.~Phys.~\andvol{77,2005,1257}.
\bibitem{WV07}
    W.~Vogelsang, J.~of.~Phys.~G~\andvol{34,2007,S149}.
\bibitem{PHENIX}
    S.~S.~Adler et al. (PHENIX Collaboration),
     \PRD{73, 2006,091102};
    \PRL{95,2005,202001}.\\
    K.~N.~Barish (PHENIX Collaboration), in Proceedings of 
    17th International Spin Physics Symposium (SPIN2006),
    Kyoto, Japan, AIP Conf.\ Proc.\ \andvol{915,2007,301}. 
\bibitem{STAR}
    B.~I.~Abelev et al. (STAR Collaboration), 
    \PRL{97,2006,252001}.\\
    B.~Surrow (STAR Collaboration), in Proceedings of 
    17th International Spin Physics Symposium,
    Kyoto, Japan, AIP Conf.\ Proc.\ \andvol{915,2007,293}. 
\bibitem{RHIC}
    G.~Bunce, N.~Saito, J.~Soffer and W.~Vogelsang, 
    Ann.~Rev.~Nucl.~Part.~Sci.\ \textbf{50} (2000), 525.
\bibitem{HERMES}
    A.~Airapetian et al.(HERMES Collaboration), 
    \PRL{94,2005,012002}.\\
    D.~Hasch (HERMES Collaboration), in Proceedings of 
    17th International Spin Physics Symposium,
    Kyoto, Japan, AIP Conf.\ Proc.\ \andvol{915,2007,307}. 
\bibitem{COMPASS}
    E.~S.~Ageev et al. (COMPASS Collaboration),
    \NPB{765,2007,31}.\\
    A.~Magnon (COMPASS Collaboration), in Proceedings of 
    17th International Spin Physics Symposium,
    Kyoto, Japan, AIP Conf.\ Proc.\ \andvol{915,2007,287}. 
\bibitem{Dutta}
    D.~Dutta et al., Letter of Intent (L15) for Nuclear and Particle Physics 
    Experiments at J-PARC, 
http://www-ps.kek.jp/jhf-np/LOIlist/LOIlist.html.
\bibitem{PAX}
    V.~Barone, et al. (PAX Collabolation), hep-ex/0505054.\\
    M.~Maggiora et al.(ASSIA Collabolation), hep-ex/0504011. 
\bibitem{RS}
    J.~P.~Ralston and D.~E.~Soper, \NPB{152,1979,109}.
\bibitem{JJ:92}
    J.~L.~Cortes, B.~Pire and J.~P.~Ralston,
    Z.\ Phys.\ C \andvol{55,1992,409}.\\
    R.~L.~Jaffe and X.~D.~Ji, \NPB{375,1992,527}.
\bibitem{BDR:02}
    V.~Barone, A.~Drago and P.~G.~Ratcliffe, \PRP{359,2002,1}.\\
    J.~Kodaira and K.~Tanaka, \PTP{101,1999,191}.
\bibitem{KLN}
    T.~Kinoshita,
\JMP{3,1962,650}.\\
    T.~D.~Lee and M.~Nauenberg,
    Phys.\ Rev.\ \andvol{133,1964,B1549}.
\bibitem{VW}
    W.~Vogelsang and A.~Weber, \PRD{48,1993,2073}.
\bibitem{CKM}
    A.~P.~Contogouris, B.~Kamal and Z.~Merebashvili,
    \PLB{337,1994,169}.
\bibitem{K1}
    B.~Kamal, \PRD{53,1996,1142}.
\bibitem{V}
    W.~Vogelsang, \PRD{57,1998,1886}.
\bibitem{MSSV}
    O.~Martin, A.~Sch\"afer, M.~Stratmann and W.~Vogelsang,
        \PRD{57,1998,3084}; ibid. \andvol{60,1999,117502}.
\bibitem{K2}
    B.~Kamal, hep-ph/9807217.
\bibitem{MSV}
    A.~Mukherjee, M.~Stratmann and W.~Vogelsang, \PRD{67,2003,114006}.
\bibitem{KKST06}
    H.~Kawamura, J.~Kodaira, H.~Shimizu and K.~Tanaka, \PTP{115,2006,667}.
\bibitem{DDT}
    Y.~L.~Dokshitzer, D.~I.~Dyakonov and S.~I.~Troyan, \PRP{58,1980,269}.
\bibitem{Sudakov}
    V.~V.~Sudakov, Sov.~Phys.~JETP \andvol{3,1956,65}.
\bibitem{PP}
    G.~Parisi and R.~Petronzio,
         \NPB{154,1979,427}.\\
    See also G.~Curci, M.~Greco and Y.~Srivastava,
         \NPB{159,1979,451}.
\bibitem{ESRW}
    S.~D.~Ellis and W.~J.~Stirling,
         \PRD{23,1981,214}.\\
    S.~D.~Ellis, N.~Fleishon  and W.~J.~Stirling,
         \PRD{24,1981,1386}.\\
    P.~E.~L.~Rakow and B.~R.~Webber,
         \NPB{187,1981,254}.
\bibitem{CS}
    J.~C.~Collins and D.~E.~Soper,
         \NPB{193,1981,381}.
\bibitem{KT}
    J.~Kodaira and L.~Trentadue,
         \PLB{112,1982,66}; Report \textbf{SLAC-Pub-2934} (1982); \PLB{123,1983,335}.
\bibitem{DWS}
    C.~T.~H.~Davies and W.~J.~Stirling,
         \NPB{244,1984,337}.\\
    C.~T.~H.~Davies, B.~R.~Webber and W.~J.~Stirling,
         \NPB{256,1985,413}.
\bibitem{AEGM}
    G.~Altarelli, R.~K.~Ellis, M.~Greco and G.~Martinelli,
         \NPB{246,1984,12}.
\bibitem{CSS}
    J.~C.~Collins and D.~E.~Soper,
         \NPB{197,1982,446}.\\
    J.~C.~Collins, D.~E.~Soper and G.~Sterman,
         \NPB{250,1985,199}.
\bibitem{W}
    A.~Weber, \NPB{382,1992,63}.
\bibitem{dG}
    D.~de~Florian and M.~Grazzini,
         \PRL{85,2000,4678}; \NPB{616,2001,247}.
\bibitem{BCFG}
    G.~Bozzi, S.~Catani, D.~de~Florian and M.~Grazzini,
         \PLB{564,2003,65}; \NPB{737,2006,73}; arXiv:~0705.3887.
\bibitem{KS}
    A.~Kulesza and W.~J.~Stirling,
         \JHEP{12,2003,056}.
\bibitem{ResBos}
    C.~Balazs and C.~P.~Yuan, \PRD{56,1997,5558}.\\
    C.~Balazs and C.~P.~Yuan, \PLB{478,2000,192}.\\
    F.~Landry, R.~Brock, G.~Ladinsky and C.~P.~Yuan, \PRD{63,2001,013004}.
\bibitem{Boer}
    D.~Boer,
         \PRD{62,2000,094029}.
\bibitem{KNV:06}
    Y.~Koike, J.~Nagashima and W.~Vogelsang,
          \NPB{744,2006,59}.
\bibitem{AEM}
    G.~Altarelli, R.~K.~Ellis and G.~Martinelli, \NPB{157,1979,461}.
\bibitem{THRESH}
    G.~Sterman,
          \NPB{281,1987,310}.\\
    S.~Catani and L.~Trentadue,
          \NPB{327,1989,323}; \NPB{353,1991,183}.
\bibitem{THRESH2}
    G.~Sterman and W.~Vogelsang,
          \JHEP{02,2001,016}.\\
    T.~O.~Eynck, E.~Laenen and L.~Magnea,
          \JHEP{06,2003,057}.
\bibitem{F:89}
    R.~D.~Field,
    {\it Applications of Perturbative QCD}, Frontiers in physics 77
    (Addison-Wesley, New York, 1989). 
\bibitem{CdG}
    S.~Catani, D.~de~Florian and M.~Grazzini,
         \NPB{596,2001,299}.
\bibitem{KLSV}
    E.~Laenen, G.~Sterman and W.~Vogelsang,
         \PRL{84,2000,4296}.\\
    A.~Kulesza, G.~Sterman and W.~Vogelsang,
         \PRD{66,2002,014011}.
\bibitem{Li}
    H.~-N.~Li, \PLB{454,1999,328}.
\bibitem{KSVKSV}
    A.~Kulesza, G.~Sterman and W.~Vogelsang,
         \PRD{69,2004,014012}.
\bibitem{Collins89}
    J.~C.~Collins, Sudakov form factors, in {\it Perturbative Quantum
	Chromodynamics} (World Scientific, Singapore, 1989) ed. A.~H.~Mueller, p.~573,
        and references therein.
\bibitem{CDT}
    S.~Catani, E.~D'Emilio and L.~Trentadue, \PLB{211,1988,335}.
\bibitem{MVV} 
    S.~Moch, J.~A.~M.~Vermaseren and A.~Vogt, \NPB{688,2004,101};\NPB{691,2004,129}.
\bibitem{EV}
    R.~K.~Ellis and S.~Veseli, \NPB{511,1998,649}.\\
    S.~Frixione, P.~Nason and G.~Ridolfi, \NPB{542,1999,311}.\\
    A.~Kulesza and W.~J.~Stirling, \NPB{555,1999,274}.
\bibitem{QZ}
    J.~W.~Qiu and X.~F.~Zhang,  \PRD{63,2001,114011}.
\bibitem{CMNT}
    S.~Catani, M.~L.~Mangano, P.~Nason and L.~Trentadue,
          \NPB{478,1996,273}.
\bibitem{ERV}
    R.~K.~Ellis, D.~A.~Ross and S.~Veseli, \NPB{503,1997,309}.
\bibitem{CTTW}
    S.~Catani, L.~Trentadue, G.~Turnock and B.~R.~Webber,
          \NPB{407,1993,3}.
\bibitem{KKT07}
    H.~Kawamura, J.~Kodaira and K.~Tanaka,
          \NPB{777,2007,203}.
\bibitem{FP}
    W.~Furmanski and R.~Petronzio, Z.~Phys.~C~\andvol{11,1982,293}.
\bibitem{BV}
    J.~Bl\"umlein and A.~Vogt, \PRD{58,1998,014020}.
\bibitem{AK}
    P.~B.~Arnord and R.~P.~Kauffman, \NPB{349,1991,381}.
\bibitem{Anselmino}
    M.~Anselmino, M.~Boglione, U.~D'Alesio, A.~Kotzinian, 
    F.~Murgia, A.~Prokudin and C.~Turk,
          \PRD{75,2007,054032}.
\bibitem{QCDSF}
    M.~G\"ockeler, et al., \PLB{627,2005,113}. \\
    M.~Diehl et al. (QCDSF Collaborations), hep-ph/0511032.
\bibitem{Wakamatsu}
    M.~Wakamatsu, arXiv:~0705.2917.
\bibitem{S}
    J.~Soffer, \PRL{74,1995,1291}. 
\bibitem{GRV98}
    M.~Gl\"uck, E.~Reya and A.~Vogt, Eur.~Phys.~J. C \andvol{5,1998,461}. 
\bibitem{GRSV00}
    M.~Gl\"uck, E.~Reya, M.~Stratmann and W.~Vogelsang, 
         \PRD{63,2001,094005}.
\bibitem{CDL06}
    M.~Contalbrigo, A.~Drago and P.~Lenisa, hep-ph/0607143 in 
        XIV International Workshop on Deep Inelastic Scattering (DIS2006),
        Tsukuba, Japan. 
\bibitem{BCCGR:06}
    V.~Barone, A.~Caferella, C~.Coriano, M.~Guzzi and P.~G.~Ratcliffe, 
        \PLB{639,2006,483}.
\bibitem{SSVY:05}
    H.~Shimizu, G.~Sterman, W.~Vogelsang and H.~Yokoya, 
        \PRD{71,2005,114007}.
\bibitem{KKT07-2}
    H.~Kawamaura, J.~Kodaira and K.~Tanaka, in preparation.
\bibitem{AM}
    X.~Artru and M.~Mekhfi, Z.\ Phys.\ C \andvol{45,1990,669}
\bibitem{HKKKM}
    A.~Hayashigaki, Y.~Kanazawa and Y.~Koike, \PRD{56,1997,7350}.\\
    S.~Kumano and M.~Miyama, \PRD{56,1997,2504}. 
\bibitem{DS04}
    L.~Dixon and M.~Schreiber, 
          \PRD{69,2004,113001}.  
\bibitem{Korch}
    G.~P.~Korchemsky,
           Mod.\ Phys.\ Lett.\ A \andvol{4,1989,1257}.
\end{thebibliography}
\end{document}